\documentclass[twocolumn]{aastex62}

\turnoffedit

\renewcommand{\bfseries}{}
\definecolor{magenta}{gray}{0}

\usepackage{amsmath}
\usepackage[all]{hypcap}

\graphicspath{{figures/}}
\hypersetup{
    pdfcreator={Naor Movshovitz},
    pdfproducer={}
}

\usepackage{silence}
\ActivateWarningFilters[pdftoc,rerun,labels,sfloat,dfloat]

\begin{document}


\newcommand{\V}[1]{\mathbf{#1}}
\newcommand{\roofs}{\rho(s)}
\newcommand{\pofs}{P(s)}
\newcommand{\bro}{\boldsymbol{\rho}}
\newcommand{\Jstar}{\mathbf{J}^\star}
\newcommand{\code}[1]{\texttt{#1}}
\newcommand{\bx}{\V{x}}
\newcommand{\mmfm}{M19}
\newcommand{\bigo}[1]{\mathcal{O}(#1)}
\newcommand{\about}[1]{\sim\!\!{#1}}
\newcommand{\unit}[1]{\;\mathrm{#1}}
\newcommand{\toJF}[2]{\textbf{\textcolor{red}{#1}}\footnote{\textcolor{red}{#2}}}
\newcommand{\sub}[1]{_{\text{#1}}} 
\newcommand{\req}{R\sub{eq}}
\newcommand{\rem}{R\sub{m}}
\newcommand{\arr}{\V{r}}
\newcommand{\arp}{\V{r'}}
\newcommand{\robg}{\rho\sub{bg}}
\newcommand{\rofg}{\rho\sub{fg}}
\newcommand{\mj}{$M_{\mathrm{J}}$} 
\newcommand{\rj}{$R_{\mathrm{J}}$}
\newcommand{\me}{$M_{\oplus}$} 
\newcommand{\re}{$R_{\oplus}$}
\newcommand{\newmank}{\citep{Mankovich2019a}}
\newcommand{\newmankt}{\citet{Mankovich2019a}}

\title{Saturn's Probable Interior:  An Exploration of Saturn's Potential Interior
Density Structures}
\author{Naor Movshovitz}
\author{Jonathan J. Fortney}
\author{Chris Mankovich}
\affiliation{Department of Astronomy and Astrophysics, University of California, Santa
Cruz, California, USA}
\author{Daniel Thorngren}
\affiliation{Department of Physics, University of California, Santa
Cruz, California, USA}
\author{Ravit Helled}
\affiliation{Institute for Computational Science, Center for Theoretical
Astrophysics \& Cosmology, University of Zurich, Switzerland}

\correspondingauthor{Naor Movshovitz}
\email{nmovshov@ucsc.edu}


\begin{abstract}
The gravity field of a giant planet is typically our best window into its interior
structure and composition. Through comparison of a model planet's calculated
gravitational potential with the observed potential, inferences can be made about
interior quantities, including possible composition and the existence of a core.
Necessarily, a host of assumptions go into such calculations, making every
inference about a giant planet's structure strongly model dependent. In this work
we present a more general picture by setting Saturn's gravity field, as measured
during the \emph{Cassini} Grand Finale, as a likelihood function driving a
Markov-Chain Monte Carlo exploration of the possible interior density profiles.
The result is a posterior distribution of the interior structure that is not tied
to assumed composition, thermal state, or material equations of state. Constraints
on interior structure derived in this Bayesian framework are necessarily less
informative, but are also less biased and more general. These empirical and
probabilistic constraints on the density structure are our main data product which
we archive for continued analysis. We find that the outer half of Saturn's radius
is relatively well constrained, and we interpret our findings as suggesting a
significant metal enrichment, in line with atmospheric abundances from remote
sensing. As expected, the inner half of Saturn's radius is less well-constrained
by gravity, but we generally find solutions that include a significant density
enhancement, which can be interpreted as a core, although this core is often lower
in density and larger in radial extent than typically found by standard models.
This is consistent with a dilute core and/or composition gradients.

\end{abstract}


\section{Introduction}\label{sec:intro}
\subsection{The Gravity Field as a Probe on the Interior}
There are a number of fundamental questions that we would like to understand about
giant planets. Do they have a heavy element core? If so, what is its mass? Is it
distinct from the overlying H/He envelope, or partially mixed into it? Is the H/He
envelope enriched in heavy elements compared to the Sun? Is the envelope fully
convective and well mixed?

Unfortunately, the vast mass of a giant planet is completely hidden from view, so
that we must use indirect methods to try to answer these questions.  Most of our
knowledge about the interiors of giant planets comes from interpreting their
gravity fields, as recently reviewed for Saturn by \citet{Fortney2018a}. Since the
planets are fluid and rapidly rotating they assume an oblate shape and their
gravitational potential differs from that of a spherically symmetric body of the
same mass. The external gravitational potential $V\sub{e}$ is a function of the
colatitude $\theta$ and distance $r$ from the center of the planet, and is
typically written as an expansion in powers of $\req/r$ where $\req$ is the
equatorial radius of the planet:
\begin{equation}\label{eq:ext_potential}
V\sub{e}(r,\theta)=-\frac{GM}{r}\left(1-\sum_{n=1}^{\infty}(\req/r)^nJ_nP_n(\cos\theta)\right).
\end{equation}
In eq.~\eqref{eq:ext_potential} $P_n$ are Legendre polynomials of degree $n$. The
coefficients $J_n$ (``the Js'') are measurable for many solar-system bodies by
fitting a multi-parameter orbit model to Doppler residuals of spacecraft on close
approach. For fluid planets in hydrostatic equilibrium, where azimuthal and
north-south symmetry holds, only even-degree coefficients are non-zero.

\begin{figure*}[tb!]
\centering
\plottwo{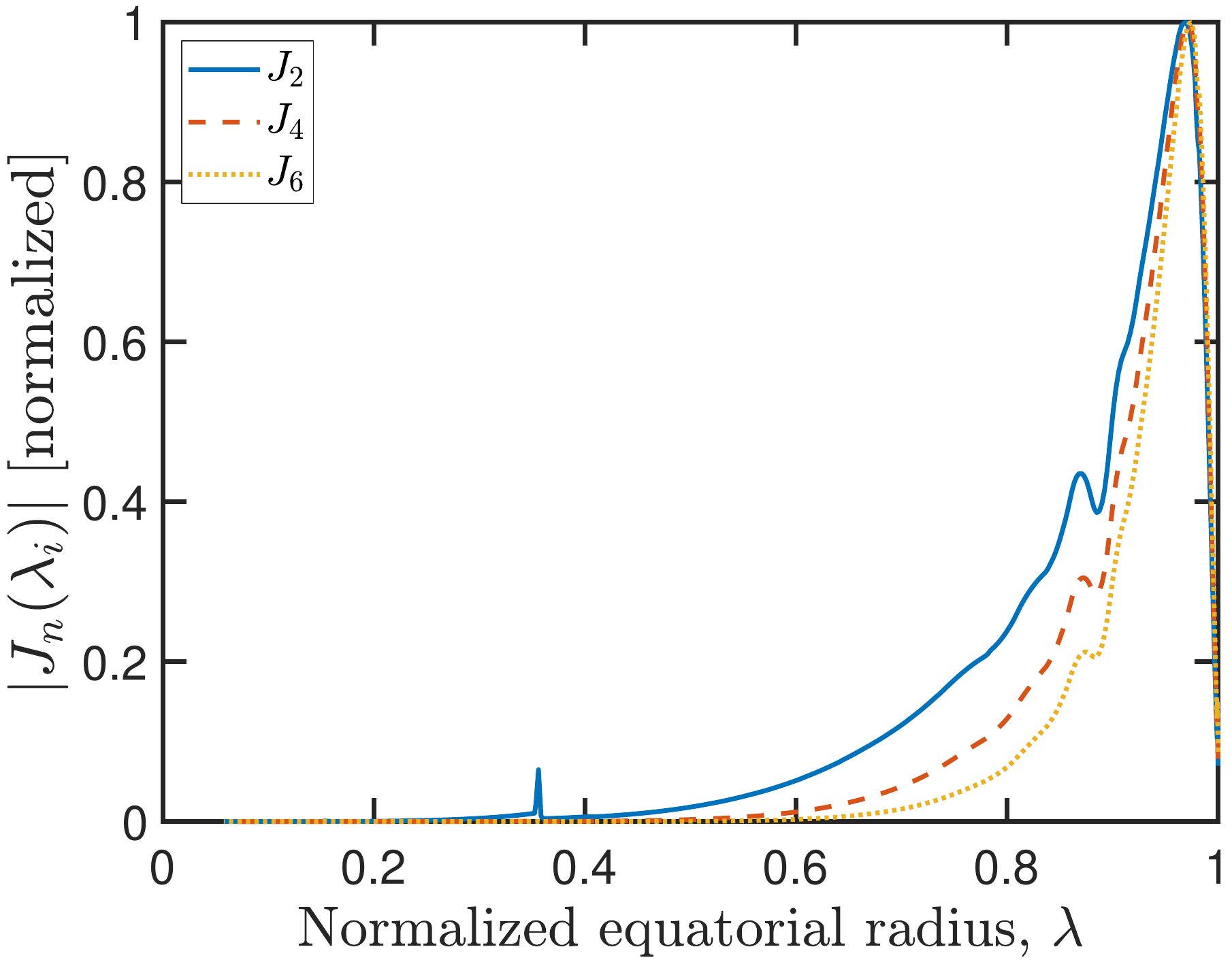}{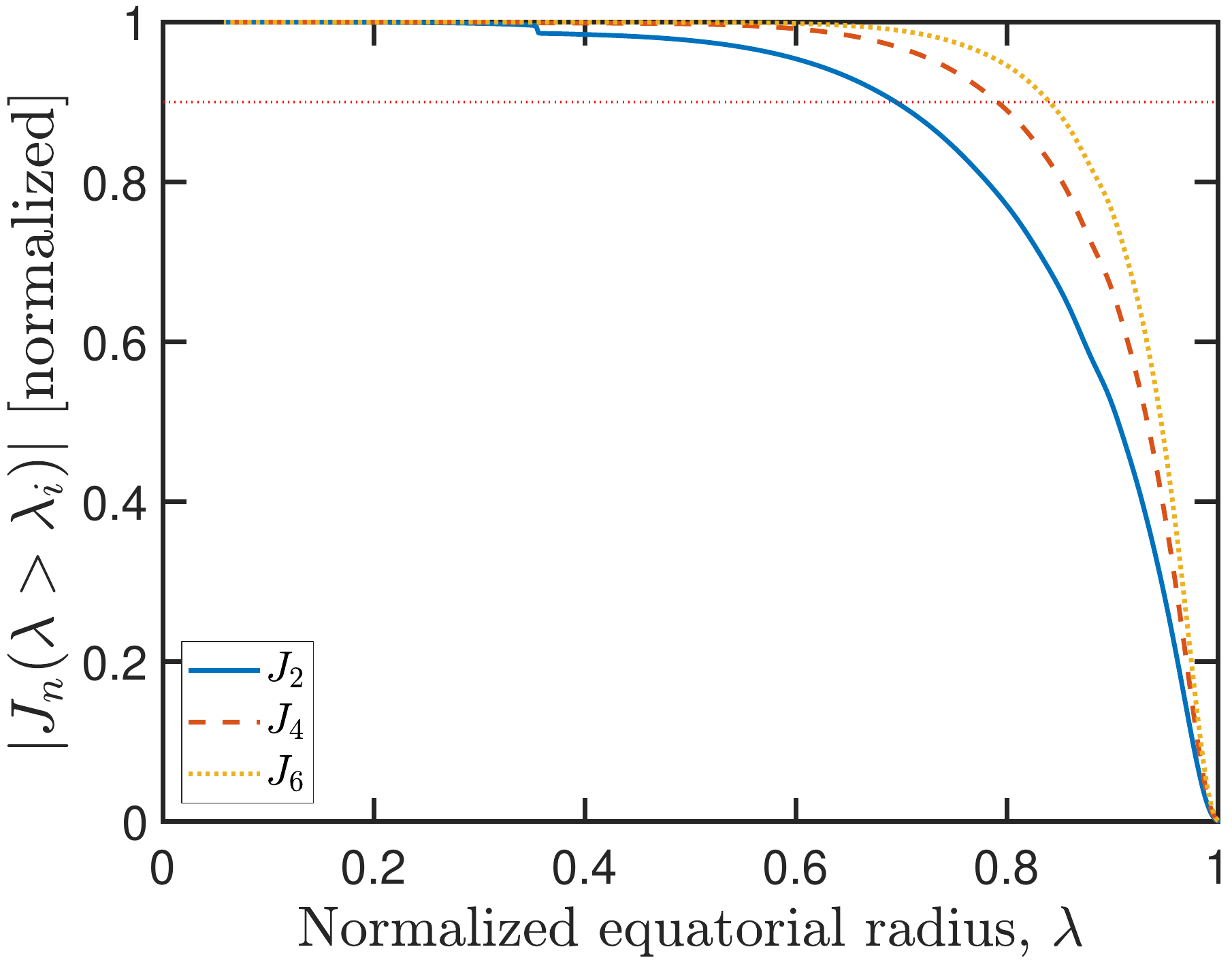}
\caption{Contribution functions of the gravitational harmonics $J_2$ (blue solid),
$J_4$ (red dashed), $J_6$ (yellow dotted) for a typical, 3-layer Saturn model. The
contribution ``density'' ($\propto{}J(r)dr$) is plotted in the left panel and the
cumulative contribution in the right panel. The horizontal line intersects the
curves at a depth where the corresponding $J$ reaches 90\% of its final
value.}
\label{fig:contribs}
\end{figure*}

When an interior model for a planet is created, the $J_n$ values are calculated as
integrals of the interior density over the planetary volume:
\begin{equation}\label{eq:J_as_rho_integrals}
M\req^nJ_n=-\int\rho(\arp)(r')^nP_n(\cos\theta')\,d\tau'.
\end{equation}
These model $J_n$ can then be compared to measured ones. As is well-known, the
different Js sample the density at different depths (with $J_2$ probing deepest)
but with significant overlap, and with most of the weighting over the planet's
outer half in radius. This point is illustrated in Figure~\ref{fig:contribs}.

The gravity field is a non-unique feature of the interior mass distribution. In
other words, different mass distributions can lead to identical gravity signals.
This complicates the process of making inferences about the interior structure
based only on the external gravity field. In principle, one should explore a wide
range of possible interior structures, of possible $\rho(\arp)$ in
eq.~(\ref{eq:J_as_rho_integrals}), to see the full range of solutions that fit the
gravity field. Initially, researchers had focused on finding a single, best-fit
solution subject to a host of assumptions, chosen for computational convenience
and not necessarily following reality. More recently there have been efforts to
explore an expanded range of interior structures, usually by making alternative
assumptions about the prototypical planet.

The main contribution of the present work is the introduction of a different
approach to the task of inferring interior structure from gravity, and the
application of this approach to Saturn. The result is a suite of interior
structure models of Saturn computed with fewer assumptions and therefore showing a
fuller range of structures consistent with observation. We describe our method in
detail and compare it with previous work of similar spirit in
sec.~\ref{sec:newermodels} and~\ref{sec:previouswork}.

\subsection{Common Assumptions in planetary interior models
\label{sec:assumptions}}
There are typically at least three significant assumptions or choices that
modelers make when constructing interior models of giant planets, thereby
implicitly constraining the possible inferences from these models.

\subsubsection{The planets have three layers}
Perhaps the most constraining assumption is the prototypical picture of three
layers, each well-mixed enough to be considered homogeneous. For Jupiter and
Saturn these are a Helium-poor \emph{outer envelope}, a Helium-rich \emph{inner
envelope}, and a heavy-element, usually constant density \emph{core}.
Investigators also adjust the abundance of heavy elements in the He-rich and
He-poor layers, with little physical motivation other than it seems to facilitate
finding an acceptable match to the gravity field.

While a core-envelope structure is certainly a plausible one, and indeed rooted in
well studied planet formation theories, the assumption of compositionally
homogeneous layers may well be a significantly limiting oversimplification.

\subsubsection{The interior pressure-temperature profile is isentropic}
A typical assumption of interior modeling is that pressure-temperature profile is
isentropic, lying on a single $(P,T)$ adiabat that is continued from a measured or
inferred temperature at 1~bar. This second assumption is likely to be true over
some of the interior, but there are good reasons to doubt that this holds
throughout the interior.

Jupiter and Saturn have an atmospheric He depletion compared to the Sun, and it
has long been suggested that this is due to He phase separation from liquid
metallic hydrogen in the deep interiors \citep{Stevenson1977,Fortney2003}. There
is likely a region with a He abundance \emph{gradient} starting between 1 to
2~Mbar in both planets \citep{Nettelmann2015,Mankovich2016}. In models that
attempt to interpret the gravity field, if such a layer is included at all it is
by interpolation between outer and inner homogeneous layers
\citep[e.g.][]{Wahl2017a,Militzer2019}, but this interpolation is unlikely to
capture fully the effects of composition gradients. Composition gradients can
inhibit large scale convection \citep{Ledoux1947} implying that heat is
transported via layered convection or radiation/conduction. This leads to higher
internal temperatures that in return allow higher heavy-element enrichment at a
given density-pressure. Indeed, non-adiabatic structures have been recently
suggested for all outer planets in the solar system
\citep[e.g.][]{Leconte2012,Vazan2016,Podolak2019}.

\subsubsection{The inferred composition relies on equation of state calculations}
As the field progresses the equations of state (EOS) used for modeling giant
planet interiors become a better representation of reality. Nevertheless, the
equations of state for all the relevant materials and mixtures are not perfectly
known. Simulations from first principles of hydrogen, helium, and their mixtures
over the conditions relevant for giant planets have been carried out and partially
validated against experimental data \citep{Nettelmann2008,Militzer2013}. These
equations of state (EOSs) for hydrogen show good agreement with data up to
$\about{1.5\unit{Mbar}}$ \citep[e.g.][]{Militzer2016}. However, the pressure at
the bottom of Saturn's H/He envelope is about 10~Mbar and for Jupiter it is about
40~Mbar, well beyond the realm of experiment. Recent structure models used EOS for
hydrogen and helium based on Density Functional Theory (DFT) simulations
\citep{Militzer2016,Nettelmann2008,Miguel2016}. Until recently, different EOS led
to different inferred compositions for Jupiter due to different approaches to
calculating the entropy. Today, there is good agreement between state-of-the-art
EOSs \citep{Nettelmann2017a}, but it should be kept in mind that DFT also suffers
from approximations \citep{Mazzola2018} and there remains an uncertainty of
$\about{2\%}$ in the hydrogen EOS, which increases significantly when it comes to
predicting hydrogen-helium demixing \citep{Morales2009}.

The heavy elements must also be represented by an EOS (typically for water or
silicates) which introduces another source of uncertainty. Therefore, the range of
possible composition and internal structure from such interior models  cannot be
taken to be the true range of allowed values, even if the parameter space of
possible EOSs, H/He/Z mixing ratios, and outer/inner envelope transition pressures
were thoroughly explored.

\subsubsection{Appreciating the complexities\label{sec:newermodels}}
The drawbacks of the assumptions discussed above have long been known and the
reality that giant planets are surely more complicated than the traditional
modeling framework allows for is generally accepted
\citep[e.g.,][]{Stevenson1985a}. More recent investigations are attempting to
allow for a more complex structure. Interior composition gradients due to remnants
of formation \citep{Leconte2012,Helled2017}, core dredge-up \citep{Militzer2016},
convective mixing of primordial composition gradients \citep{Vazan2016,Vazan2018},
and He sedimentation \citep{Nettelmann2015,Mankovich2016} have been considered and
were found to lead to different structures. Additionally, some investigators have
begun using what may be referred to as ``empirical'' models. In this context an
empirical model is one that is focused on the more direct connection between
gravity and density \citep[e.g.][]{Helled2009} or gravity and equilibrium shape
\citep{Helled2015}, without invoking the compositional and thermodynamical origin
of these structures.

The work we present here is in the spirit of empirical models. We explore
systematically, in a Bayesian inference framework, the possible density profiles
of Saturn. We limit our assumptions as much as possible, in order to find the
widest range of interior structures with their probability distribution based on
their gravitational potential matching the observed field.

\section{Composition-independent interior density calculation}\label{sec:method}
The premise of removing some assumptions and deriving composition-free interior
density profiles (sometimes referred to as empirical models) is simple and in fact
has been pursued in previous work
\citep[e.g.,][]{Marley1995a,Podolak2000,Helled2009}. (We discuss similarities and
differences with these works in sec.~\ref{sec:previouswork} below.) The only
information that is needed to calculate a gravity field is the density everywhere
inside the planet, $\rho(\arr)$, and so this is the only quantity we will directly
vary. In fact, hydrostatic equilibrium produces \emph{level surfaces}, closed
surfaces of constant density, pressure, and potential, and therefore a
one-dimensional description of the mass distribution is sufficient: we can use
$\rho(\arr)=\rho(s)$ where $s$ is the volumetric mean radius of the unique level
surface of density $\rho$.

All other properties of the planet will be inferences, rather than input
parameters. Since there is unavoidable uncertainty associated with the measurement
of the gravity field (and also with its theoretical calculation from interior
models; see~\ref{sec:fastof}), this means that there must be a continuous
distribution of possible density profiles that fit the gravity solution, and we
must base our inferences on the entire distribution. In practice, since we can
only ever consider a finite number of solutions, this means that we must base our
inferences on a \emph{random sample} from this unknown distribution of allowed
solutions.

In this section we describe the process of obtaining this random sample, as
applied to Saturn. For the most part the same process would apply equally well to
the other giant planets. We mention in places modifications that may be needed if
the same method is to be applied to Jupiter, Uranus, or Neptune.

\subsection{Overview}\label{sec:overview}
Formally, the distribution we are after is the \emph{posterior probability}
$p(\bro|\Jstar)$, the probability that the planet's interior density follows
$\bro=\roofs$ given that the gravity coefficients were measured as $\Jstar$. This
consists of several subtasks. First we must find a suitable parameterization of
$\roofs$. This parameterization should be able to represent all the physically
reasonable $\roofs$ curves without undue loss of generality, but this is not
particularly difficult. It is also necessary that the range and behavior of the
numeric values of all parameters is such that they can be efficiently
\emph{sampled}, e.g., with a Markov Chain Monte Carlo (MCMC) algorithm. This is
easier said than done, and the best parameterization may be different for different
planets.

For Saturn we find that a piecewise-quadratic function of density as a function of
normalized radius works best for the bulk of the planet, with a quartic (degree 4)
polynomial required to represent the uppermost region (for
$P\lesssim{2}\unit{GPa}$). We describe this parameterization in detail in
section~\ref{sec:params}. Note that this is one place where modifications might be
needed before applying the same procedure to Jupiter or the ice giants.

To drive the sampling algorithm we need a way to evaluate the relative likelihood
of two model planets and we do this by comparing how well they match Saturn's
observed mass and gravity field. The details of this calculation are given in
section~\ref{sec:lossfunction}.

The likelihood calculation requires that we know the equilibrium shape and gravity
field of a given density profile, to sufficient accuracy. Note that in
eq.~\eqref{eq:J_as_rho_integrals} the integrand is known but the integration
bounds are unknown. We need to first determine the planet's equilibrium shape. The
shape is determined by a balance between the centrifugal acceleration of the
rotating planet and the gravitational acceleration. This is therefore a circular
problem, requiring an iterative calculation to converge to a self-consistent
solution.

We use an implementation of $4^{th}$-order Theory of Figures (ToF) using the
coefficients given in \citet{Nettelmann2017a} and employ optimization techniques
that allow us to solve the hydrostatic equilibrium state to desired precision very
quickly. The details are given below in the section~\ref{sec:fastof}.

The emphasis on speed is necessary, as the next subtask is to employ a suitable
MCMC algorithm to draw a large sample of possible $\bro$. There is no generally
agreed upon method of predicting the number of sampling steps required for
convergence\footnote{Or even of being sure that convergence was reached.}. By
experimentation we find that our Saturn parameterization requires tens of
thousands of steps to become independent of its seed state and has a long
auto-correlation time, requiring a large number of steps following convergence to
obtain the desired effective sample size. Producing a valid sample required the
computation of about ten million model planets in total. We give the details of
our sampling method and convergence tests in sec~\ref{sec:mcmc} \edit1{and
appendix}~\ref{app:chains}.

The last step is calculating some derived physical quantities of interest, based
on the obtained $\bro$ sample. Given the gravity field, the pressure on each level
surface can be computed from the hydrostatic equilibrium equation. And with
knowledge of the pressure and density at each level we may begin to estimate other
quantities of interest, e.g.,~the helium fraction, the heavy element content, etc.
These quantities are not determined directly by the gravity field but can be
inferred, with additional assumptions. We discuss the results of this analysis, as
applied to Saturn, in section~\ref{sec:results}.

\subsection{Parameterization of $\roofs$}\label{sec:params}
Our goal is to sample from a space of $\roofs$ curves that is as general as
possible, making a minimum of assumptions about $\roofs$ while still restricting
the sample to physically meaningful density profiles and, importantly, keeping the
number of free parameters small, for sampling efficiency. These competing
requirements are not easy to satisfy and it may be that the best parameterization
depends on the planet being studied as well as on the available sampling
algorithms and computing resources.

\begin{figure}[tb!]
\centering
\plotone{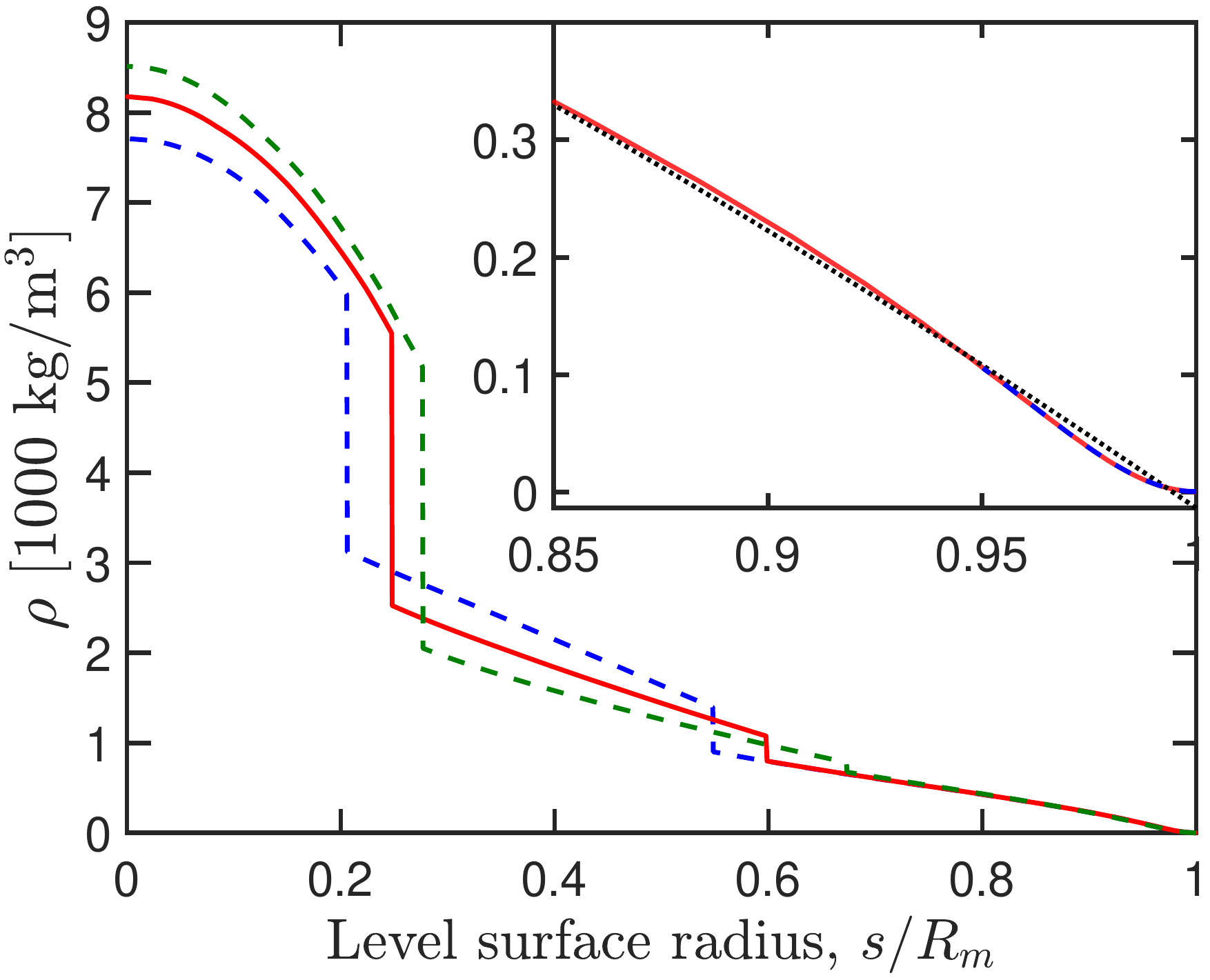}
\caption{Three representative Saturn density profiles from \mmfm. These profiles
were derived using the standard, three-layer assumption, and thus represent only a
subset of possible profiles. On the other hand they are known to be in strict
agreement (by construction) with theoretical EOS throughout the interior. The
inset shows a zoomed-in view of the top part of the envelope. The red solid line
is the same curve as in the full scale figure; the black dotted line is a
quadratic fit, a good approximation of the upper envelope overall, and the blue
dashed line is a quartic fit to the segment $s/\rem>0.94$, a much better fit
there (appendix~\ref{app:quartic}).}
\label{fig:mank-typical}
\end{figure}

When looking for a good parameterization of $\roofs$ we were guided by previously
published work on Saturn's interior. Traditionally derived models are less general
than we would like but they are physically sound. Examining them exposes the major
features expected of a $\roofs$ curve representing Saturn's interior.
Figure~\ref{fig:mank-typical} shows the density profiles of several Saturn models
recently published by \citet[][hereafter \mmfm]{Mankovich2019}. \edit1{These
models assume a three-layer structure for Saturn along the lines of what was
considered by \cite{Nettelmann2013}. They consist of a homogeneous outer envelope
with helium mass fraction $Y=Y_1$ and water mass fraction $Z=Z_1$, a homogeneous
inner envelope with $Y=Y_2$ and $Z=Z_2$, and finally a central core with $Z=1$.
These models assume an additive-volume mixture of hydrogen, helium, and water as
described by the \cite{Saumon1995} and \cite{French2009} EOSs, and are assumed to
have adiabatic temperature profiles throughout the envelope with an isothermal
core.}

The general feature is a monotonic and piecewise-smooth function in three
segments. This is not surprising, as these models were all derived with the
assumption of three layers of homogeneous composition, commonly thought of as an
upper envelope, lower envelope, and core. While we do not wish to make such a
strong assumption, we find it necessary to make the much weaker assumption that
$\roofs$ is a monotonic, piecewise-smooth function, with no more than (but
possibly fewer than!)\ two density discontinuities. Further, between
discontinuities the density appears to follow very smooth curves, suggesting that
it may be well approximated by a quadratic function of $s/R_m$ for each segment,
where $\rem=58232\unit{km}$ is Saturn's volumetric mean 1-bar radius
\edit1{\citep{Lindal1985}}. By experimentation, we find no advantage
in using higher order polynomials to approximate any of the main segments.

This piecewise-continuous model should not be confused with the traditional
3-layer one. The density being piecewise continuous is a much less strict
assumption than the composition being piecewise constant, even if they lead to
visually similar plots. Nevertheless, it would be even better to allow more
discontinuities or, better yet, a variable number of them. While this may seem
like a relatively straightforward generalization, it would in fact greatly
increase the computational cost of sampling the parameter space. To understand why
consider that each additional discontinuity in $\roofs$ not only introduces four
additional parameters (the three parameters required to describe the quadratic
plus the location of the additional break point), these parameters will also be
highly \emph{correlated} with the rest. As it turns out, this correlation is
already evident with just two discontinuities. Informally, each of the two density
``jumps'' can substitute for the other in the large subset of models where only a
single pronounced discontinuity appears. This evident ``redundancy'' is by no
means proof that there cannot be more than two sharp density jumps in Saturn's
interior. But it helps us accept, at least temporarily, a compromise between
maximum generality and minimum CPU hours.

\edit1{When we examine more closely the very top of the density curves in
Fig.~\ref{fig:mank-typical} we find that the uppermost part of the envelope (by
radius, $r{}\gtrsim{}r_a=0.94\rem$) does not follow the same quadratic as the rest
of the upper envelope. Instead, it is more similar to a quartic polynomial. This
is demonstrated visually for one density profile in the inset of
Figure~\ref{fig:mank-typical}, and in more detail in appendix~\ref{app:quartic}.
In this low-pressure region the physical models are based on well-tested EOSs of H
and He, and the assumption of an adiabatic gradient is appropriate, so we would be
well advised to constrain our profiles to make use of this information. In
appendix~\ref{app:quartic} we explain how we derive a one-parameter family of
quartic functions that keeps us grounded to realistic density values in the region
above $r_a$, while still allowing variation by letting the value of
$\rho_a=\rho(s=r_a)$ be sampled.}

It is important to note that to date all EOS-based models of Saturn find
solutions consistent with the measured gravity field that predict a concentration
of heavy elements in the envelope of at most a few times the protosolar value
\citep[e.g.][]{Nettelmann2013,Helled2013,Militzer2019} while Saturn's atmospheric
spectra indicate a higher value, perhaps as high as 10 times the protosolar
metallicity \citep{Atreya2016}. In principle, atmospheric enrichment might not
represent the bulk composition of the outer envelope, as was recently suggested
for Jupiter \citep{Debras2019}. Nevertheless this demonstrates that, while we wish
to be guided by physical models, our parameterization must not be overly
constrained by them.

To summarize, we arrive at the following parameterization, using $z=s/\rem$:
\begin{equation}\label{eq:sat-params}
\rho(z)=
\begin{cases}
\rho\bigl(z, [\V{q}_1,\V{q}_2,\V{q}_3,z_1,z_2]\bigr), &0<z<z_a\\
\rho\bigl(z, [\V{Q}(\rho_a)]\bigr), &z_a<z<1.
\end{cases}
\end{equation}

Here $\V{q}_1$ are the three parameters defining the first (outermost) quadratic
segment, $\V{q}_2$ are the three parameters defining the second (middle) quadratic
segment, and $\V{q}_3$ are the three parameters defining the third (deepest)
quadratic segment. \edit1{There is more than one way to let three numbers define a
quadratic and although they are all equivalent, the associated range of values and
degree of correlation make some choices better suited for MCMC sampling. The
precise definition of $\V{q}_i$ that we find, by trial-and-error, to work well in
this case is given in appendix~\ref{app:fullparam}.} The transition between the
first and second segments is at normalized radius $z_1$ (which we let vary from
0.35 to 0.9 in normalized radius) and the transition between the second and third
segments is at $z_2<z_1$ (which we let vary from 0.1 to 0.4).
\edit1{The top of the upper envelope is defined by the quartic polynomial
$\V{Q}(\rho_a)$ for $z>z_a=0.94$. The values in $\V{Q}$ and their definition are
given in appendix~\ref{app:quartic}.} The quartic segments are uniquely determined
by the density at $z_a$, itself already determined by the coefficients $\V{q}_1$.
We thus have 11 free parameters -- three each for the three quadratic segments,
plus the two ``floating'' transition radii.

{\bfseries \color{magenta} \subsection{Comparing model and observation}
\label{sec:lossfunction}
MCMC sampling works by comparing, at every iteration, the likelihood of a proposed
vector of parameter values, $L(\V{y})$, with that of the current vector of
parameter values, $L(\V{x})$, and accepting or rejecting the proposed values with
probability proportional to the relative likelihoods. If the likelihood function
is itself proportional to the desired (unknown) posterior probability, in our
notation, if $L(\V{x})\propto{}p\bigl(\bro(\V{x})|\mathbf{OBS}\bigr)$, then the
resulting Markov chain will converge, in the long run, to a sample from that
posterior. For $\mathbf{OBS}$ we substitute any number of observed quantities that
may differ from those calculated in the model.

A likelihood function that is proportional to the desired posterior is the
function $L(\V{x})=p\bigl(\mathbf{OBS}|\bro(\V{x})\bigr)p\bigl(\V{x}\bigr)$. It is
proportional to the posterior as a consequence of Bayes' rule. The \emph{prior
probability} $p(\V{x})$ is necessary. It is our informed, subjective assessment of
what values the model parameters $\V{x}$ are expected to take, and typically it is
a simple product of the individual prior probabilities of the independent
variables $x_i$, which in turn are either uniform or normal inside a region of
reasonable values\footnote{A full definition of our chosen prior is given in
appendix~\ref{app:chains}.}. What we need then is to provide the MCMC algorithm
with a function that evaluates the relative goodness of the match between the
observed properties of the planet and the values of the same properties as
calculated for the model, taking uncertainties from both model and observation
into account.

In this context the planetary properties that our models need to match are the
gravity coefficients $\Jstar$ and the planet's mass $M\sub{Sat}$.

First, the gravity. We assume that the observed values $\Jstar{}$ are normally
distributed about the true, unknown, mean values, and calculate a distance:
\begin{equation}\label{eq:lossfunc}
    D_J^2 = \left(\frac{J_2-J^\star_2}{\sigma_{J_2}}\right)^2 +
            \left(\frac{J_4-J^\star_4}{\sigma_{J_4}}\right)^2 +
            \left(\frac{J_6-J^\star_6}{\sigma_{J_6}}\right)^2,
\end{equation}
where the $\sigma_{J_i}$ are measures of the uncertainty in either the measured or
the computed values, or both\footnote{Depending on the source of uncertainties
$\sigma_{J_i}$ they may be correlated. In that case the definition of $D_J$ would
involve a covariance matrix but the rest of the calculation would remain
unchanged.}.

In the work presented here only $J_2$, $J_4$, and $J_6$ were considered for the
purpose of calculating the likelihood function. Higher order coefficients
$J_8$--$J_{12}$, as well as non-vanishing odd-indexed coefficients $J_3$ and
$J_5$, have been measured for Saturn, to impressive precision, by the Cassini
Grand-Finale gravity experiment \citep{Iess2019}. But it seems clear that these
reflect an increasingly large contribution from an asymmetric and/or time-varying
field, deriving either from planet-scale differential rotation or from deeply
rooted zonal winds, or both \citep{Galanti2017,Kaspi2018,Galanti2019,Iess2019}.
These phenomena are important in themselves and offer a promising avenue for
studying further Saturn's dynamic nature, but for the purpose of constraining the
bulk interior structure their net result is to increase the \emph{effective
uncertainty} of the low-order Js ascribed to solid-body rotation
\citep{Guillot2018}. Studies of differential rotation on Saturn demonstrate that
their contribution to the low-order even harmonics can be significant
\citep{Hubbard1982,Galanti2017}. As our goal is to capture the widest range of
probable interior structures, we compute eq.~(\ref{eq:lossfunc}) for every model
by assuming solid-body rotation and setting $J^\star_2=16290.573\times{10}^{-6}$,
$J^\star_4=-935.314\times{10}^{-6}$, and $J^\star_6=86.340\times{10}^{-6}$
\citep{Iess2019}, with uncertainties $\sigma_{J_2}=1.5\times{10}^{-5}$ and
$\sigma_{J_4}=\sigma_{J_6}=5\times{10}^{-6}$. The values adopted for the
uncertainties come from interpreting the largest contribution from winds found in
\citet[][their fig.~4]{Galanti2017} as a symmetric, two-sigma range.

We cannot simultaneously hold fixed both the total planetary mass and the surface
radius while also specifying the density at all radii. Since we use the density
$\rho_a$ as one of the sampled parameters, the converged hydrostatic interior
profiles can be scaled to fix $R\sub{eq}$ or $M$ precisely but not both. Saturn's
mass and radius are known to comparable precision\footnote{$GM$ is measured with
exquisite accuracy but $G$ is known to $\about{10^{-4}}$ precision (CODATA 2014,
\url{https://physics.nist.gov}).}, and it is convenient to fix all models to
$R\sub{eq}=60268\unit{km}$, Saturn's measured equatorial radius at the 1-bar level
\citep{Lindal1985}. The calculated mass of a converged and scaled density profiled
will therefore exhibit a small spread around a nominal value, leading to another
distance term:
\begin{equation}
D_M^2 = \left(\frac{M-M\sub{Sat}}{\sigma_M}\right)^2,
\end{equation}
with $M\sub{Sat}=568.336\times{10}^{24}\unit{kg}$ and
$\sigma_M=0.026\times{10}^{24}\unit{kg}$
\citep[][\url{https://ssd.jpl.nasa.gov}]{Jacobson2006}. Then, assuming that
$\sigma_M$ and $\sigma_J$ are uncorrelated, we use $D^2=D_J^2 + D_M^2$ as a
measure of a model's fit to observation and a natural likelihood function is
\begin{equation}\label{eq:like}
L\propto\exp{\Bigl(-\frac{1}{2}D^2\Bigr)}p(\bx).
\end{equation}
We need not worry about a normalizing constant since the sampling algorithm
evaluates only ratios of likelihood.

A final minor modification of eq~\eqref{eq:lossfunc} is worth mentioning. Since
the uncertainty values that define the $\sigma_{J_i}$ are due in large part to the
contribution from non-rigid rotation, and since this contribution, while unknown
in detail is very likely to be non-zero, it seems unwarranted to ``privilege'' the
point $\Jstar$ as the Gaussian likelihood~\eqref{eq:like} does. Instead, we
measure the distance of a model's gravity not to the center, $\Jstar$, but to the
nearest corner of the cube defined by $J^\star_i\pm\sigma_{J_i}$. Models inside
this ``1-sigma cube'' are considered equally likely. This likelihood seems to us
more physically justified. It turned out to have negligible effect on the derived
samples however. }

\subsection{Fast Calculation of Gravity Coefficients}\label{sec:fastof}
The main computational effort involved in the sampling, and thus the prime
candidate for optimization, is the calculation of the gravity
coefficients\footnote{Higher order Js can be used when appropriate; the
computation time is independent of how many Js are sought.}
\edit1{$\V{J}=[J_2,J_4,J_6]$} given a particular $\roofs{}$.

The calculation of the $J_i$ for fluid planets has a long and rich history. In
modern times the choice is between two algorithms. The faster but less precise
method is the Theory of Figures \citep[ToF;][]{Zharkov1978}. When carried to
fourth order in powers of the small parameter $m=\Omega^2R_m^3/GM$, where $\Omega$
is the uniform rotation rate and $GM$ is the total gravitational mass, the
theoretical truncation error is $|\delta{J_2}/J_2|\lesssim10^{-4}$ and
$|\delta{J_4}/J_4|\lesssim10^{-3}$. We use the shape-function coefficients given
by \citet{Nettelmann2017a} to $\bigo{m^4}$ and confirm her findings, that this
level of precision is also achievable in practice.
\edit1{For Saturn, the Cassini mission's Grand Finale orbits provided gravity
coefficients to much better precision\footnote{The same would be true for Jupiter,
with gravity obtained during the Juno mission, whereas for Uranus and Neptune the
measurement uncertainty would still be dominant \citep{Hubbard1995}. }, but as
discussed above the measured values include a potentially large contribution from
dynamic flow (winds), greatly increasing the effective uncertainty in the portion
of the gravity field attributed to the underlying density structure. For $J_2$ the
wind contribution becomes the dominant source of uncertainty, while for $J_4$ and
$J_6$ winds and the uncertainties associated in the ToF calculation are comparable
in magnitude and are therefore added, in quadrature, to define
$\sigma_{J_i}$.\footnote{The values of $\sigma_{J_i}$ given in
sec.~\ref{sec:lossfunction} include both sources.}}

\edit1{The solid-body rotation period for Saturn is itself still somewhat
uncertain. The rotation period measured long ago by Voyager as 10h 39min 24s
\citep{Desch1981} is now commonly understood to be much too slow to represent the
bulk planetary rotation. More recently several estimates of a faster rotation rate
have been proposed, based on a few independent methods that seem to point to a
period of 10h and between 33 and 34 minutes
\citep{Read2009,Helled2015,Mankovich2019,Militzer2019}, but an exact rotation rate
is not available. The uncertainty in rotation rate can be used to estimate a
corresponding correction to the already large gravity uncertainty, but there is a
better way.}

\edit1{We can let the rotation parameter $m$ be itself a sampled variable, guided
by a suitable prior as always. Adding an extra variable to a sampling problem is a
risky proposition but in this case it turned out to have minimal performance cost,
because the rotation parameter is uncorrelated with the other sampled variables
and because the likelihood function is not strongly sensitive to this variable, at
least within the range of values implied by the prior. We use a relatively strong
prior of $m$ normally distributed centered on $m^\star=0.14224$ (10h 33min 30s)
with $\sigma_m=4.5\times{10}^{-4}$ ($\about{1}\unit{min}$).}

The second option for calculating the Js is the Concentric Maclaurin Spheroids
method \citep[CMS;][]{Hubbard2012,Hubbard2013a} which allows for calculation of
$J_i$ of any order and to arbitrary precision, but at the cost of a much slower
computation. \edit1{The CMS method was developed in anticipation of the
extraordinarily precise data expected from the Cassini Grand Finale orbits (and
similarly precise measurements of Jupiter's gravity by Juno). But although the
radio science indeed determined Saturn's gravity to very high precision
\citep{Iess2019} as discussed above the presence of non uniform rotation leads to
effective uncertainty much higher than the measurement uncertainty. The large
uncertainty associated with deep zonal winds means that the faster, ToF method is
adequate for the purpose of calculating the rigid-body Js. In this work we
therefore let ToF do the majority of the calculation, including all of the
computation embedded in the sampling process. We use CMS for validation and to
compute a subset of some tens of high likelihood models.}

Both CMS and ToF can benefit from the following optimization. To achieve the
theoretical level of precision the integrals involving the mass distribution
$\roofs{}$ must be computed with higher accuracy than that required by the rest of
the algorithm. In general this means that $\roofs{}$ must be resolved on a fine
enough grid in normalized radii, $z_i$, for the numerical integration to properly
converge \citep[e.g.][eq.~B.9]{Nettelmann2017a}. It is not necessary, however, to
carry out the computationally expensive solution of non-linear equations for the
shape functions, in the case of TOF, or the root finding of potential as function
of latitude in the case of CMS, on such a fine grid. Since the shape of the planet
deviates only slightly from spherical even for a fast rotator such as Saturn, the
shape of a level surface, $r(z,\theta)$, is a very smooth function in both $z$ and
colatitude $\theta$. Taken as a function of $z$ for fixed $\theta$ the function
can be interpolated with excellent precision from only a handful of known values
between $z=0$ and $z=1$, using a spline interpolant.

This affords us a significant reduction in the time required to compute the shape
and gravity for a single model. For example, we find that we can achieve expected
theoretical precision of fourth-order ToF with $\roofs{}$ resolved on $N=2048$
levels but with the shape equations solved on only $n=64$ intermediate levels, and
then interpolated onto the full set. Our implementation then returns a candidate
model's gravity coefficients in under one second running on a single CPU core.
This is a key optimization that allows the sampling procedure to be completed on
modest hardware.

\edit1{The same optimization can be implemented for CMS with even better results,
as demonstrated in \citet{Militzer2019}. Unfortunately, for a sampling problem of
this scope, this speedup is not enough to mitigate the speed disadvantage of CMS
compared with ToF.}

\subsection{MCMC Sampling of Parameters}\label{sec:mcmc}
There is a wide variety of MCMC sampling algorithms; all fundamentally seek to
sample the posterior distribution by a sequence of random steps through parameter
space. The main difficulty is constructing an appropriate random-stepping
algorithm, called a \emph{proposal distribution}, to efficiently explore a
high-dimensional parameter space.

MCMC can often benefit from parallel execution. A variant that has proved very
useful for this work is the parallel stretch-move algorithm \citep{Goodman2010},
as implemented in the \code{emcee} Python package \citep{Foreman-Mackey2013}. In
this algorithm the proposal distribution is automatically constructed by taking a
step along the line segment (in parameter space) connecting the current position
of two ``walkers'' in an ensemble that explores parameter space simultaneously.
This greatly simplifies the most difficult task of MCMC but if the walkers in the
ensemble are run in serial the computation time would be too long. Luckily this
approach can benefit from parallelization with minimal overhead, and is thus
perfectly suitable to run on a large supercomputer. The sampling calculations for
this work were run on NASA's Advanced Supercomputing facility at the Ames Research
Center.

A critical consideration in the application of any MCMC algorithm is the issue of
convergence. Simply put, we must decide when it is safe to stop the sampling run
and use the obtained draws to calculate anything of interest, trusting that the
sample distribution is similar enough to the underlying posterior. Theoretical
considerations offer only loose bounds on the variance of sampled parameters and
are rarely useful in practical work. A number of diagnostic schemes have been
suggested that attempt to either hint at convergence or to warn of a failure to
converge \citep[e.g.~review by][]{Cowles1996}. But even this more limited task is
still an open problem in statistics and the decision to accept a sample as
``converged'' still involves case-by-case, subjective judgment.
\edit1{Appendix~\ref{app:chains} includes a discussion of the mixing and burn-in
length of our samples.}

In our case, examining the traces, autocorrelations, and joint posteriors of
partial samples, we find that we can significantly accelerate convergence by
separating our \edit1{12}-dimensional parameter set into two subsets that are
sampled in hierarchical fashion. Recall that of the 11 parameters needed to define
a $\roofs{}$ curve (eq.~\ref{eq:sat-params}), two are the normalized radii
locating the points of possible density discontinuity; their values have a
straightforward, physical meaning. The other nine parameters, defining the
geometry of the quadratic segments, take values whose highly nonlinear effect on
the density is entirely dependent on the value of the first two parameters. In
statistical terms, we have two proper subsets of parameters with very high
correlation between sets but low correlation within each one. Consequently, fixing
values for the transition radii, we can sample the \emph{conditional joint
posterior} of the nine geometric parameters efficiently. Of course we must repeat
this sampling many times, on a fine grid of values for the transition radii, and
finally combine the conditional probabilities to a full joint posterior. But the
gain in sampling efficiency provided by this hierarchical approach is such that we
still come out ahead in terms of CPU hours and overall length of simulation.
\edit1{There is more than one way to combine conditional joint probabilities to a
single joint posterior. We use the Bayes Information Criterion, defined fully in
appendix~\ref{app:chains}.}

\edit1{A final minor optimization is worth mentioning. In hydrostatic equilibrium,
the condition $dP/dr=-\rho{g}$ requires that the pressure gradient go to zero at
the center of the planet. For a continuous thermal profile this implies that the
density gradient likewise vanishes at the center, in our notation:
\begin{equation}
\lim_{s\to{0^+}}\frac{d\rho(s)}{ds}=0.
\end{equation}
Since $\roofs$ is in our case piecewise quadratic, the linear term of the
innermost quadratic segment should vanish, or equivalently, any three parameters
used to define the quadratic are correlated, such that only two independent
parameters are needed. This not only results in more realistic density profiles
but also helps by reducing the dimensionality of the sample space -- always a good
idea.}

\subsection{Relation to previous work}\label{sec:previouswork}
While in previous sections we discussed the drawback of ``standard'' approaches,
here it is worth discussing how our work compares to previously published
alternative approaches.

\citet{Helled2011b,Helled2009} investigated models for Saturn, Uranus, and Neptune
where the interior $\roofs$ profile was parametrized as a high-order polynomial. A
single best-fit polynomial was found, given the gravity field, and the results
were interpreted by comparison with physical EOS for H, He, ices, and rock. In
other studies a large range of density profiles was considered allowing for
different core masses and radii, with the core being represented by a constant
density \citep{Helled2011a,Helled2011c,Kaspi2013}. Our work has a similar spirit
but we determine the statistical distribution of the empirical models while also
allowing a more general structure, and more than one density discontinuity, which
is favored by the gravity solution.

Another approach was that of \citet{Leconte2012}, who investigated Jupiter and
Saturn structure models that were super-adiabatic throughout most of the interior,
due to an ad-hoc composition gradient in the planetary interior. These models
yield significantly different interior structures (that were much richer in heavy
elements than standard models) but there was little exploration of a range of
models. \citet{Vazan2016, Vazan2018} ran evolution models of Jupiter and Saturn
with composition gradients, and helium settling for Saturn, and several models
have been presented, not aimed at a statistical description.

Another approach to interior modeling that is quite similar to ours in spirit but
very different in practice was previously attempted by \citet{Marley1995a} and
\citet{Podolak2000}. As a means to forgo as many assumptions as possible the
authors studied a number of randomly generated interior density profiles for
Uranus and Neptune, matching only the constraints of mass, radius, $J_2$, and
$J_4$. Their model generation was truly random, not based on a sampling algorithm.
Naturally this algorithm, while simple, has a very low success rate, i.e.~the
number of valid models per $n$ models generated was quite low and the authors were
forced to restrict the parameter space in some arbitrary ways, the most important
was forcing a single value for the core radius and a small range of radii for a
secondary density jump in the envelope.

Even with these restrictions, the investigation was able to produce only a small
number of valid models for each planet, much too small to draw statistical
conclusions from. Particularly as this set of empirical density curves was not
constructed to be a representative sample. Nevertheless, the models thus obtained
were different from models generated by the traditional approach in interesting
ways. Most importantly, the derived pressure-density relation for both Uranus and
Neptune implied a gradual composition gradient in the outer shells of both planets
\citep[][their Figure 2]{Marley1995a}.


\section{Saturn's density profile and inferred properties}
\label{sec:results}
\begin{figure*}[tb!]
\plottwo{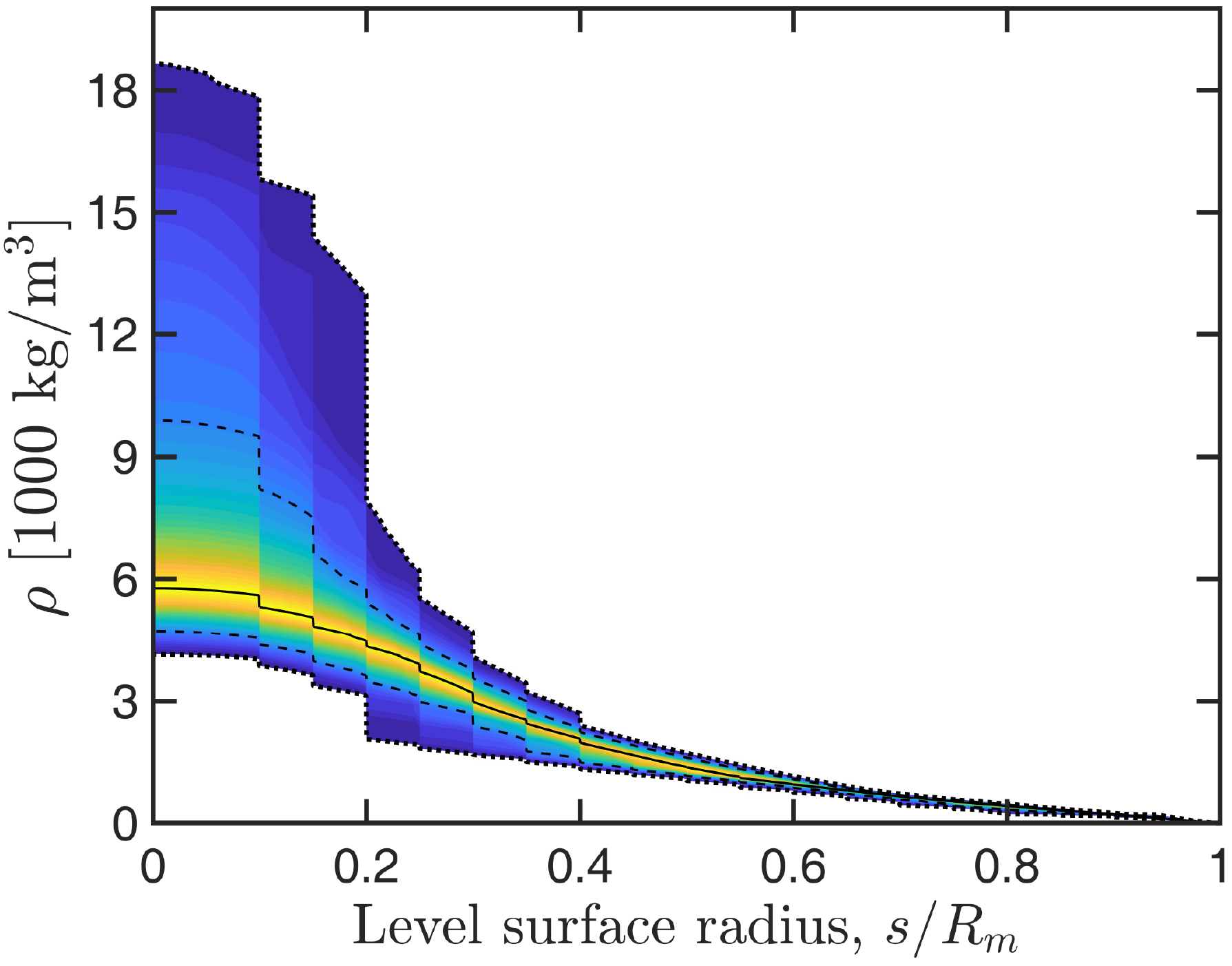}{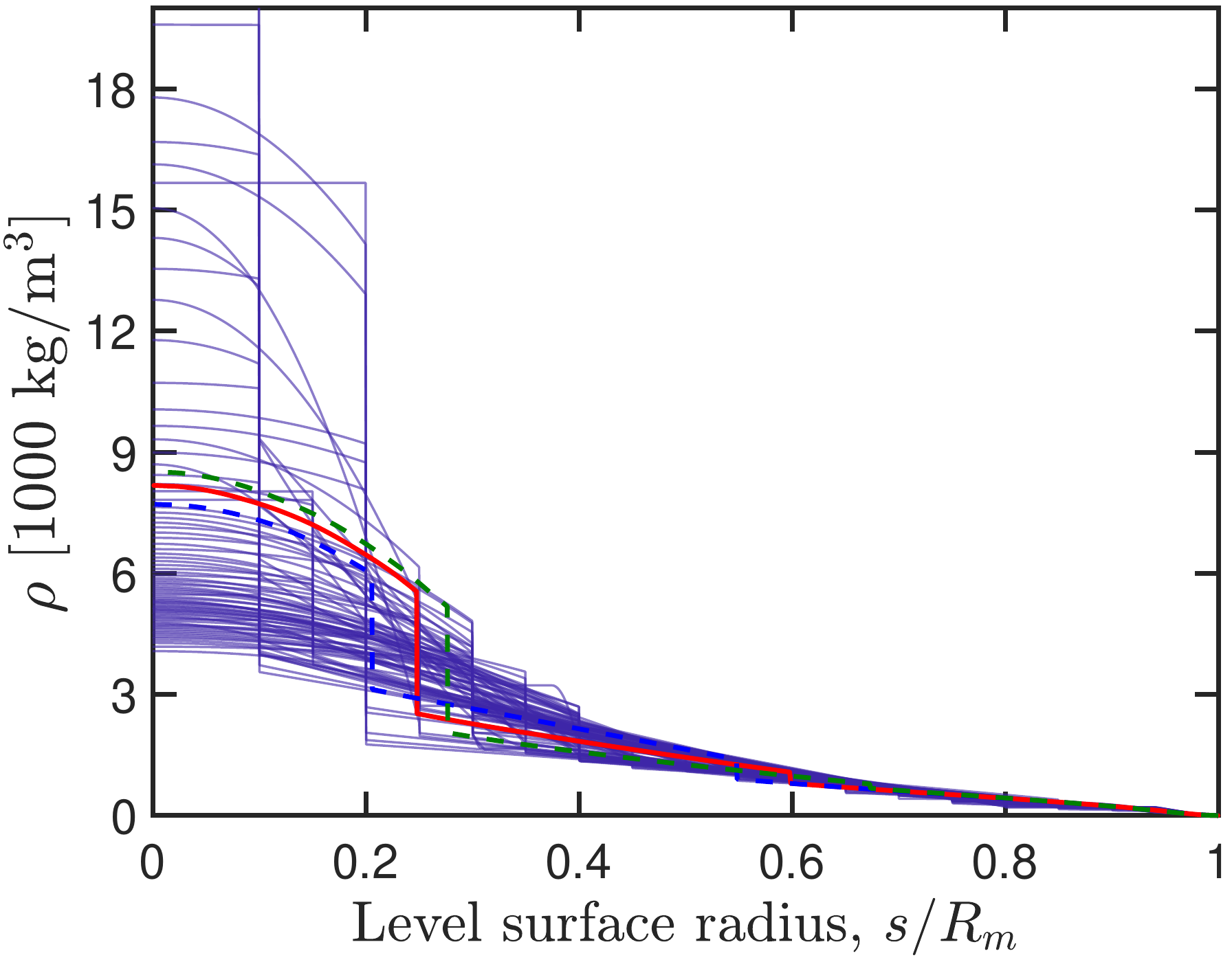}
\caption{Visualization of the posterior probability distribution of
Saturn interior density profiles. \emph{Left}: The thick black line is the
sample-median of density on each level surface. The dashed lines mark the the 16th
and 84th percentiles and the dotted lines mark the 2nd and 98th percentiles;
between the lines percentile value is indicated by color. \emph{Right}: Several
hundred profiles covering the sampled range. By nature of the MCMC algorithm
regions of the figure where lines are closer together correspond to high
likelihood areas of parameter space. For comparison, three profiles derived by
physical models with a pure H$_2$O core \citep[][same profiles as in
fig.~\ref{fig:mank-typical}]{Mankovich2019} are overlaid.}
\label{fig:rhoposterior}
\end{figure*}

After obtaining an independent random sample from the posterior in parameter space
we examine the resulting distribution of density profiles.
Figure~\ref{fig:rhoposterior} is a view of the sample distribution. Density is
plotted against the normalized level-surface radius. In the left panel, the thick
black curve is the ensemble-median density at each radius and the shaded regions
indicate the width of the distribution. In the right panel a subset of the entire
sample is plotted, selected to illustrate the sample range. Regions of higher line
density (where the lines are closer together) correspond to high-likelihood areas
in parameter space, by the nature of the MCMC algorithm. The EOS-based profiles
from Figure~\ref{fig:mank-typical} (from M19) are overlaid for comparison.

\begin{figure}[b!]
\plotone{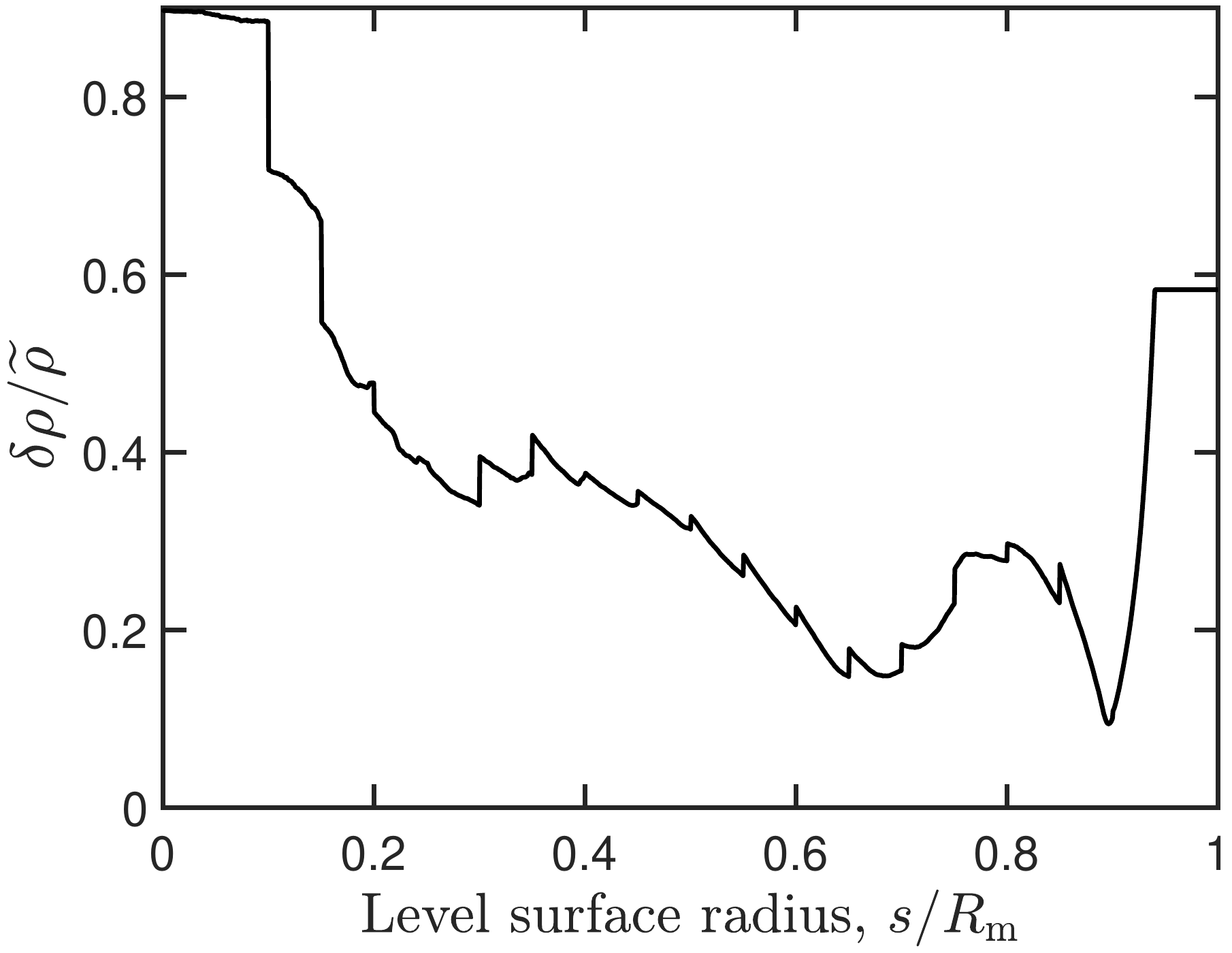}
\caption{Width of the distribution of density values found in the posterior
sample at each radius. The quantity $\delta\rho$ is the difference of 84th and
16th percentile values, giving the equivalent of a 2-sigma spread;
$\widetilde{\rho}$ is the sample-median density. The flat region near $s/\rem=1$
is a consequence of the relatively strong prior imposed in that region (see
appendix~\ref{app:chains}).}
\label{fig:sigmarho}
\end{figure}

Two insights are possible by inspection of Figure~\ref{fig:rhoposterior}. First,
from the left panel, the observed gravity can constrain the top half of the planet
much more strongly than it can the bottom half. This was expected (see
Figure~\ref{fig:contribs}) but it is worth emphasizing again that it is a
fundamental limitation of using gravity to probe the interior. This limitation is
with us to stay; it will not be completely removed by increasing the accuracy of
measurement or the precision of calculations. The same point is illustrated
quantitatively in Figure~\ref{fig:sigmarho} where the sample-spread of density
values is shown for each radius.

The second interesting feature, easier to spot in the right panel, is the
existence of density discontinuities. Recall that our parameterization allowed up
to two discontinuities; it did not \emph{require} any. Indeed many profiles in the
ensemble lack one or both discontinuities, the interpretation being that they lack
a sharp composition or phase boundary.

\begin{figure*}[tb!]
\plottwo{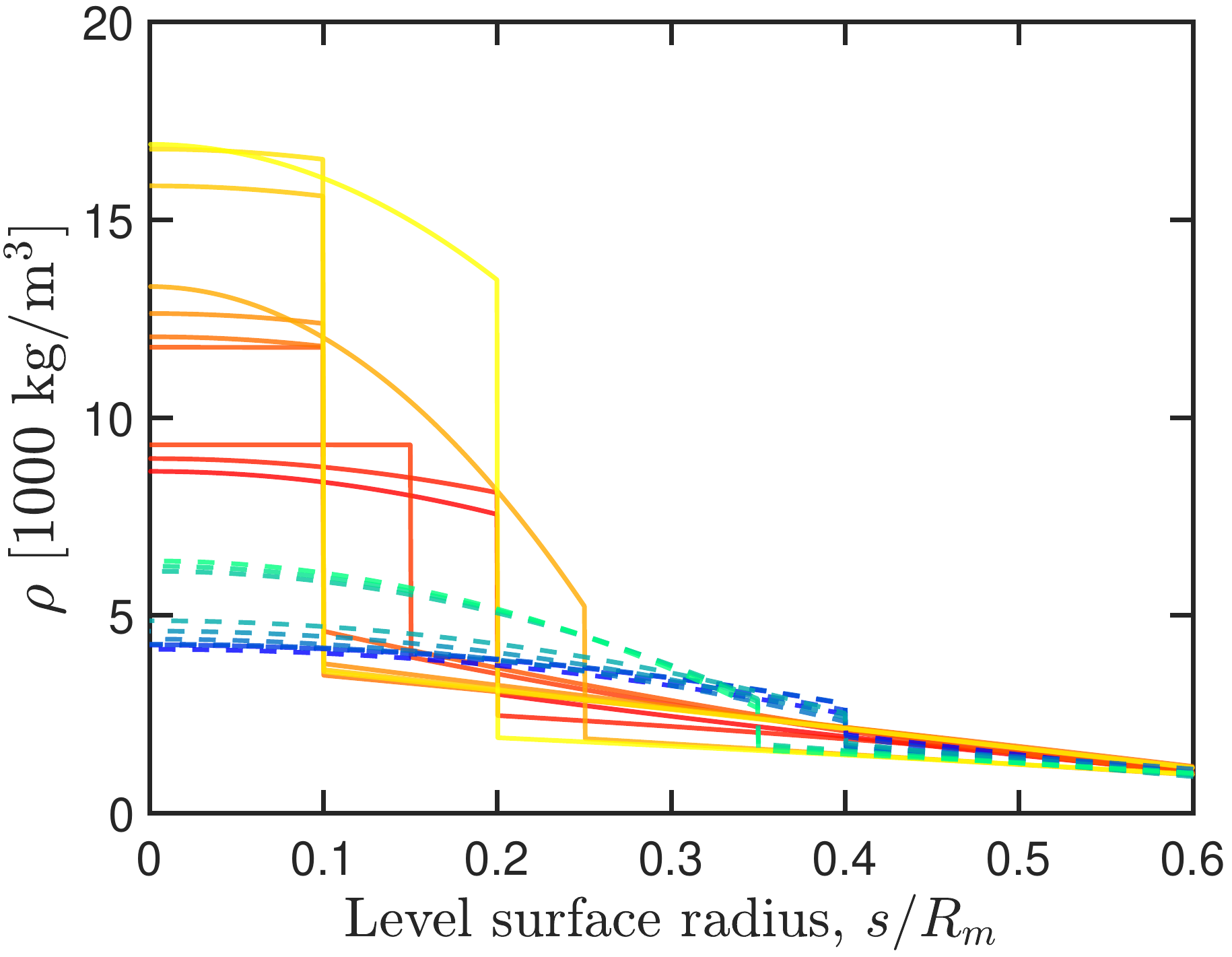}{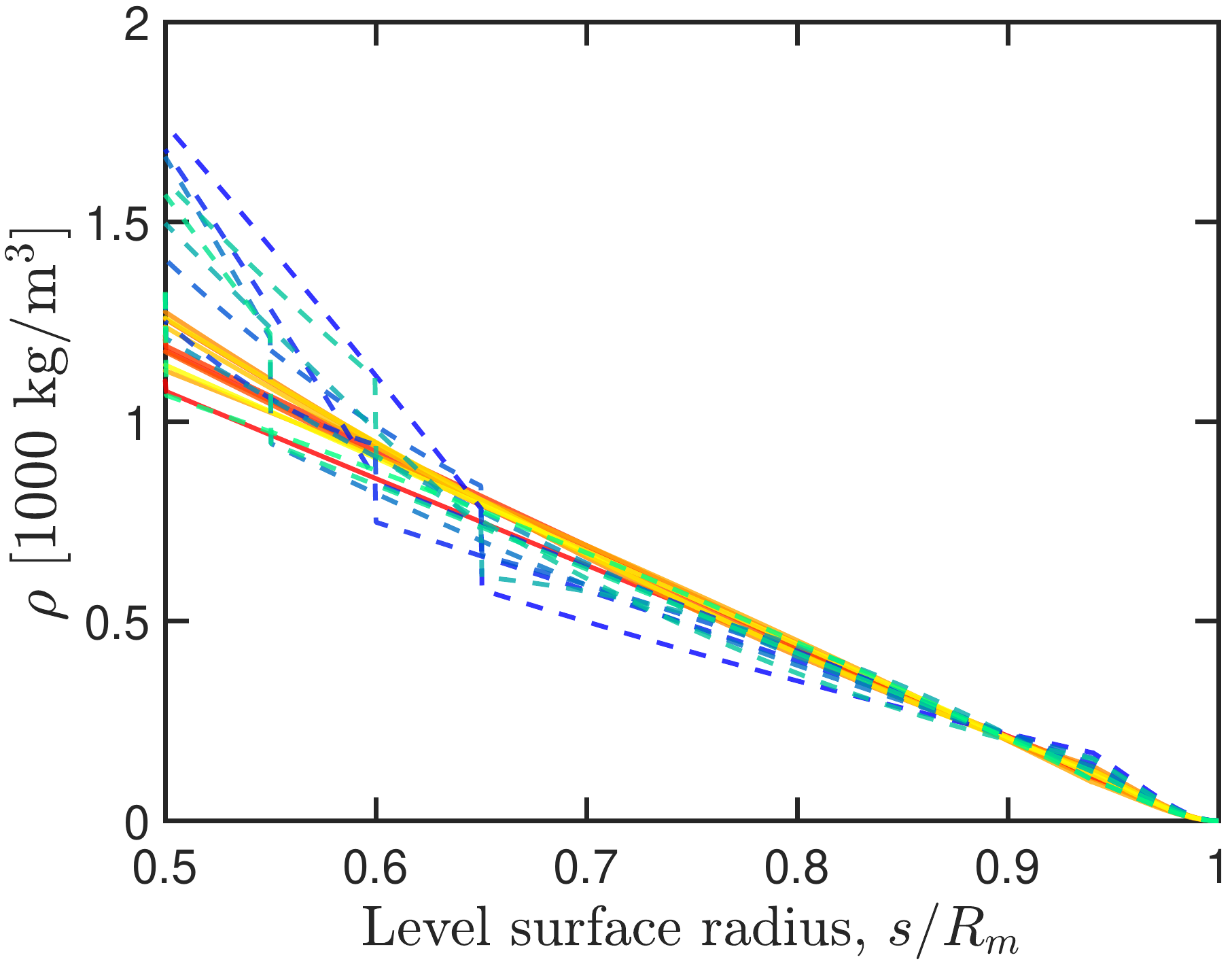}
\caption{\emph{Left}: A subset of profiles from the posterior distribution chosen
to illustrate the idea of compact (solid lines) versus diluted (dashed lines)
core. All have comparable likelihood values. A precise value of $\Delta\rho/\rho$
marking the difference between compact and diluted cores is hard to define (see
discussion in text). \emph{Right}: A subset of profiles from the posterior
distribution chosen to illustrate the possibility of continuous He abundance in
the envelope (solid lines) as well as the traditional idea of Helium rain
separating He-poor and He-rich layers (dashed lines). Again, likelihood values of
both subsets are comparable and, again, a precise cutoff below which the curve is
considered continuous is not obvious.}
\label{fig:core_envelope_configs}
\end{figure*}

\begin{figure}[b!]
\plotone{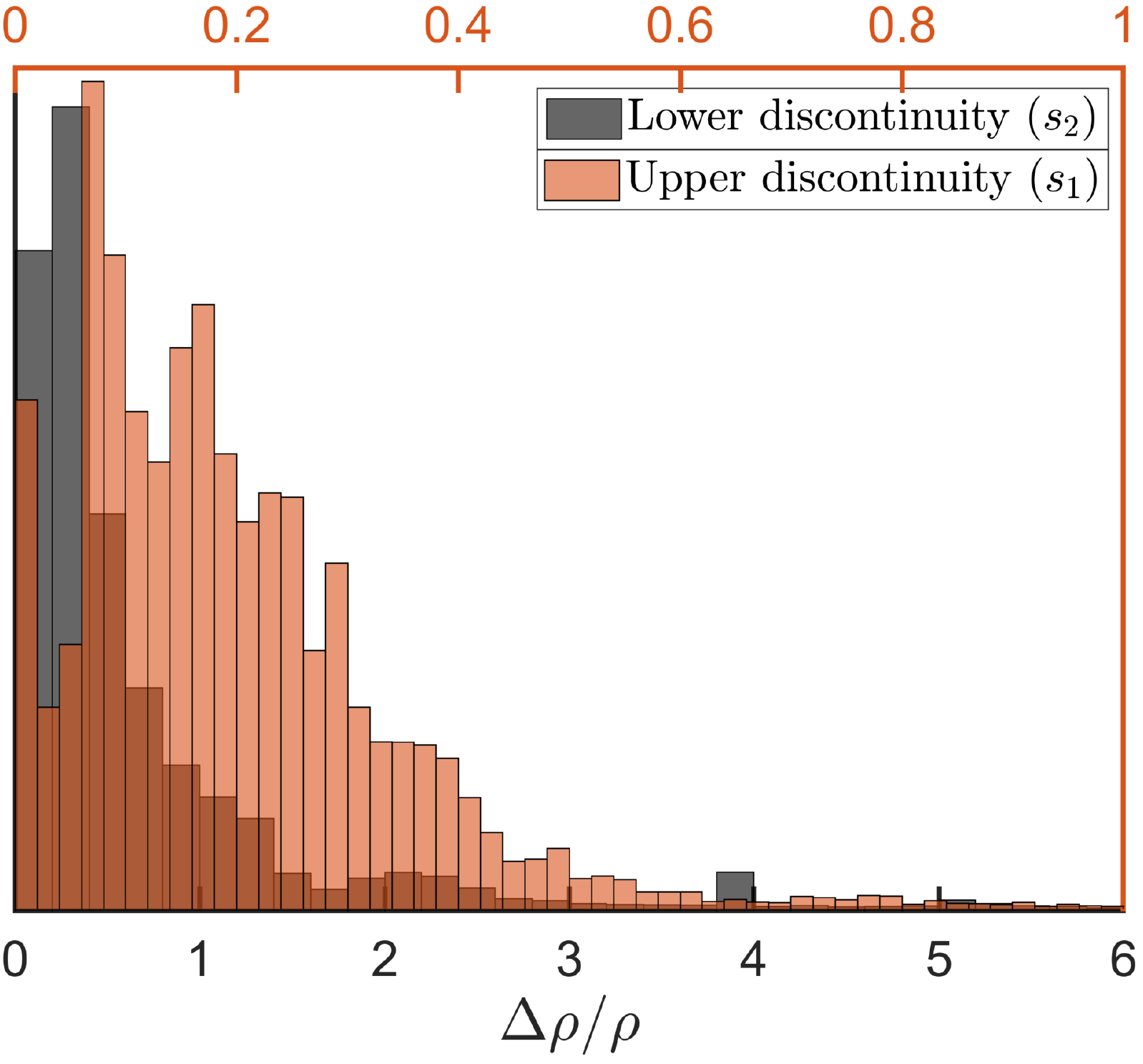}
\caption{Histograms of the density increase at the inner (bottom axis) and outer
(top axis) discontinuities, perhaps representing a phase or composition change.}
\label{fig:drho}
\end{figure}

The inner discontinuity, at $s=s_2$, was meant to represent the possibility of a
distinct core. Many density profiles indeed show a discontinuity pronounced enough
to clearly indicate a transition to a heavy-element core, while in many others a
much smaller density jump is observed instead, indicating a more subtle
composition change, consistent perhaps with the idea of a fuzzy/dilute core
\citep{Helled2017} or compositional gradients \citep{Leconte2012}. For
illustration, subsets from the sample with and without a pronounced discontinuity
are shown in Fig.~\ref{fig:core_envelope_configs} (left panel). To put a
probability value on the existence of a heavy-element core we can look at the
distribution of $\Delta\rho/\rho$ at $s=s_2$, shown in Figure~\ref{fig:drho}, but
it is not clear what ``cutoff'' value should indicate the core/no-core property.
For reference we can look at previously published, EOS-based models where a core
was explicitly assumed. In such models the relative density jump at the core
boundary exhibits a wide range, from as low as $\Delta\rho/\rho\approx{0.3}$ to
more than tripling the density \citep[e.g.][Mankovich et al.~in
review.]{Vazan2016,Mankovich2019}. With this in mind perhaps the most precise
statement to make is that at least half the density profiles in our sample show a
discontinuity pronounced enough to be consistent with a heavy-element core
transition.

The outer discontinuity, at $s=s_1$, was meant to represent the possibility of an
abrupt change in density in the envelope, where the He mass fraction changes from
depleted (relative to protosolar values) to enriched. This transition was expected
based on theoretical considerations about the miscibility of He in H, in the
region of phase space where hydrogen undergoes a molecular-to-metallic phase
transition \citep{Stevenson1975}. An abrupt change in He mass fraction, $Y$, is
often explicitly included in interior models, usually as a free parameter. However
this two-layered envelope is only one possible arrangement among many, including a
continuous $Y$ gradient. For example, if Saturn's interior is sufficiently cold
for He phase separation to occur in the first place then its true helium
distribution is determined by the precise solubility of helium throughout the
metallic interior, quantitative predictions of which have been made from first
principles simulations \citep{Schottler2018}. Applying these predictions
self-consistently to Saturn interior models, \newmankt{} find equilibrium profiles
wherein helium abundance increases continuously with depth inside
$P\approx{2}\unit{Mbar}$ with the exception of a single deep discontinuous jump in
density connecting the helium gradient region with a deeper pool of undissolved
helium-rich material.

The sampled profiles include both continuous-density envelopes as well as
those with small density jumps at $s_1$. While a density jump does not uniquely
correspond to a jump in He abundance, a continuous $\rho(s)$ does imply continuous
$Y(s)$. As seen in Figure~\ref{fig:drho}, both possibilities (illustrated in the
right panel of Fig.~\ref{fig:core_envelope_configs}) are consistent with the
observed gravity.

\begin{figure}[b!]
\plotone{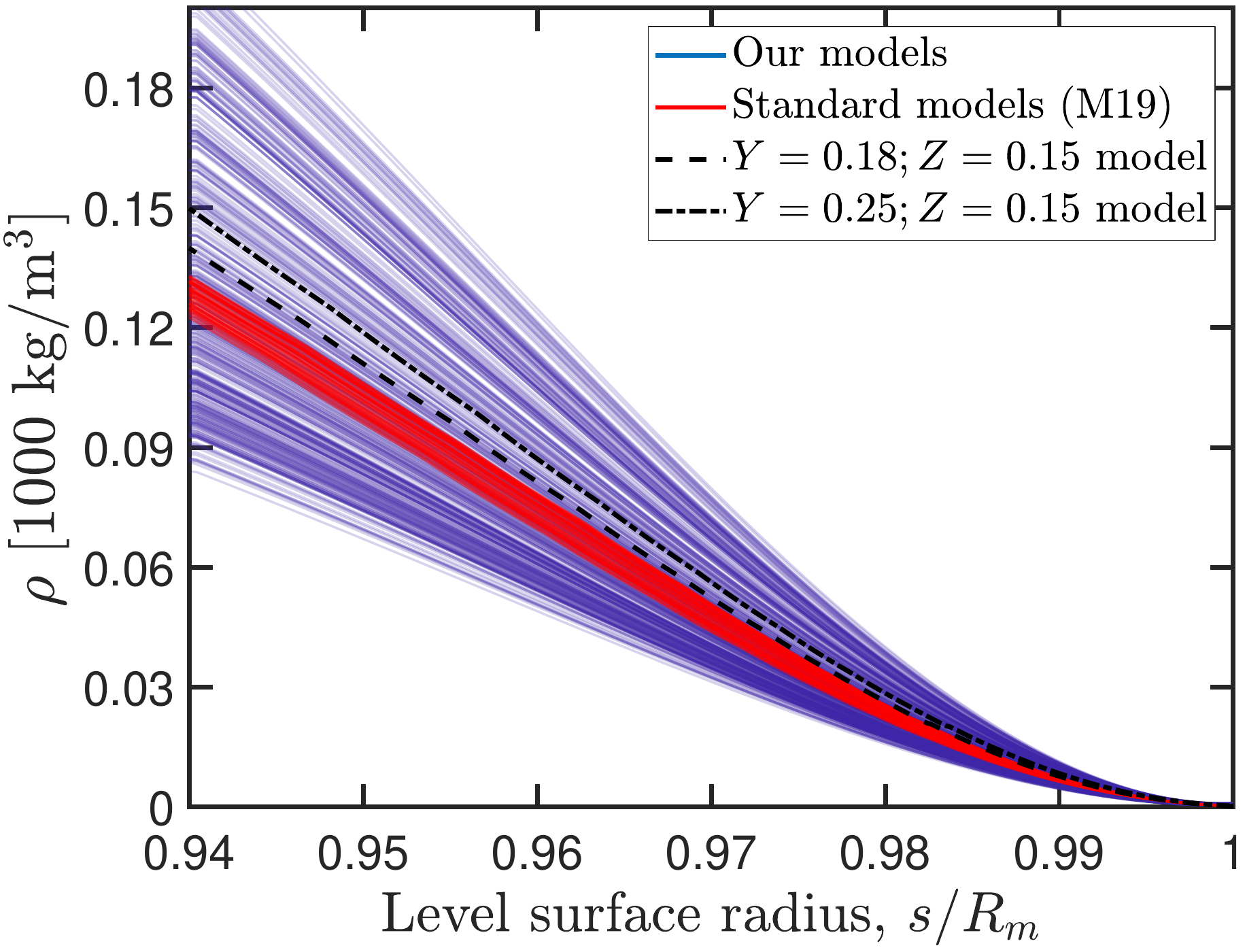}
\caption{Density profiles in the upper envelope derived from our
composition-agnostic sample (purple), traditional three-layer models with standard
values of $Z\lesssim{0.05}$ (\mmfm, red), and two three-layer models that have a
much higher value of $Z=0.15$ consistent with atmospheric abundances (black dashed
and dot-dashed) \edit1{but do not fit the observed gravity field}.}
\label{fig:zrich}
\end{figure}

Perhaps the most useful aspect of the empirical-model approach is the possibility
of finding unexpected solutions that can never arise where explicit composition
modeling is used. Figure~\ref{fig:zrich} takes a closer look at the density
solutions, this time focusing on the low pressure region above
$\about{2\unit{GPa}}$. A long-standing point of tension in Saturn modeling is that
Saturn's atmosphere is known to be enriched in heavy elements \citep{Atreya2016},
showing about ten times the solar abundance for C, P, S (seen in CH$_4$, PH$_3$,
and H$_2$S). That is, a ``metals'' mass fraction of $Z\approx{0.15}$ for the H/He
envelope. However, modern Saturn models, even post Grand Finale, find a fit to the
gravity field only with a much lower $Z<0.05$ in the outer H/He envelope
\citep{Iess2019,Mankovich2019,Nettelmann2013}. Traditional models cannot match all
of the atmospheric constraints, suggesting that we do not have a complete picture
of Saturn's interior. In contrast, we find that a natural outcome of our
composition-agnostic approach is density-enhanced outer layers. For comparison we
show two traditionally calculated models with $Z=0.15$ in their envelopes and they
fall nicely inside our posterior sample. The density profiles in our sample
(purple lines in Fig.~\ref{fig:zrich}) fit the measured gravity field while the
traditional model cannot, with such high $Z$ fraction, because of quite different
deeper interior profiles (Figure~\ref{fig:rhoposterior}).

\subsection{Inferences on possible composition\label{sec:composition}}
\begin{figure*}[tb!]
\plottwo{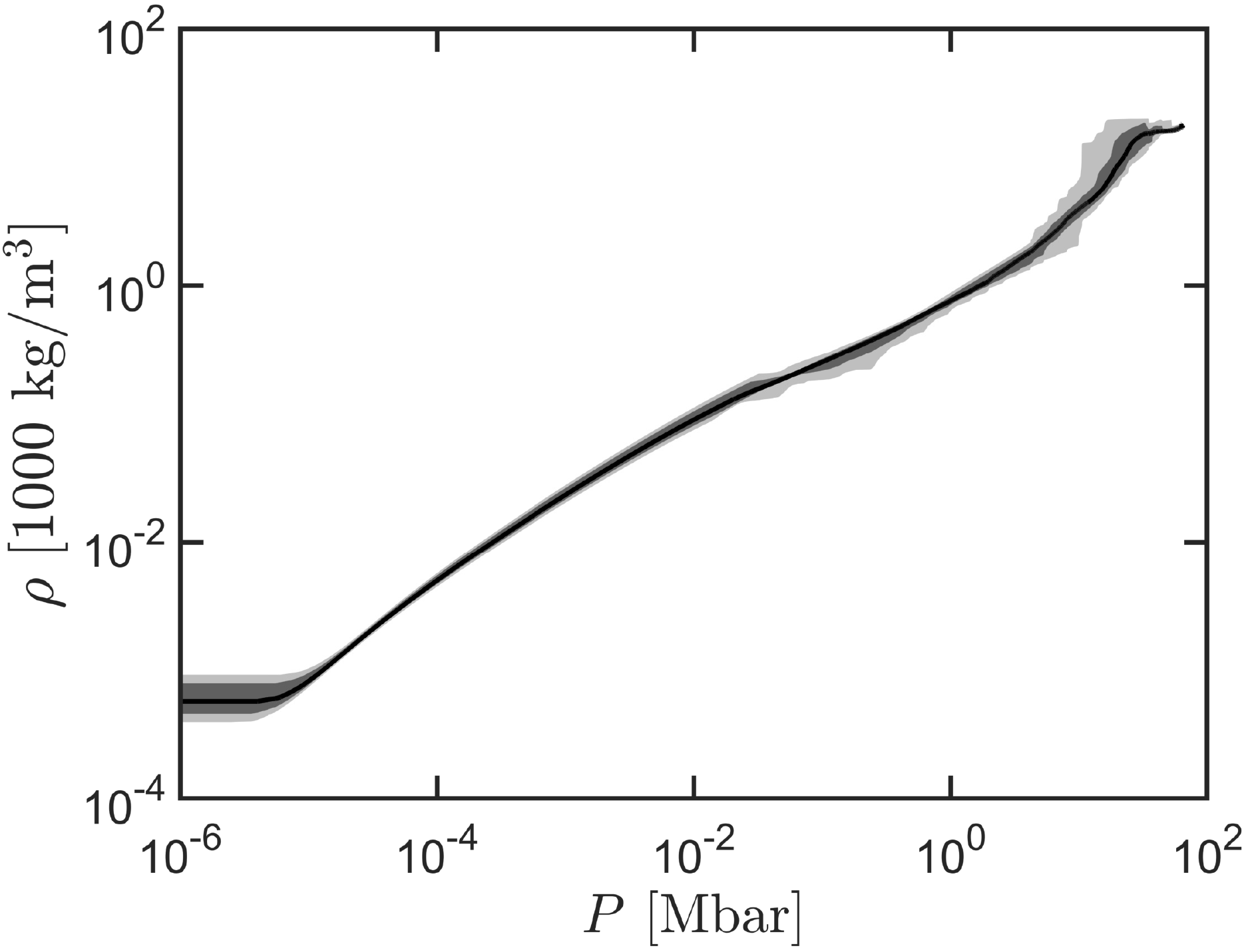}{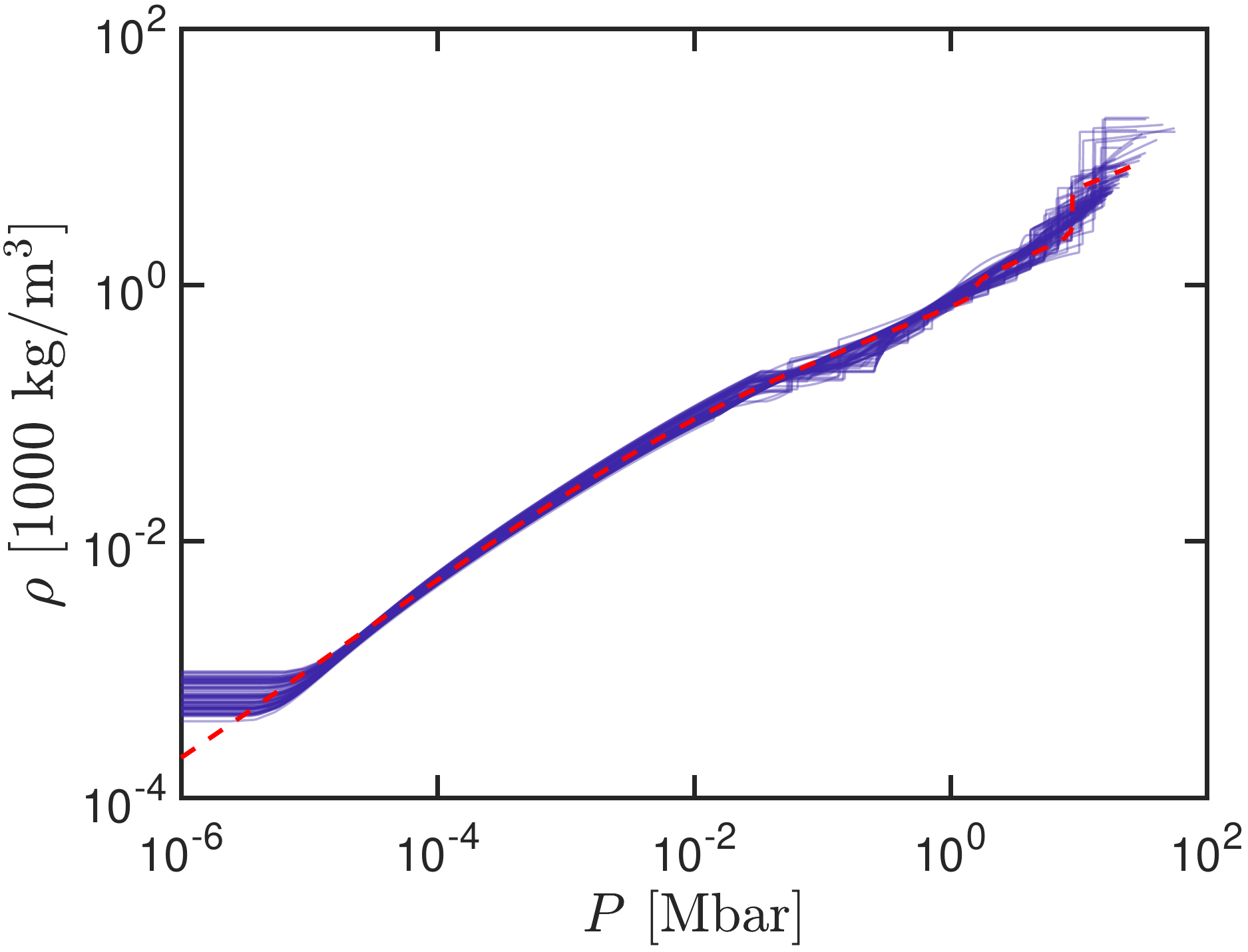}
\caption{Visualization of the posterior distribution of empirical Saturn
barotropes (pressure-density relations). \emph{Left}: The sample median (thick
black line), 16th to 84th percentile range (dark gray shaded), and 2nd to 98th
percentile range (light gray shaded). \emph{Right}: Thinned subset of sampled
barotropes. The median barotrope implied by the physical models of \mmfm{} (red
dashed line) overlaid for comparison.}
\label{fig:rhoofp}
\end{figure*}

With each $\roofs{}$ profile is associated a corresponding pressure profile,
$\pofs$, by the assumption of hydrostatic equilibrium. Combining the two profiles
to eliminate the radius variable results in a unique pressure-density relation,
often called a \emph{barotrope}. The posterior distribution of Saturn barotropes
implied by our sample is shown in Figure~\ref{fig:rhoofp}. By itself the barotrope
distribution does not provide much new insight, however it serves as the basis for
the derivation of implied constraints on composition, by comparison with known
equations of state, described next.

So far we have focused our attention on what the gravity implies \emph{directly}
about the interior, avoiding additional assumptions. We now wish to see what can
be inferred about the planet's composition; some assumptions and approximations
become necessary. The reason is that the density and pressure are not determined
solely by composition; the thermal structure is a separate, and unknown, variable.
Although the 1-bar temperature (to be used as a boundary condition) can be
determined by observation, the interior thermal profile is unknown unless we make
the strong and not entirely justified assumption of a single adiabatic profile
extending at least some fraction of the way down into the planet
(sec.~\ref{sec:assumptions}).

A possible approach is to compare the empirical barotropes obtained above to some
reference barotrope and examine the ``residual'' density for possible constraints
on composition. Deviations of the density in the sampled profiles from this
reference are due to a combination of the actual composition being different from
the assumed reference and of the real temperature profile being different from
adiabatic.\footnote{And if the reference barotrope was constructed using a
theoretical equation of state than of course there is an additional source for the
deviation -- the accuracy of the underlying EOS.} This degeneracy means that we
can only hope to estimate \emph{bounds} on composition, rather than a nominal
value.

In detail the calculation is this:  Given the density $\rho$ and pressure $P$ on a
level-surface with mean radius $s$ we can compute $\robg=\rho\sub{bg}(P)$ using a
background (bg) EOS and an assumed thermal gradient to compute a background
barotrope. The residual density, $\rho-\robg$, is already instructive, but we can
further compute $\rofg=\rho\sub{fg}(P)$ using a foreground (fg) barotrope
for heavy elements (water or rock) with the same pressure and temperature as the
background. The heavy element mass fraction $Z$ then follows from the additive
volume formula,
\begin{equation}\label{eq:volumeadd}
\frac{1}{\rho}=\frac{1-Z}{\rho\sub{bg}} + \frac{Z}{\rho\sub{fg}}.
\end{equation}
The mass fraction $Z$, calculated with different choices for the foreground EOS,
can be used to constrain the heavy element content consistent with the sampled
density profiles.

\begin{figure}[b]
\plotone{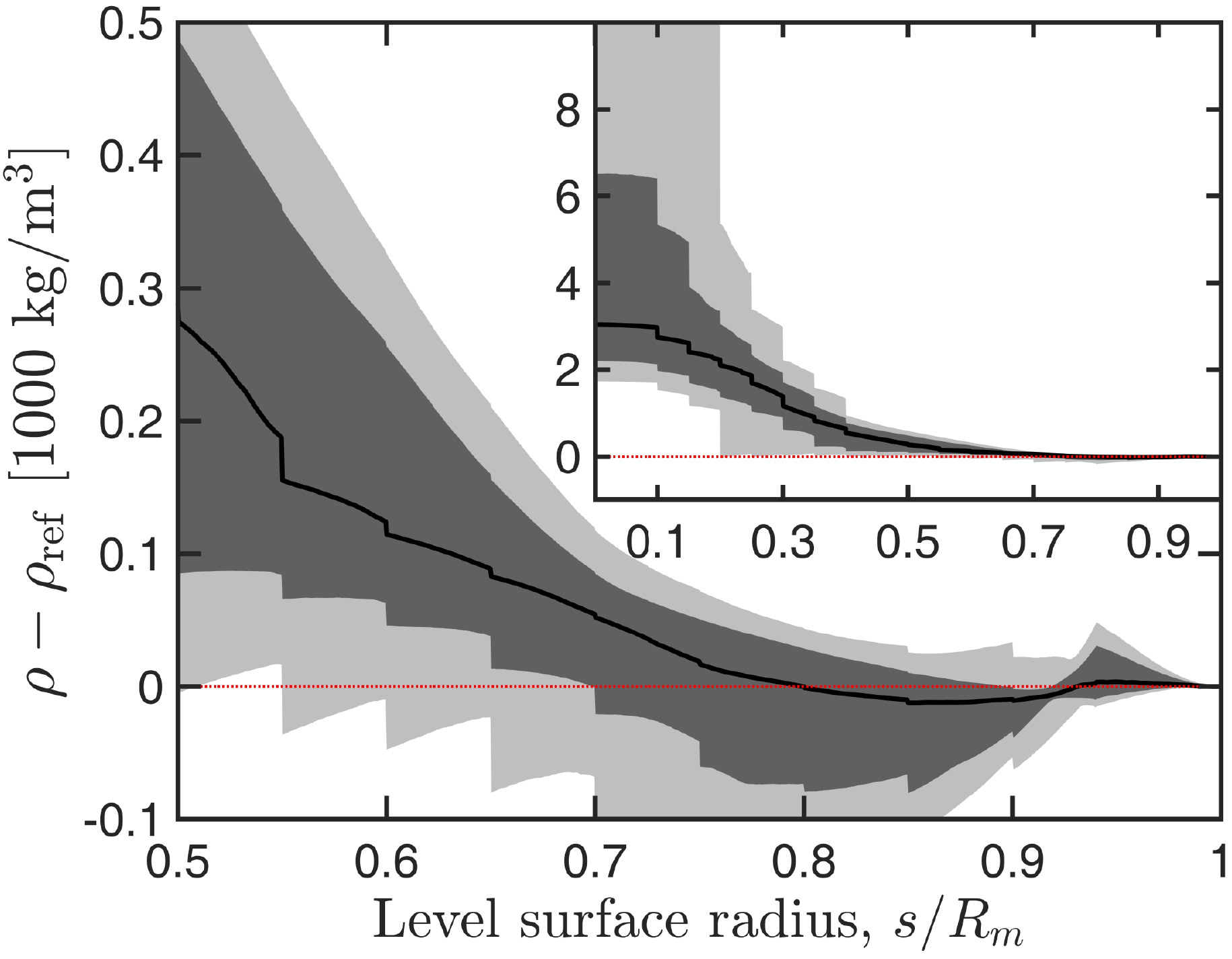}
\caption{Residual density above a background derived from a reference adiabat
calculated for a H/He mixture with He mass fraction $Y=0.275$ and
$T(1\unit{bar})=140\unit{K}$. The thick curve is the sample median and the dark
and light shaded regions include 68\% and 96\% of the sample,
respectively.}
\label{fig:resrho_Y275}
\end{figure}

In the simplest case our background can be a mixture of only hydrogen and helium
in protosolar mass fraction with an adiabatic temperature gradient. We use the EOS
of \citet{Saumon1995} to generate pressure-density points for H ($X=0.725$ by
mass) and He ($Y=0.275$ by mass) with constant entropy corresponding to a
temperature $T=140\unit{K}$ at a pressure of $P=1\unit{bar}$. The residual density
of the sampled profiles relative to this background is shown in
Fig.~\ref{fig:resrho_Y275}. Clearly, there is an excess density compared to the
adiabat in the regions of the planet below $70\%$ of the planet's radius, which
becomes extreme in the inner $30\%$. If a lower $Y$ reference adiabat were chosen
in the outer layers, larger density excess would be needed.

\begin{figure}[b]
\plotone{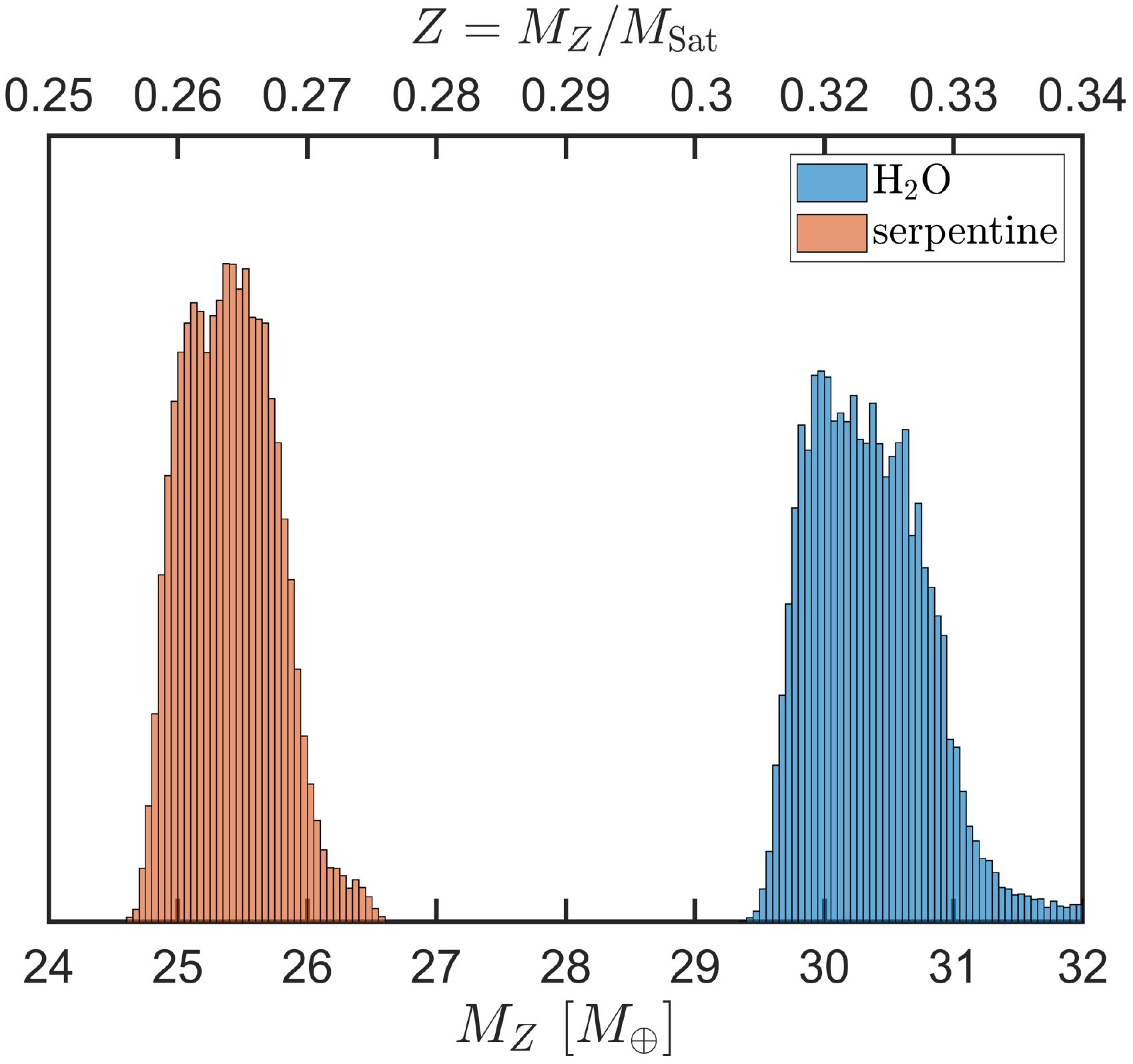}
\caption{Residual mass in heavy elements and corresponding residual bulk
metallicity assuming either pure H$_2$O ice or pure serpentine rock EOS. These are
end members of what is likely a mixture of both materials in unknown ratio. The
figure shows the mass in heavy elements inferred with a reference density based on
a H/He adiabat with $Y=0.275$ extending throughout the planet, and should be
interpreted as an upper bound since it excludes the possibility of a \emph{pure}
heavy-element core.}
\label{fig:ZMASS}
\end{figure}

Next, using a foreground EOS for either pure water ice
\citep{French2009,Thompson1990} or pure rock \citep{Thompson1990} we apply
eq.~(\ref{eq:volumeadd}) to each of the sampled profiles. What we obtain is an
empirical probability distribution of the heavy element content in Saturn's
interior. In Figure~\ref{fig:ZMASS} we plot a histogram of this distribution,
which should be taken as an estimated upper bound rather than a precise
distribution, given the assumptions underlying this calculation. These values are
typically higher than those from standard models because the excess heavy
elements, even at high pressure where one might expect a pure-heavy-element core,
are here always determined as an excess density over that of the lower density
H/He.

\begin{figure}[b]
\plotone{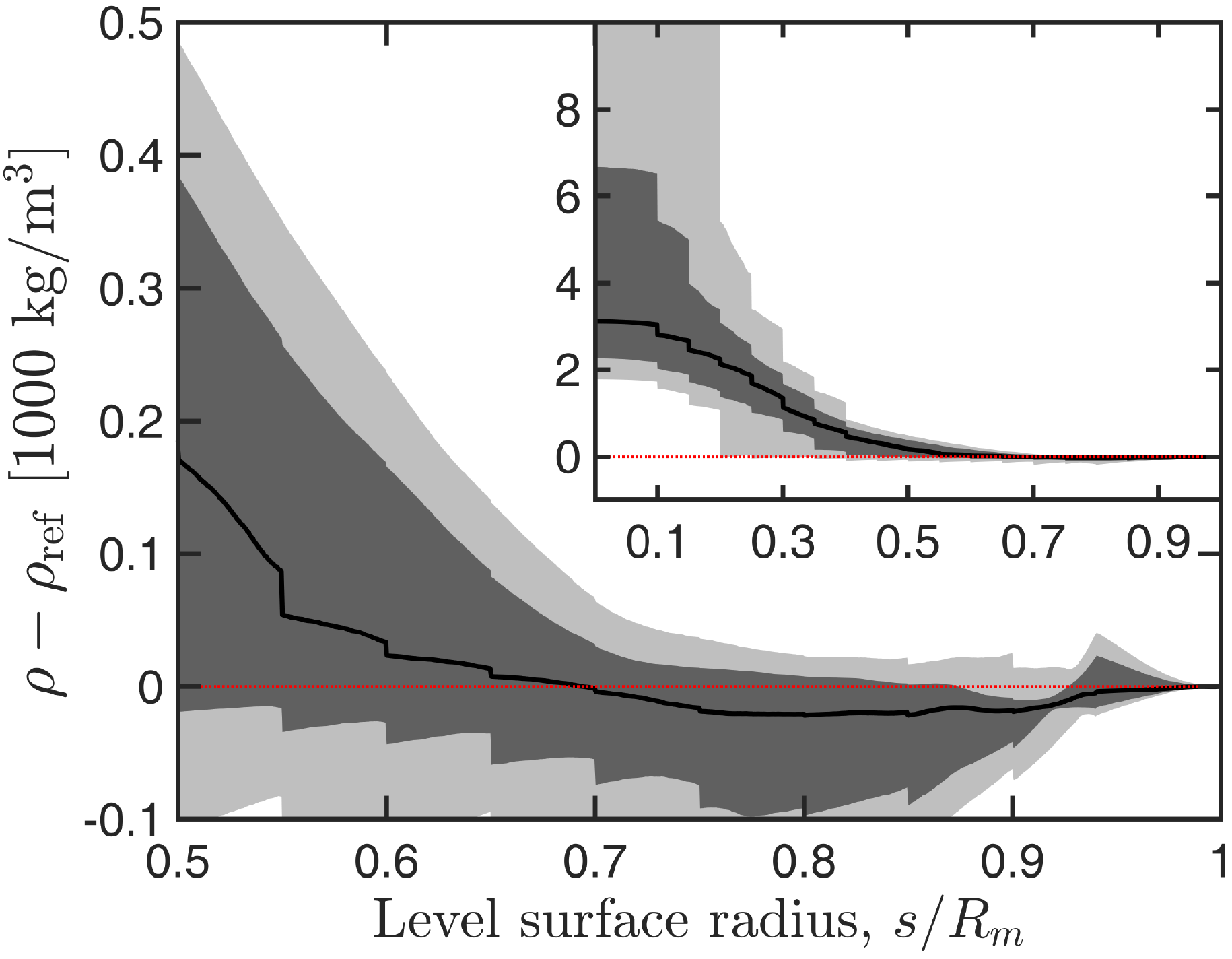}
\caption{Same as Fig.~\ref{fig:resrho_Y275} but background density derived from
adiabat calculated for $Y=0.1$ and $Z=0.135$.}
\label{fig:resrho_Y100_Z135}
\end{figure}

\edit1{The same calculation can be repeated for different internal thermal
structures or with different choices of background and foreground EOS. There is no
need to repeat the time consuming task of sampling the density profiles. As a
second example, Figure~\ref{fig:resrho_Y100_Z135} shows the residual density
relative to a background adiabat with a lower value of $Y=0.1$ and with heavy
elements mixed in with a ratio $Z=0.135$ in line with atmospheric constraints at
$\about{9}\times$ solar enrichment. This adiabat was calculated using the
\citet{Militzer2013} and \citet{Saumon1995} EOSs as combined by \citet{Miguel2016}
to treat arbitrary H-He mixtures, and ANEOS \citep{Thompson1990} for water ice.}

\edit1{The median density of the empirical models is consistent with the adiabatic
density down to $r/\rem\approx{0.95}$, is somewhat lower down to $r/\rem
\approx{0.7}$ then climbs again. But the median is not the distribution. To  a
``1-sigma'' level the adiabatic density profile is consistent with the empirical
samples down to at least $r/\rem=0.35$.}

\begin{figure}[tb!]
\plotone{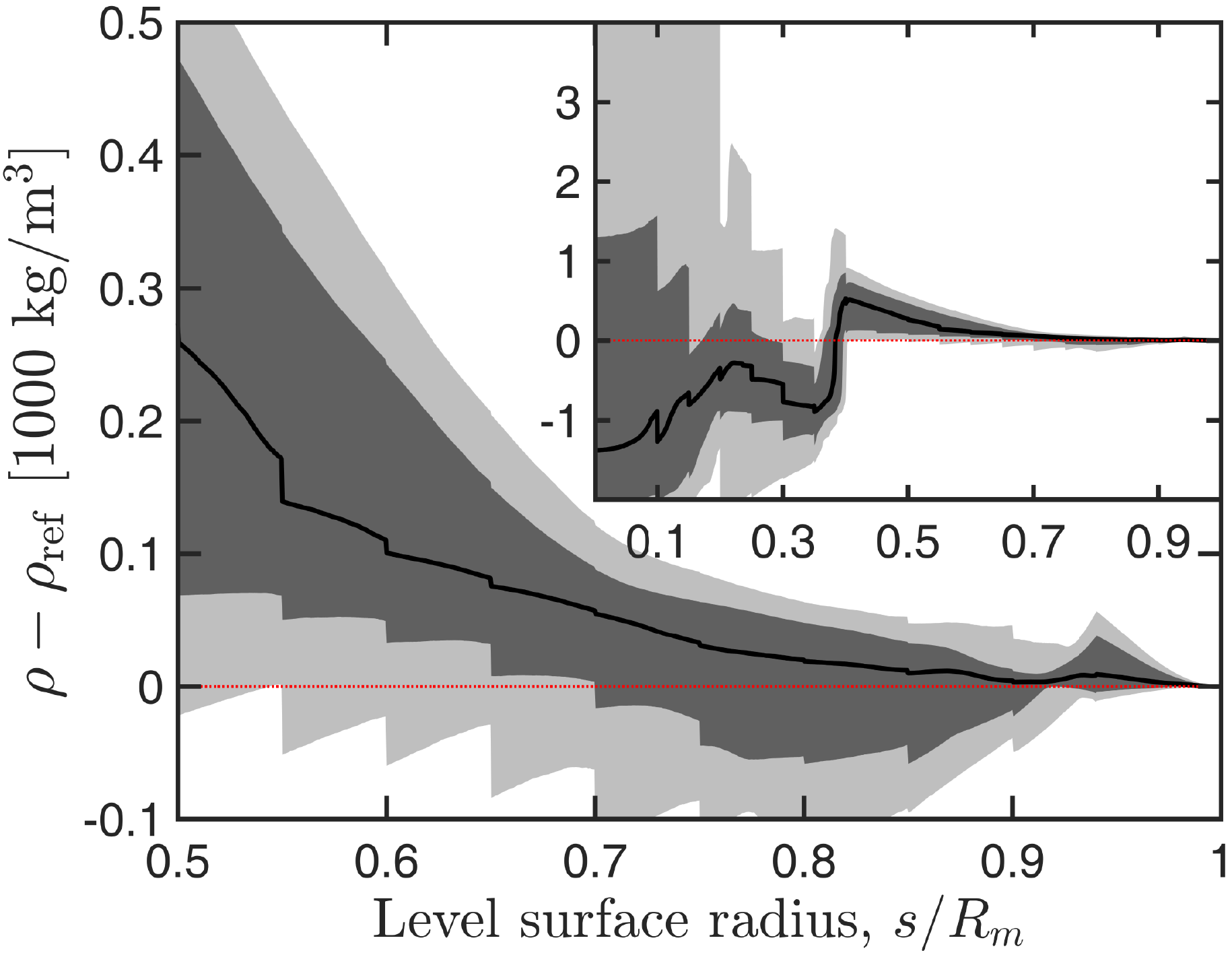}
\caption{Same as figure \ref{fig:resrho_Y275} but with the background density
defined by the end state of an evolution model \newmank{}.}
\label{fig:evol-resrho}
\end{figure}

As a last example we use as our reference background the end state of a recent
Saturn thermal evolution model \newmank{}. This structure derives from calculation
of the cooling of Saturn's interior, including the phase separation of He from H
in the interior. This leaves the molecular part of Saturn's \edit1{envelope
depleted in He (to $Y=0.07$) and the inner regions extremely He-enriched
($Y\gtrsim0.9$ inside $0.24\lesssim{}s/\rem\lesssim{0.37}$). The model includes a
uniform metallicity $Z=0.048$ in the envelope, with a dense $Z=1$ core below
$s/\rem=0.24$.} The model matches Saturn's present-day radius and intrinsic
luminosity but does not attempt to match the observed gravity field. Subtracting
this background density we again examine the residual density in the sampled
profiles (Figure~\ref{fig:evol-resrho}). \edit1{Compared with this particular
evolution model, a majority of our gravity solutions produce quite consistent
densities throughout the interior of the planet. That the density residual is
consistent with zero virtually everywhere in the planet indicates (1) that this
rather extreme level of helium depletion in the molecular envelope is permitted by
Saturn's observed gravity field; (2) that the overdensity of our models at depth
($s/\rem\lesssim{0.2}$) compared to constant-composition adiabats
(figures~\ref{fig:resrho_Y275} and \ref{fig:resrho_Y100_Z135}) can indeed be
provided by a central core of dense material, as expected; and (3) a helium-rich
shell surrounding such a core is also consistent with the low-order gravity
field.} These observations are at the 1--2$\sigma$ level, i.e., solutions also
exist that do not follow these trends.

%

\section{Discussion and Conclusions}
\label{sec:discussion}
In this paper we presented an empirical approach to using gravity data to explore
the interior structures of fluid planets and applied it to Saturn using data from
\emph{Cassini}'s Grand Finale orbits. Here we wish to summarize our findings for
Saturn, and about planetary interior modeling in general, and to consider the
strengths and weaknesses of our ``density first'' approach, versus traditional,
composition-based modeling.

First, a point that was already made above but bears repeating: \emph{Gravity data
alone offers robust but loose constraints.} The great variety of density profiles
included in our sample may seem surprising and counter-intuitive but it is an
unavoidable consequence of using an integrated quantity, in this case the external
potential, to study the spatial distribution of local quantities, in this case,
the interior density and all properties of the planet that derive from it. Without
imposing additional constraints we necessarily obtain non-unique solutions, and
this is a separate and more fundamental limitation than the problem of uncertainty
in the data and/or calculation.

As a result, the main finding we can report on, with respect to Saturn, is to
confirm the well-known but often underappreciated suspicion that solutions to
Saturn's gravitational potential field exist that do not conform to a simple model
of a few compositionally homogeneous and thermally adiabatic layers. While this
may not be a surprise, it is nevertheless a previously unproven result. We could
not know, a priori, whether the non-uniqueness of gravity solutions would
translate to a narrow range of allowed interior structures or to a wide variety,
as appears to be the case.

We can contrast this with the seemingly more informative but less robust outcomes
from traditional models. These are often able to report narrow ranges for a
number of key quantities (typically core mass, bulk metallicity, H/He envelope
metallicity, atmospheric helium depletion) that were the free parameters in the
chosen model. The trade off for these precise, straightforward estimates is their
unknown validity, being tied to very particular and often very simple \emph{a
priori} modeling framework for the planet. Conversely, the results we report on
here are of much wider validity, with the trade off of being much less specific
and more difficult to interpret.

Finally, our inferred heavy-element mass for Saturn relied on the SCVH EOS for
H-He. This widely-used EOS has been recently updated to be more thermodynamically
consistent \citep{Chabrier2019}. In the updated version, hydrogen is found to be
denser under Jupiter and Saturn conditions, in agreement with DFT calculations.
Therefore the heavy-element masses listed here are likely to represent upper
bounds. Clearly, a more detailed investigation of that topic in the future is
desirable.

\subsection{Narrowing down the posterior distribution}
It is certainly possible that a subset of the sampled density profiles can be
``disqualified'' based on other physical considerations, and indeed we consider
this a natural avenue for future work. Any reduction of the allowed solution space
will be an improvement, as it narrows down the probable actual structure of
Saturn. However any such reductions must be considered carefully, so that they do
not rely too strongly on implicit assumptions of the exact kind we decided to
avoid in the first place. Such low-hanging fruit as disqualifying unphysical
density inversions or density extremes had already been picked by passing an
appropriate prior probability function, $p(\V{x})$, to the MCMC sampler
(appendix~\ref{app:chains}). For instance, that is why the posterior sample does
not contain profiles with stationary points or with central densities much higher
than ${2}\times{}10^4\unit{kg/m^3}$. More subtle constraints, e.g., looking for
convective instabilities or checking pressure-density pairs against known
equations-of-state, require knowledge of the thermal state and inevitably require
additional assumptions.

A second and unrelated way to narrow the predicted distribution somewhat is to
``sharpen'' the likelihood function by including higher order coefficients and/or
with tighter uncertainties. Recall that $J_2$ and $J_4$ are known for Saturn with
better accuracy than was assumed in eq.~(\ref{eq:lossfunc}). The same is true for
Jupiter, and higher-order coefficients are also known, with decreasing accuracy,
for both planets. More precise calculation of the Js for a given density profile
can reach this level of accuracy, with the only down side being increased
computation time. \edit1{While this would be a worthwhile improvement it would
only be appropriate if and when the actual rotation state of Saturn is known,
including any dynamical and/or non rigid body components, to sufficient accuracy
from independent measurements. That would allow matching models of rigid
rotation with an adjusted gravity measurement reflecting a known correction due
to differential rotation.}

What we have accomplished is an understanding of a much fuller range of interior
density profiles for Saturn that are allowed by the planet's gravity field as
determined by the \emph{Cassini} Grand Finale, a data set that will likely not be
surpassed for some decades.  We hope that the allowed density distributions are a
long-lived data product that other workers may find useful as new ideas about
planetary formation, structure, and evolution emerge.  Such ideas can be compared
against the allowed interior density distributions that we have found here. To
facilitate this we archive the data products and analysis tools used in this
study, documented in sufficient detail to allow reuse and alternative analysis.
The archive can be found at
\href{https://doi.org/10.7291/D1P07G}{https://doi.org/10.7291/D1P07G}.


\acknowledgments
We would like to thank Dan Foreman-Mackey, Nadine Nettlemann, Bill Hubbard,
Burkhard Militzer, Sean Wahl, Daniele Durante, Luciano Iess for helpful advice on
several aspects of this work. We thank Tristan Guillot for his thorough and
thoughtful review. JJF acknowledges the support of NASA Cassini Participating
Science grant NNX16AI43G and the University of California grant A17-0633-001 to
the Center for Frontiers in High Energy Density Science. Resources supporting this
work were provided by the NASA High-End Computing (HEC) Program through the NASA
Advanced Supercomputing (NAS) Division at Ames Research Center, as well as the lux
supercomputer at UC Santa Cruz, funded by NSF MRI grant AST 1828315.

\software{Emcee \citep{Foreman-Mackey2013}}
\appendix

\section{A single-parameter description of low pressure region}\label{app:quartic}
As explained in sec.~\ref{sec:params}, when choosing a parameterization our goal is
to find the best compromise between a simple description, with a small number of
parameters suitable for MCMC sampling, and a general description, letting the
resulting $\roofs{}$ curves explore all reasonable profiles. Our choice of
parameterization by piecewise-quadratic functions was guided by, but much more
general than, previously published models that were based on physical EOSs and an
adiabatic temperature gradient \citep{Mankovich2019}. We found that, for the bulk
of the planet, a piecewise-quadratic $\roofs{}$ is able to capture the profiles
derived with a physical EOS and flexibly explore beyond them.

\begin{figure}[tb!]
\centering
\plotone{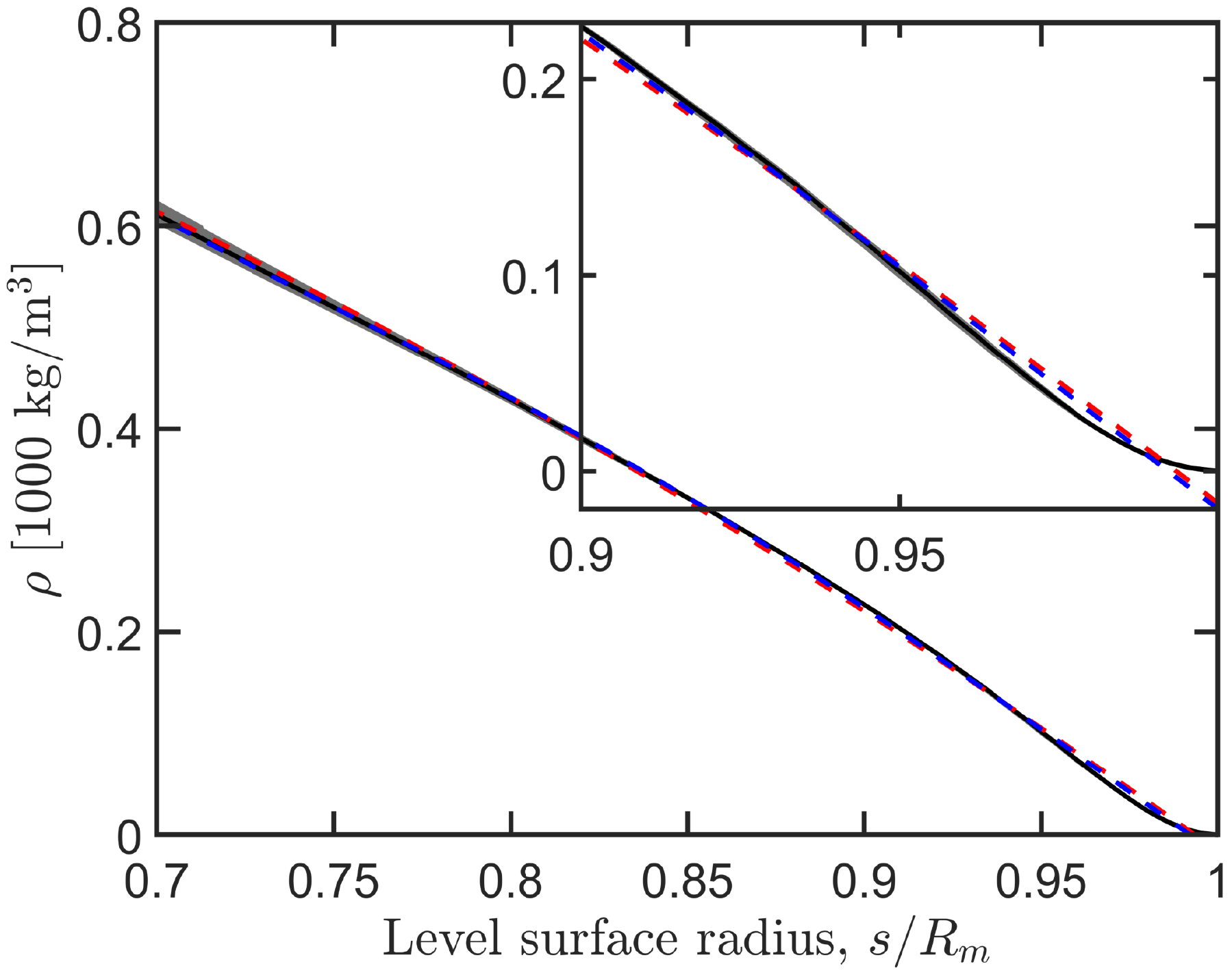}
\caption{A close look at the upper envelope of traditional Saturn models
\citep{Mankovich2019}, the same models seen in fig.~\ref{fig:mank-typical} in the
main text. The solid black curve is the ensemble median density at each radius,
with the light gray band denoting the 1-$\sigma$ variation. The red and blue
dashed lines are best-fit polynomials of degree 2 and 4 approximating the density
profile in the upper envelope as a whole. Neither is a good approximation in the
small region where $s\gtrsim{0.95}R_m$.}
\label{fig:quartic}
\end{figure}

However, empirical $\roofs{}$ profiles derived from this parameterization
inevitably exhibit a small but significant deviation from profiles derived by
physical models, in a small region at the top of the upper envelope.
Figure~\ref{fig:quartic} illustrates the problem. An inflection is seen in all the
EOS-based $\roofs$ curves, always in the neighborhood of $s/R_m\approx{0.95}$, and
this inflection cannot be captured if a single polynomial is used to approximate
the entire upper envelope (typically extending down to at least
$s/R_m\approx{0.65}$). Above the inflection point is a small region where $\roofs$
seems to follow a different curve. And yet this small region of the upper envelope
is one where physical models are most reliable, at a pressure and temperature
region where equations of state are well tested and where an adiabatic temperature
gradient is expected to exist. Closely matching the EOS-based models in this upper
region of the planet is an important way by which to constrain empirical models.

The obvious solution is to add an additional segment to the piecewise-polynomial
parameterization but unfortunately this cannot be implemented. The problem is not
simply that this would require 5-6 additional parameters and greatly complicate
the sampling process. More seriously, the small region in question contains
relatively little mass. Small changes in density in this region do not make a big
enough difference in the $J$ values, at our level of precision, to effectively
``drive'' the likelihood function. There is no reason to expect then that profiles
from the resulting posterior would be any more like the EOS-based ones.

Instead, we use a more explicit constraint, ad-hoc in nature, which achieves the
desired result of keeping the top of the envelope in empirical models similar to
EOS-based models while retaining enough flexibility to mimic varying composition.

\begin{figure}[tb!]
\centering
\plottwo{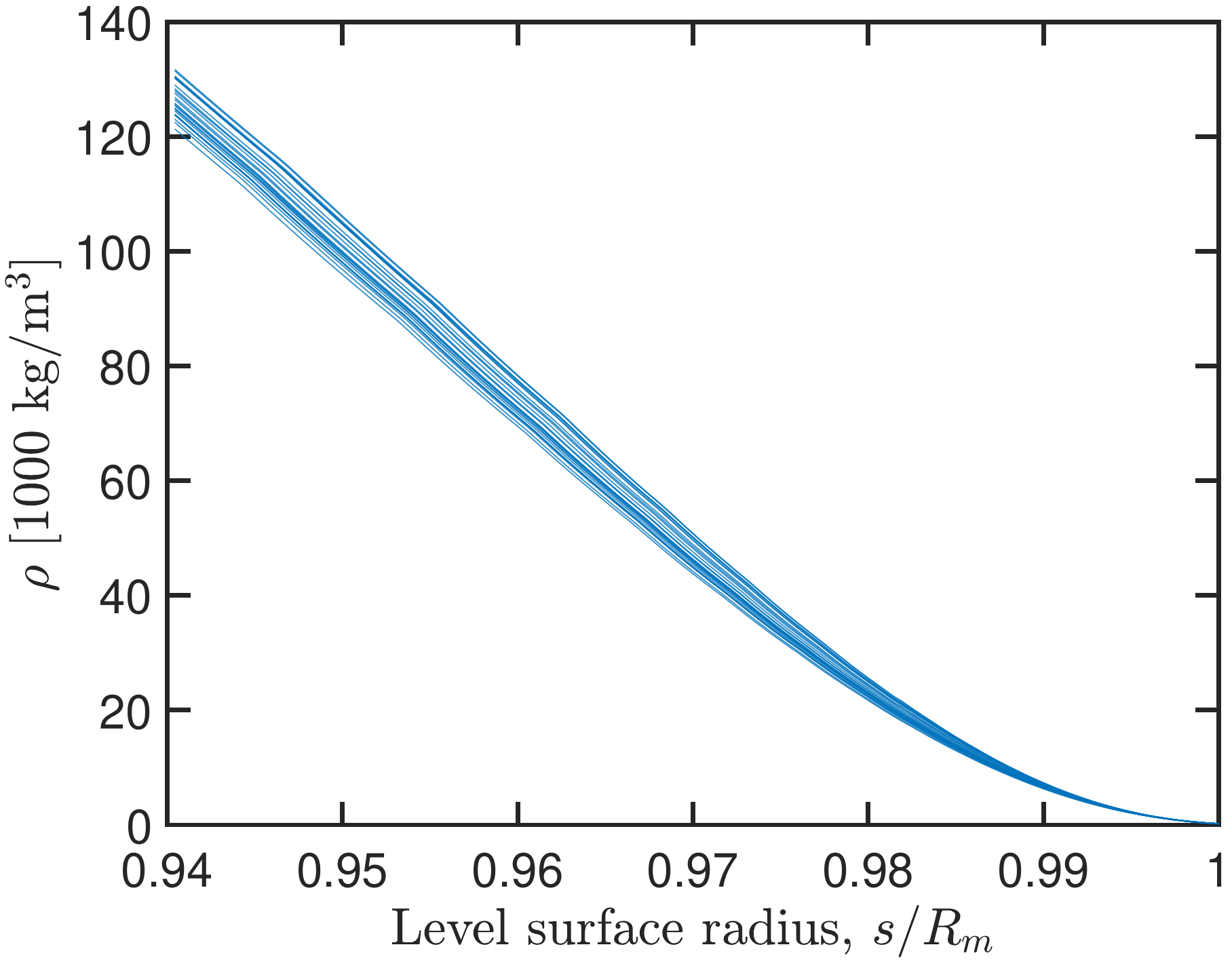}{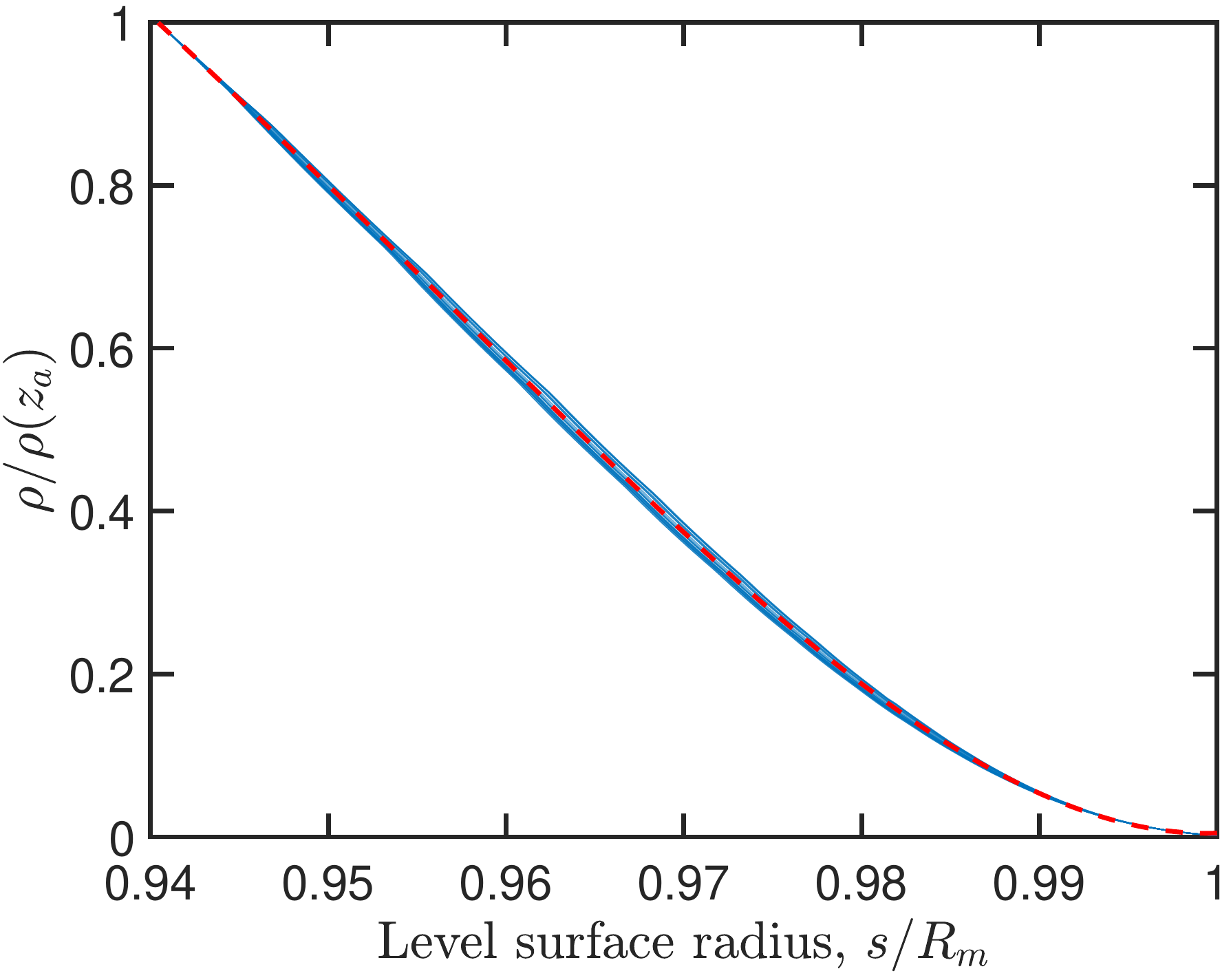}
\caption{The same profiles as in fig.~\ref{fig:quartic} truncated at
$s/R_m=0.94$, slightly below the inflection point. The apparently similar
curvature motivates us to fit them all with a \emph{single} quartic polynomial in
$z=s/R_m$ by normalizing to the value $\rho_a=\rho(z_a=0.94)$. This polynomial,
eq.~\eqref{eq:Q}, together with a posterior distribution of $\rho_a$ generate
density profiles that resemble the physical models in shape but are free to
explore ``around'' them.}
\label{fig:quartfamily}
\end{figure}

We examine the shapes of the density profiles of \mmfm{} in the region above
$z_a=s/R_m=0.94$ (fig.~\ref{fig:quartfamily}a). We choose this fixed point,
slightly below the inflection seen in the models, to make sure we always capture
the slope accurately. For $z<z_a$ we use the main parameterization by
piecewise-quadratics (eq.~\eqref{eq:sat-params} and appendix~\ref{app:fullparam}).
Above $z_a$, we find that all profiles can be fit by fourth-degree polynomials
(quartics) to excellent agreement. Further, if we denote $\rho_a=\rho(z_a)$ we
find that, for $z_a\le{z}\le{1}$, the curves $\rho(z)/\rho_a$ are equally well fit
by quartics (not surprising), and in fact that they can all be adequately
approximated by \emph{the same} quartic polynomial:
\begin{equation}\label{eq:Q}
\frac{\rho(z_a\le{z}\le{1})}{\rho(z_a)} \approx Q(z) =
(3\times{10}^{4})z^4 - (1.128\times{10}^{5})z^3 + (1.587\times{10}^{5})z^2 -
(9.914\times{10}^{4})z + (2.323\times{10}^{4}),
\end{equation}
shown in fig.~\ref{fig:quartfamily}b.
The profiles in fig~\ref{fig:quartfamily}a can be recovered, approximately but
with high fidelity, by multiplying the polynomial~\eqref{eq:Q} by a particular
value of $\rho_a$.

In other words, in the region $z\ge{z_a}$ the physical, EOS-based models form a
one-parameter family of quartic functions. We do not see special physical meaning
here. It is simply that the variation in density that originated from making
different choices about composition (i.e., the envelope's helium mass fraction and
metallicity) under the severe but, in this region, well-justified adiabatic
assumption, can be empirically captured by varying the value of
$\rho_a=\rho(s/R_m=0.94)$. To make sure that our empirical profiles are similar to
but not overly constrained by EOS-based models in the region $z\ge{z_a}$ all we
have to do is set an appropriate prior on the parameter $\rho_a$. Guided again by
the physical models we choose a uniform prior in the range
$100\unit{kg/m^3}\le{\rho_a}\le{200}\unit{kg/m^3}$ with an exponentially decaying
probability outside this range.

\section{Complete definition of parameters sampled by MCMC}\label{app:fullparam}
As explained in sec.~\ref{sec:params} of the main text, our choice of
parametrization of empirical density profile is a piecewise quadratic function.
There are two breakpoints, at normalized radii $z_1$ and $z_2$, where a jump
discontinuity is explicitly allowed (but not required) and between them are three
quadratic segments each defined by three parameters, for a total of 11 free
parameters required to define a density profile $\roofs{}$.

There is more than one way to let three numbers define a quadratic function. In
principle all are equivalent but in practice MCMC sampling works best (i.e.
converges fastest) when the parameters are minimally correlated and the likelihood
function is a smooth function of their numerical values. It is especially
important to avoid likelihood ``cliffs'', where a small change in one parameter
value results in a sudden drop in the likelihood value, perhaps because a
physically motivated prior condition has been violated. This is a real danger and
often leads the most intuitive and simple parametrizations to fail.

For example, defining the quadratic segments by
\begin{equation}
\rho(z)=
\begin{cases}
a_1z^2 + b_1z + c_1, & z_1 < z,\\
a_2z^2 + b_2z + c_2, & z_2 < z \le z_1,\\
a_3z^2 + b_3z + c_3, & z \le z_2,
\end{cases}
\end{equation}
would not do. The 9 parameters $a_i,b_i,c_i$ are highly correlated, meaning a
small change in the value of one usually requires a simultaneous and
``coordinated'' change in several others to prevent the resulting density profile
from changing too much and landing in a low-likelihood region. Worse, the locus of
parameter values that yield physically permissible density profiles (without
negative density or any density inversions) form distinct islands in parameter
space, with zero-likelihood regions between then that are practically impossible
for MCMC algorithms to cross.

By trial and error we arrive at the following alternative parameterization;
admittedly complicated, but effective. In addition to $z_1$ and $z_2$, the 9
parameters defining the quadratic segments are:
\begin{equation}\label{eq:pnames}
\begin{split}
\V{x} = \bigl[
&a_1,\ \rho_{10},\ y_{11}=\log(\rho_{11}-\rho_{10}),\\
&a_2,\ y_{21}=\log(\rho_{21}-\rho_{11}),\ y_{22}=\log(\rho_{22}-\rho_{21}),\\
&a_3,\ y_{32}=\log(\rho_{32}-\rho_{22}),\ y_{33}=\log(\rho_{33}-\rho_{32})
\bigr].
\end{split}
\end{equation}
The parameters $a_i$ control the curvature of segment $i$ and the $\rho_{ij}$ are
the densities at the segment ends. The segments are numbered from top to bottom:
segment 1 includes $z_1<z\le{z_a=0.94}$, segment 2 includes $z_2<z\le{z_1}$, and
segment 3 includes $0<z\le{z_2}$. (See appendix~\ref{app:quartic} for why the top
segment extends up to $z_a$ instead of $z=1$.)

Next, $\rho_{10}=\rho(z_a)$  (top of segment 1) and
$\rho_{11}=\lim_{z\to{}z_1^+}\rho(z)$ is the \emph{right-limit} density at $z_1$
(i.e. bottom of segment 1). Similarly, $\rho_{21}=\lim_{z\to{}z_1^-}\rho(z)$ is
the \emph{left-limit} density at $z_1$ (top of segment 2) and
$\rho_{22}=\lim_{z\to{}z_2^+}\rho(z)$ is the \emph{right-limit} density at $z_2$
(bottom of segment 2). Finally, $\rho_{32}=\lim_{z\to{}z_2^-}\rho(z)$ is the
\emph{left-limit} density at $z_2$ (top of segment 2) and $\rho_{33}=\rho(z=0)$ is
the density at the bottom of segment 3, the center of the planet.

The use of curvature-and-endpoints description is less familiar but more intuitive
than the well known polynomial coefficients representation. Notice that the
endpoint density values are defined implicitly, the actual parameter values are
the log of difference of neighboring density values. This transformation is a
common MCMC ``trick.'' It allows the sampled parameters to have values in the
range $[-\infty,\infty]$ and keeps the corresponding physical parameters in their
meaningful range. All values of $y_{ij}$ are permissible and lead to physical,
monotonically decreasing density profiles. Larger values of $y_{21}$ and $y_{32}$
lead to more pronounced density jumps between segments, while more negative values
result in the jumps disappearing and the segments merging into one. Thus all
possibilities from the canonical, sharp envelope-envelope and core-envelope
transitions to a completely smooth density profile throughout are representable
and reachable by continuous variation of parameter values.

The density profile itself is constructed from the parameters by solving for the
polynomial coefficients that reproduce the end-point densities:
\begin{equation}\label{eq:actual}
\rho(z)=
\begin{cases}
a_1(z^2 - z_a^2) +
\bigl[\frac{\rho_{11} - \rho_{10}}{z_1 - z_a} - a_1(z_1 + z_a)\bigr](z - z_a) +
\rho_{10}, & z_1<z\le{z_a},\\
a_2(z^2 - z_1^2) +
\bigl[\frac{\rho_{22} - \rho_{21}}{z_2 - z_1} - a_2(z_2 + z_1)\bigr](z - z_1) +
\rho_{21}, & z_2<z\le{z_1},\\
a_3(z^2 - z_2^2) +
\bigl[\frac{\rho_{33} - \rho_{32}}{0 - z_2} - \,a_3(0 + z_2)\;\bigr](z - z_2) +
\rho_{32}, & \;0<z\le{z_2}.
\end{cases}
\end{equation}

The prior probabilities set for the above parameters and the resulting posterior
chains are given in appendix~\ref{app:chains}.


\section{Sampling procedure}\label{app:chains}
The full list of parameters we need to explore is:
\begin{equation}\label{eq:plist}
\V{x} = \{m\sub{rot},a_1,y_{10},y_{11},a_2,y_{21},y_{22},y_{32},y_{33},z_1,z_2\}.
\end{equation}
See appendix~\ref{app:fullparam} for the meaning of these parameters. Notice that
$a_3$ is apparently missing from the list above. In fact, as explained in
sec.~\ref{sec:mcmc}, the requirement that $\lim_{s\to{0^+}}d\roofs/ds=0$
constrains the innermost segment of $\roofs$ such that only two parameters are
independent. The curvature of that segment follows:
\begin{equation}\label{eq:a3}
\begin{split}
\rho_{10} &= y_{10},\\
\rho_{11} &= \rho_{10} + \exp(y_{11}),\\
\rho_{21} &= \rho_{11} + \exp(y_{21}),\\
\rho_{22} &= \rho_{21} + \exp(y_{22}),\\
\rho_{32} &= \rho_{22} + \exp(y_{32}),\\
\rho_{33} &= \rho_{32} + \exp(y_{33}),\\
a_3 &= \frac{(\rho_{32} - \rho_{33})}{z_2^2}.
\end{split}
\end{equation}

In sec.~\ref{sec:mcmc} we explain that the high degree of correlation between the
variables in~\eqref{eq:plist} makes it very difficult to sample from the full
posterior simultaneously. We find it necessary to sample instead from the
conditional probabilities, $p_\V{z}=p(\V{x'}|\V{Z=z})$, where $\V{Z}=\{z_1,z_2\}$
and $\V{x'}=\V{x}\setminus{\V{Z}}$. In words: we fix values for the radii $z_1$ and
$z_2$ and sample the remaining 10 parameters, resulting in a conditional
distribution. We repeat this for many values of $z_i$ to build a picture of the
full posterior.

\subsection{Prior probabilities of sampled parameters}
The prior for $\V{x}$ is a product of independent priors for each component. The
rotation prior is $m\sub{rot}\sim\mathcal{N}\,(0.14224,4.5\times{10^{-4}})$. The
mean corresponds to a rotation period of 10h:33min:30s and the deviation is about
1 minute.

The curvature parameters take a uniform prior
$a_i\sim\mathcal{U}\,(-3\times{10^6},3\times{10^6})$. These limits do not have a
special physical meaning, they are reasonable bounds we find by experimentation.

The parameter $y_{10}=\rho_a$ has a particularly important prior. Recall that this
is a density at a reference point $\rho_a=\rho(z_a=0.94)$ that we use to keep the
density in the low-pressure region of the envelope compatible with values derived
in traditional, EOS-based models. Guided by the models presented in
\citep{Mankovich2019} we set\footnote{Happily we never have to worry about
normalizing the probability as only probability ratios (actually log-probability
differences) are ever used.}
\begin{equation}
\log{p(y_{10})}\sim -\frac{1}{2}\left(
\frac{100 - \min(y_{10},100)}{10} + \frac{\max(y_{10},200) - 200}{10}
\right)^2,
\end{equation}
and the numerical values are in $\mathrm{kg}/\mathrm{m}^3$. In words: it is a
uniform probability inside the 100 to 200 $\unit{kg/m^3}$ range with exponentially
decaying probability outside of it with an e-folding distance of
$10\unit{kg/m^3}$.

The other $y_{ij}$ parameters are logarithms of density differences. They can take
positive or negative values, and the values get exponentiated and added to define
the densities at the end points of the quadratic segments, $\rho_{ij}$. It is
natural to define the prior on the actual density values, say a uniform prior in
the 0 to 30,000 $\unit{kg/m^3}$ range (merely a guess as to the highest density
achievable in Saturn). We need to be careful though. The transformation from
$\rho_{ij}$ to $y_{ij}$ involves a transformation of the probability; the prior on
$y_{ij}$ is \emph{not} uniform. Instead it follows from conservation of
probability mass in equivalent parts of the distribution:
$p_\rho(\rho')\,d\rho'=p_y(y')\,dy'$. The answer is
$y_{ij}\sim{}e^{y'}\mathcal{U}(-\infty,\infty)$, but it helps to cut off the
uniform probability outside of a reasonable range. \footnote{There are, after all,
a lot of numbers available between, say, $-20$ and $-\infty$ that as logarithms
all mean simply: $\Delta\rho=0$.} The final prior therefore is
\begin{equation}
\log{p(y_{ij})}\sim
\begin{cases}
y_{ij} & -20<y_{ij}<12,\\
-\infty & \text{otherwise}.
\end{cases}
\end{equation}

There is no prior on $z_1$ and $z_2$ because they are not MCMC sampled.

\subsection{Sampling from the conditional distributions}
We obtain a sample from $p_\V{z}$ for each pair
$\{z_1,z_2\}\in\{0.35,0.4,0.45,\ldots,0.9\}\times\{0.1,0.15,0.2,\ldots,0.4\}$
subject to the condition $z_1>z_2$. There are 81 pairs and thus 81 separate MCMC
runs to produce samples from the different conditional distributions. We use the
implementation of ensemble sampling in \code{emcee}, with the default stretch move
algorithm, and run 78 walkers for 60000 steps each.

\begin{figure}[tb!]
\gridline{\fig{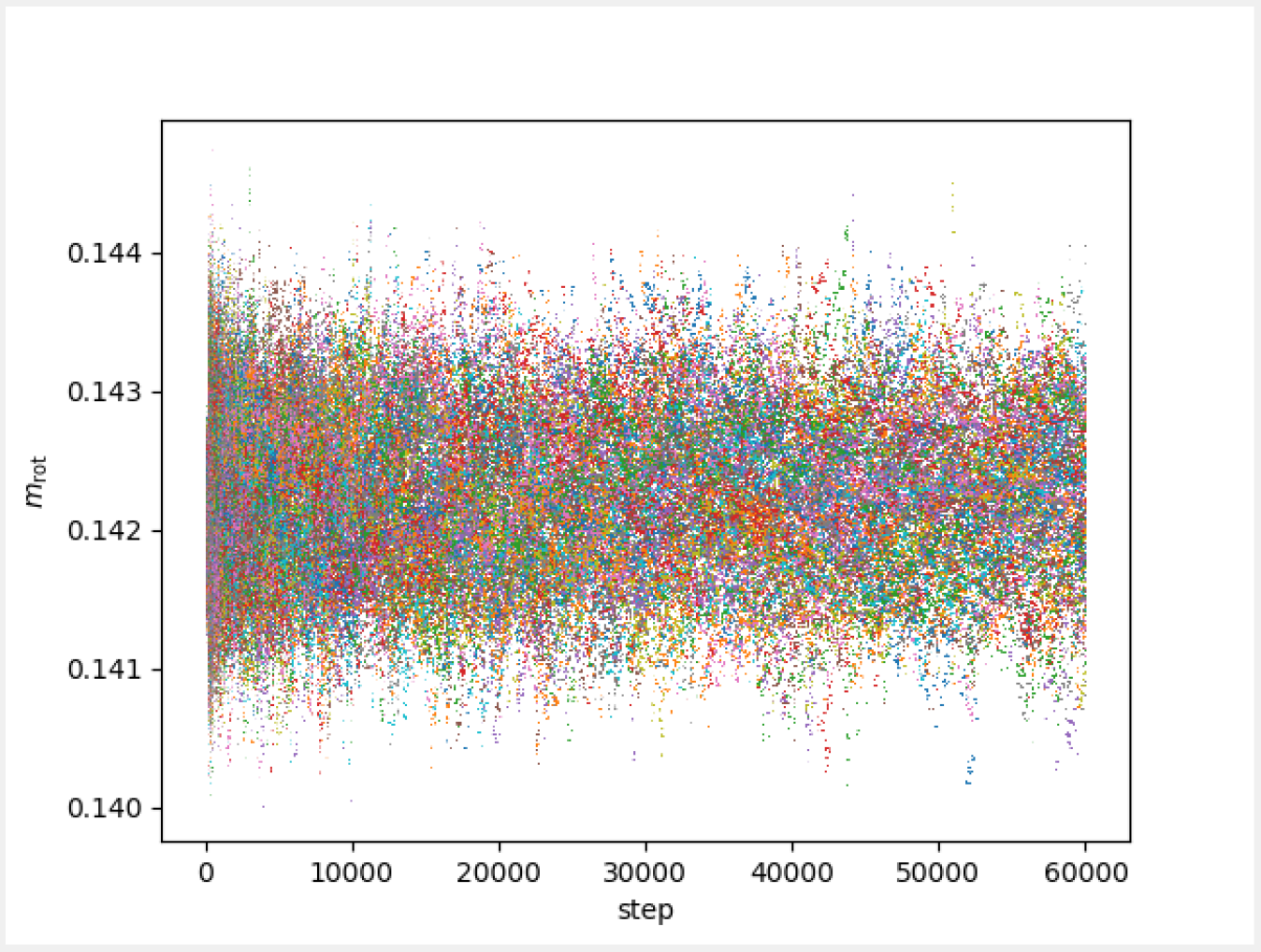}{0.3\textwidth}{}
          \fig{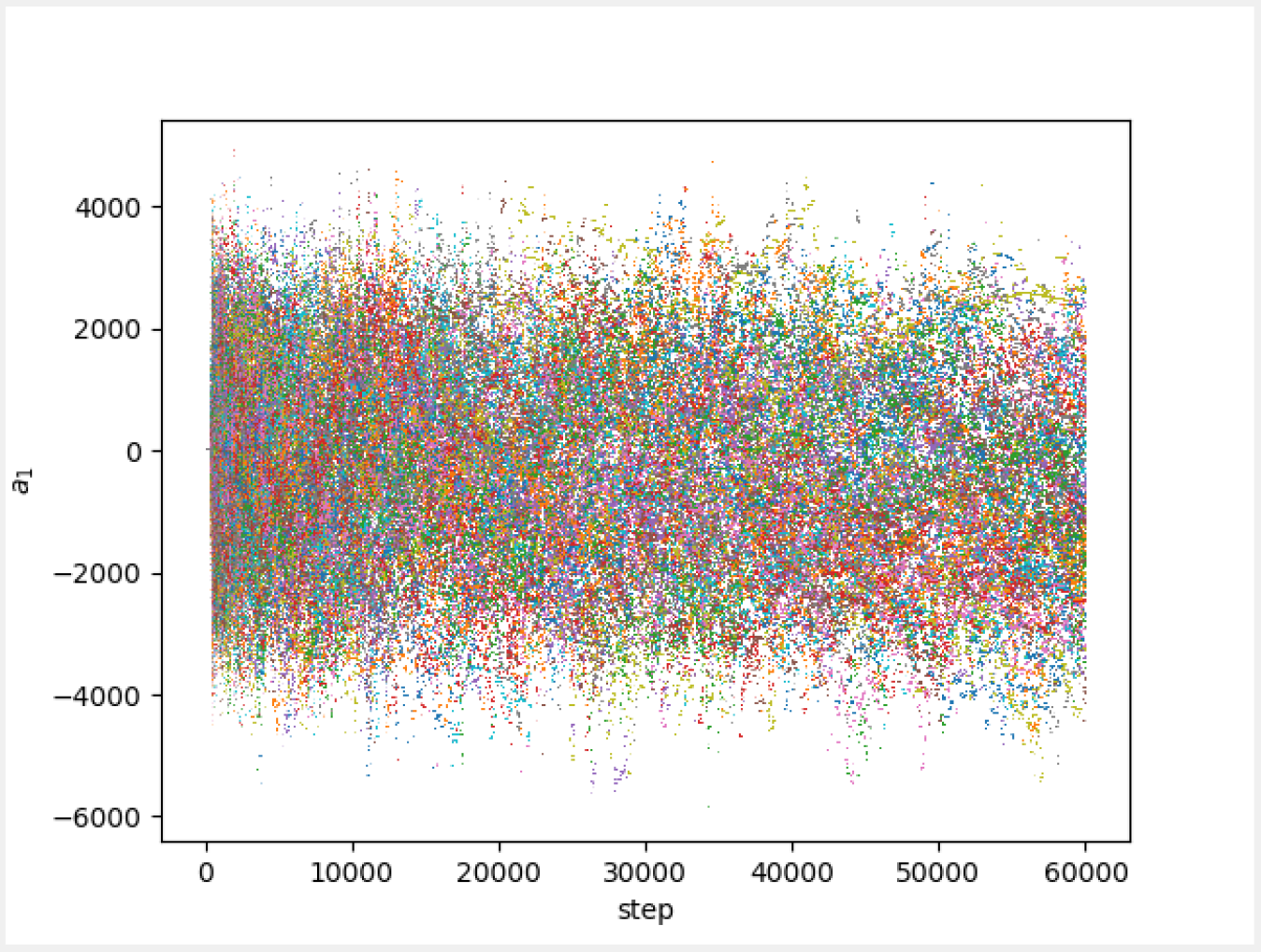}{0.3\textwidth}{}
          \fig{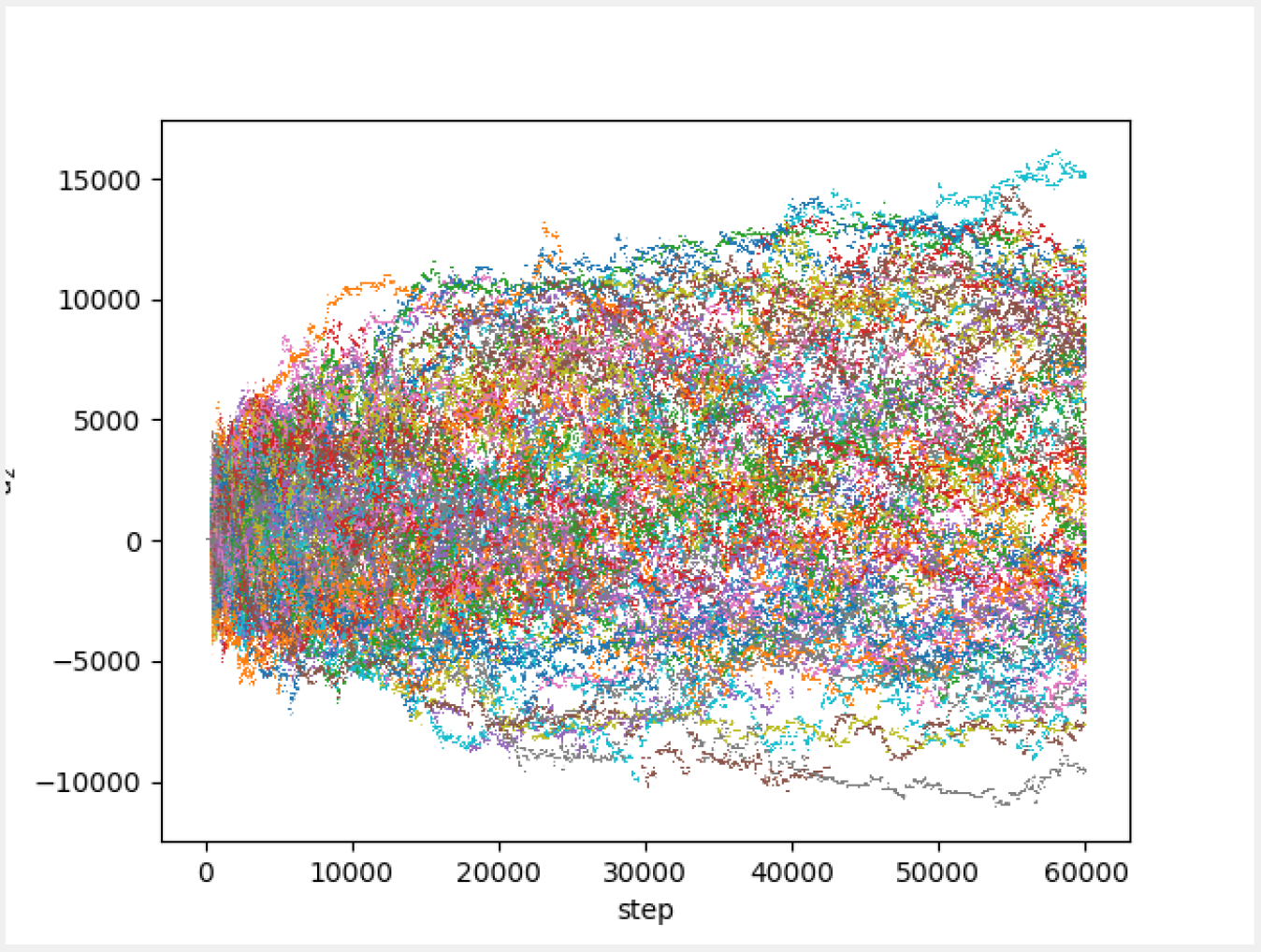}{0.3\textwidth}{}}
\gridline{\fig{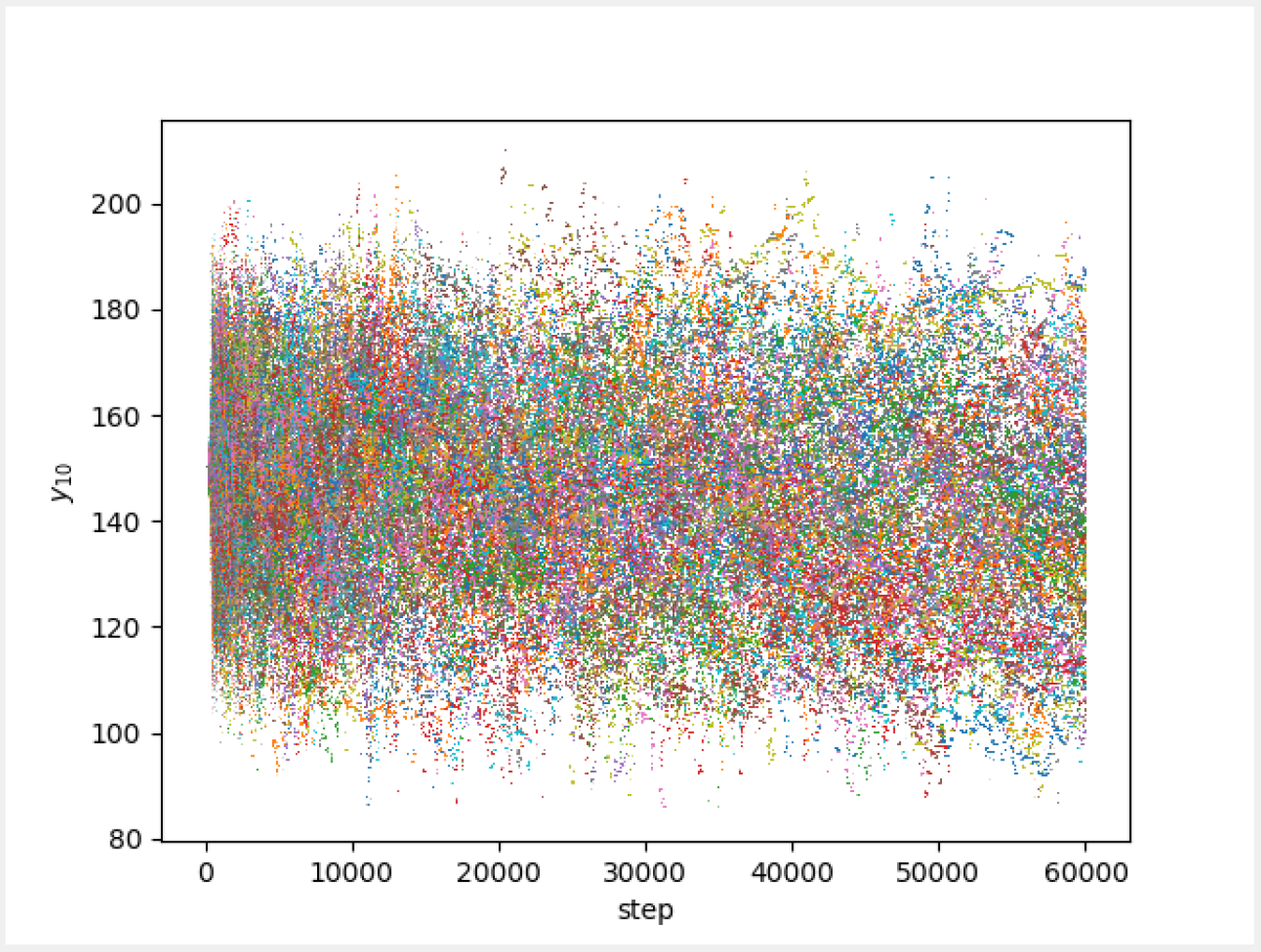}{0.3\textwidth}{}
          \fig{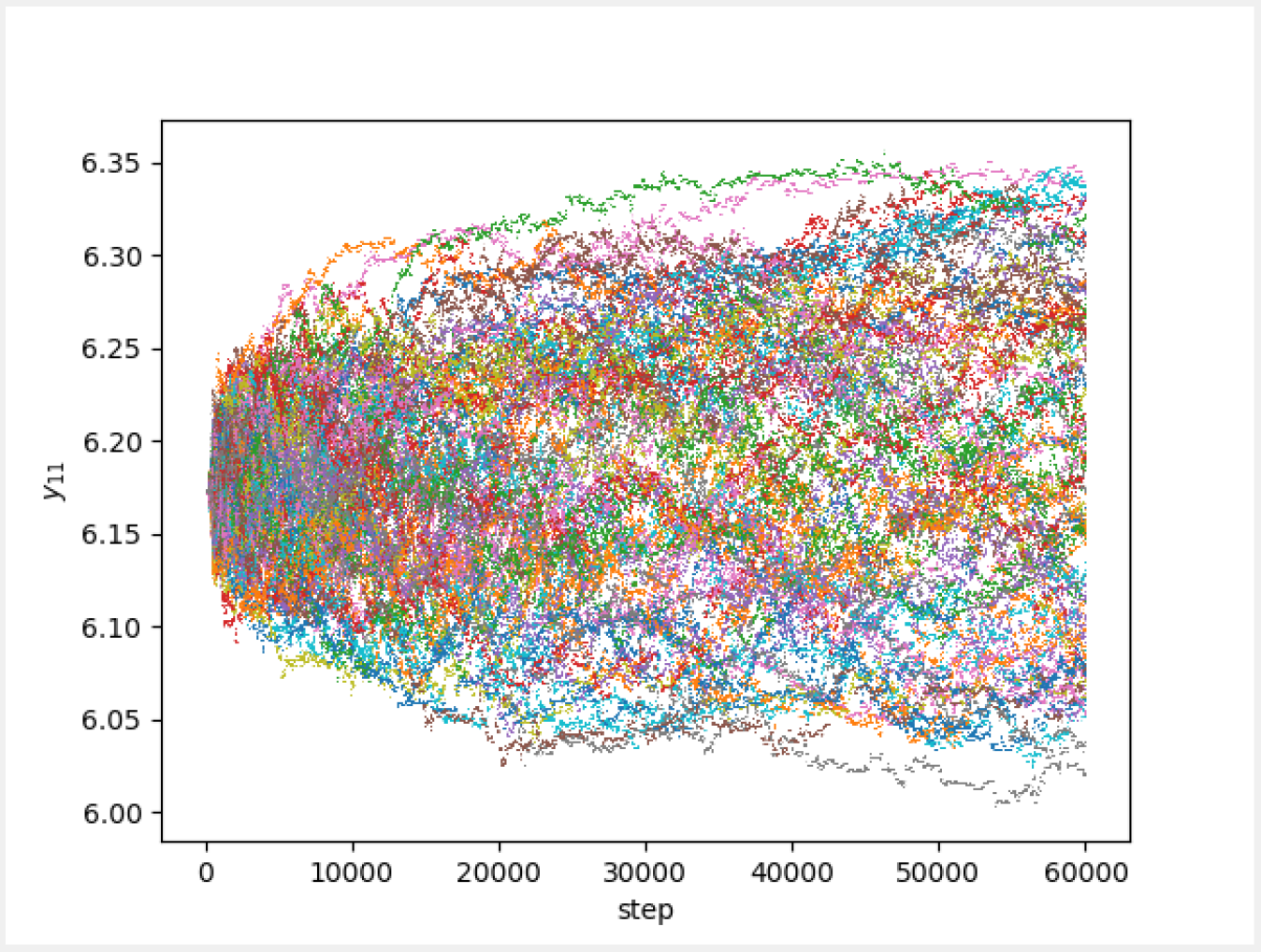}{0.3\textwidth}{}
          \fig{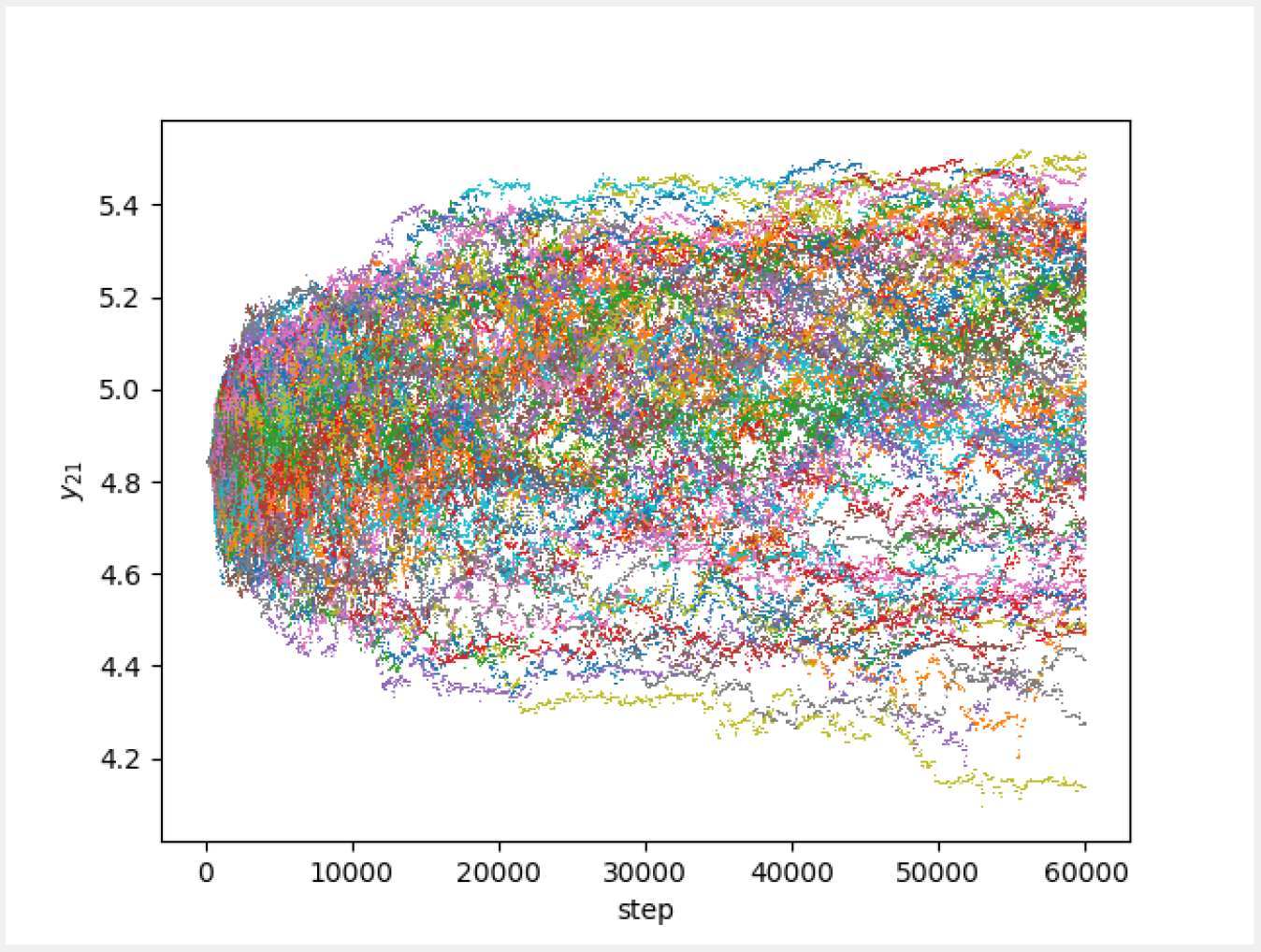}{0.3\textwidth}{}}
\gridline{\fig{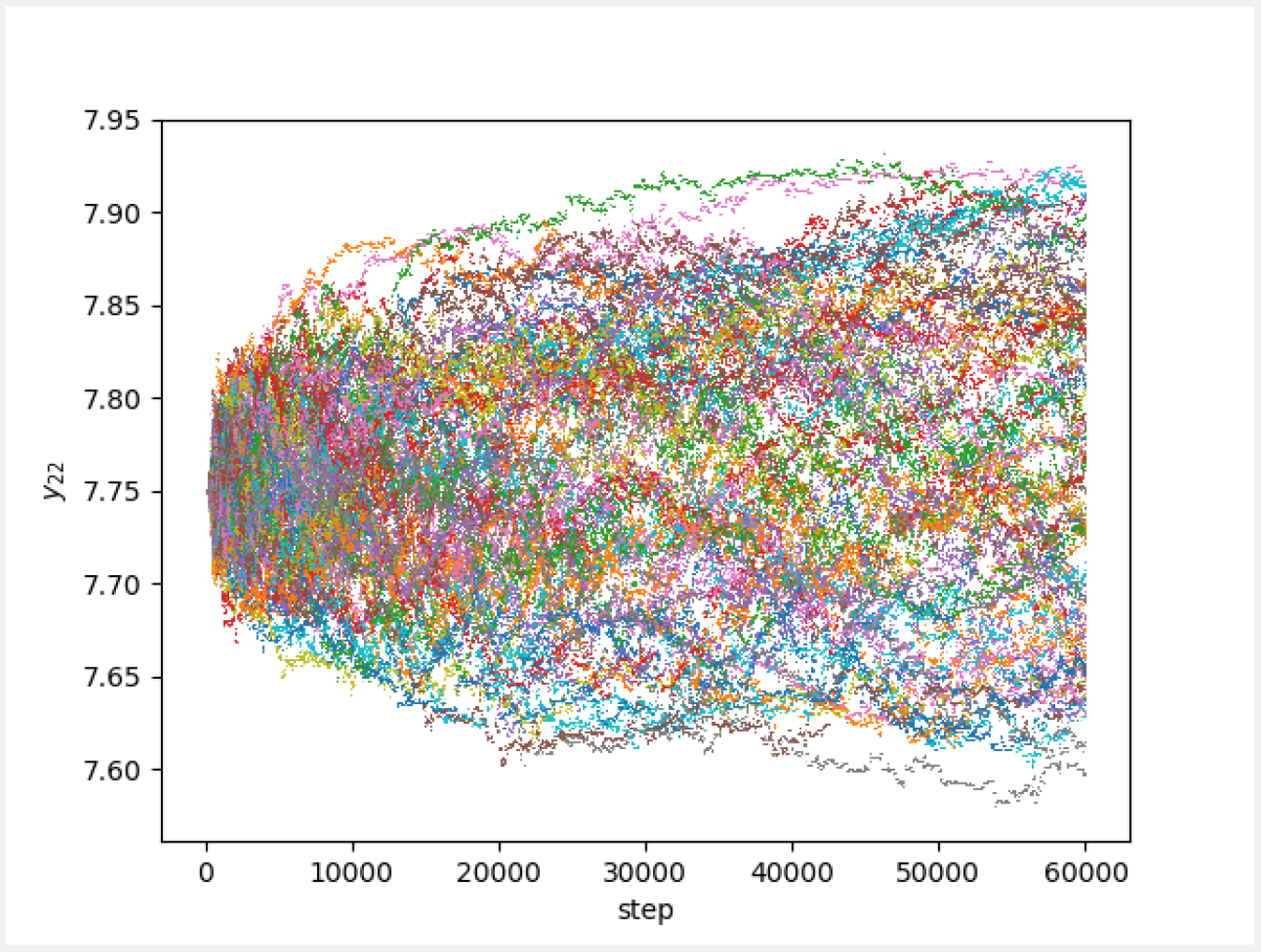}{0.3\textwidth}{}
          \fig{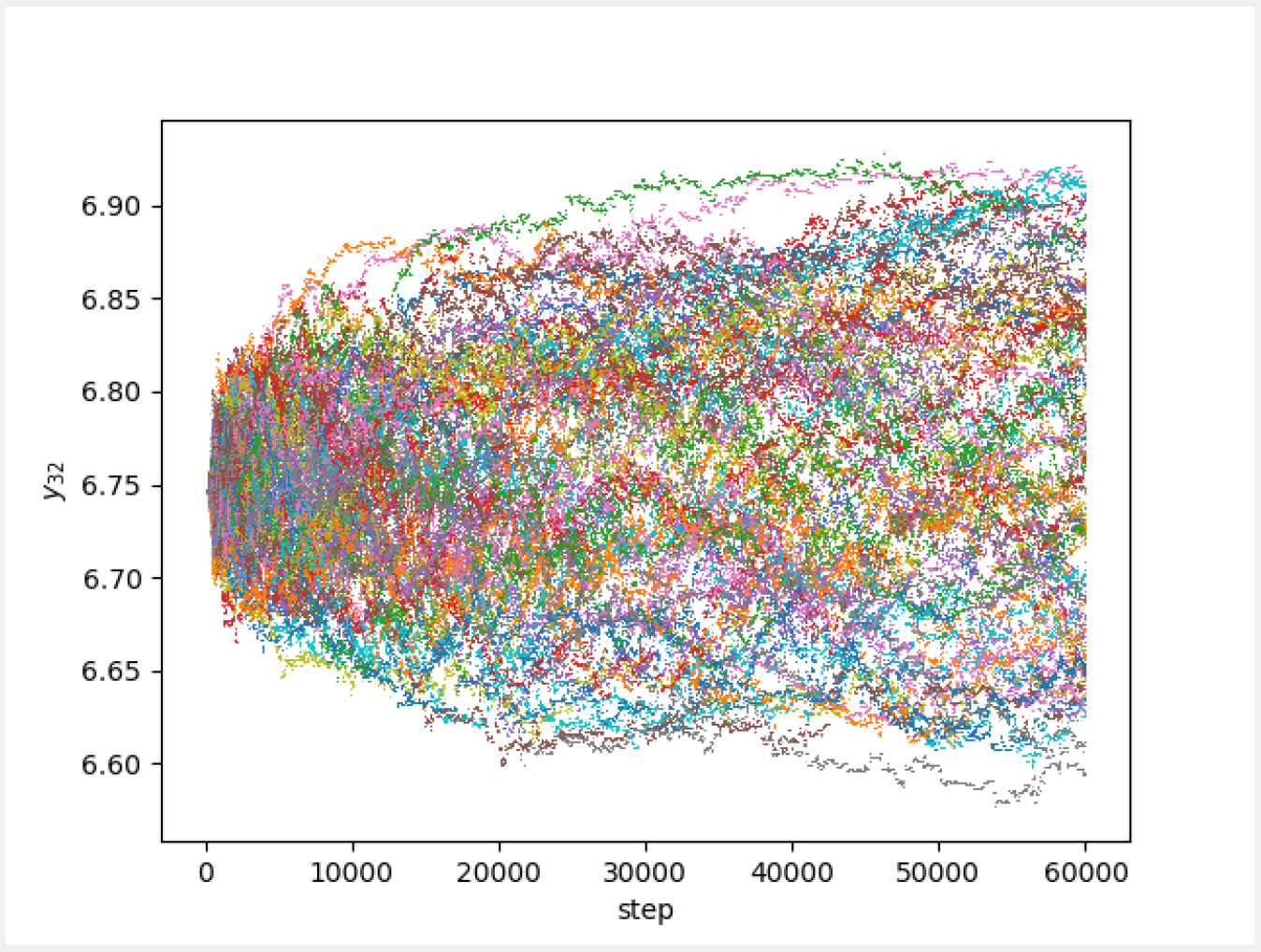}{0.3\textwidth}{}
          \fig{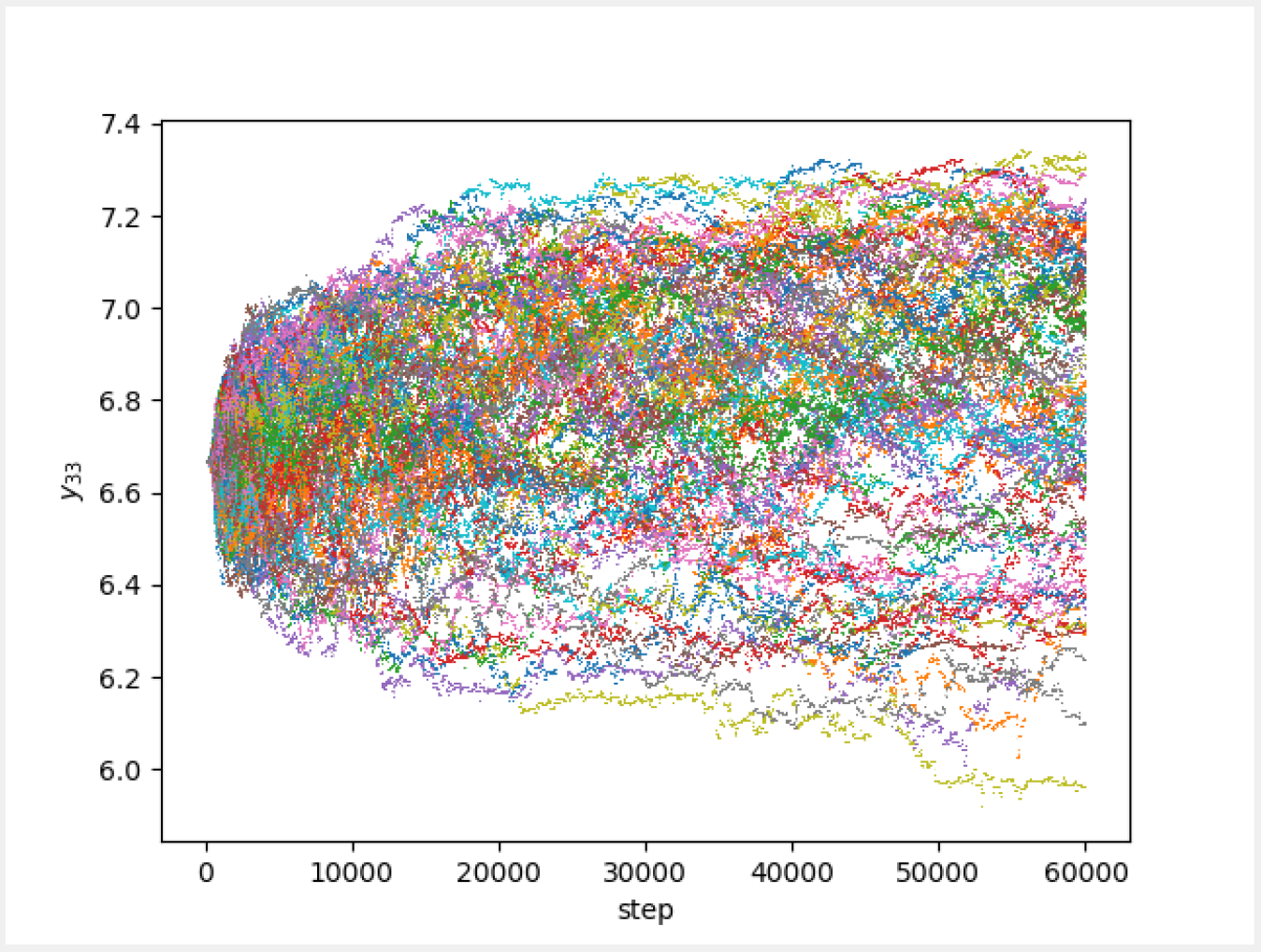}{0.3\textwidth}{}}
\caption{Trace plots from MCMC run with $z_1=0.65$ and $z_2=0.2$.}
\label{fig:traces}
\end{figure}

Trace plots for one such MCMC run are shown in Figure~\ref{fig:traces}, similar
behavior is exhibited in all runs. Visual inspection of trace plots is one method
of deciding what part of the MCMC chain we can use to take independent samples
from. Inspection of figure~\ref{fig:traces} reveals why we had to use many walkers
for so many steps. Several of the parameters exhibit slow mixing, taking more than
30000 steps to fully forget their seed state. Even worse than the long burn-in
time is the low acceptance rate, which leads to quite long autocorrelation in many
dimensions. In other words, successive steps are not independent, requiring about
200 steps to become uncorrelated. This means that an MCMC run evaluating more than
4.5 million candidate models produces only about 10,000 usable ones.

It is common practice to display the results of MCMC sampling in a series of
series of two-dimensional histograms of parameter pairs. This visualization, often
called a corner plot, is a convenient way to quickly make sense of the
distribution of parameters including the relationships between them. In our case
the parameters are too far removed from a physical meaning for us to derive any
useful insight from their pair-wise histograms. We include the corner plot for one
MCMC run anyway, in Figure

\begin{figure}[tb!]
\plotone{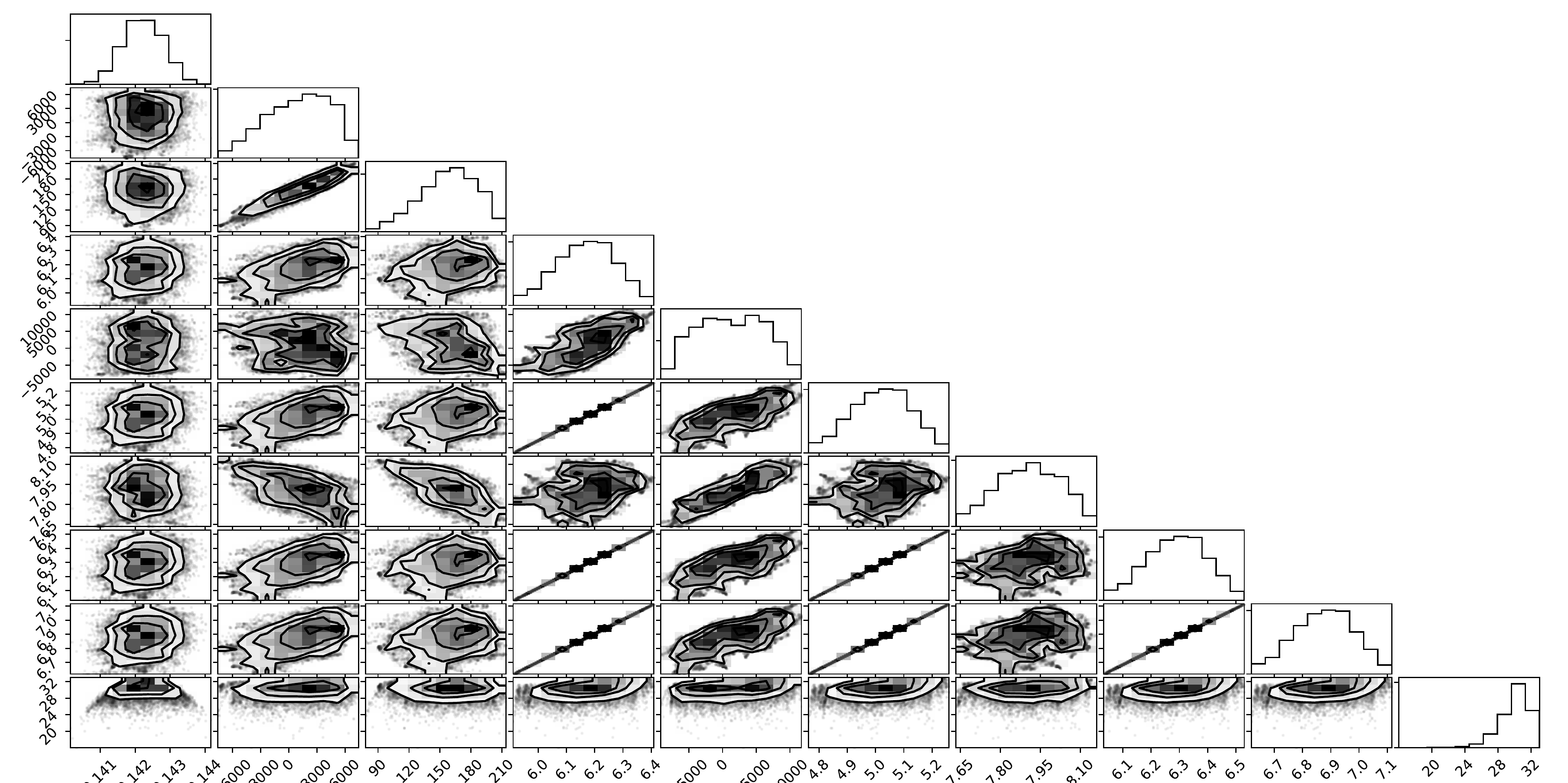}
\caption{Corner plot of parameters sampled for the $z_1=0.65$ $z_2=0.2$
conditional probability (same run as Fig.~\ref{fig:traces}), after discarding the
first 30,000 steps from each walker and thinning the rest by keeping one in every
200 steps. The subplot in the $i$-th row and $j$-th column is the two-dimensional
histogram of parameters $i$ and $j$, as ordered in eq.~\eqref{eq:plist}.}
\label{fig:corner}
\end{figure}

\subsection{Combining the conditional probabilities into a single posterior.}
After culling the MCMC chains we have what we hope are independent samples from
the conditional probabilities $p_\V{z}$. Next we need to combine subsets from
these samples in a way that approximates a sample from the full posterior,
$p(\V{x})$. This task is similar to the \emph{model selection} problem of Bayesian
inference. We have found parameter distributions for different statistical models,
and we wish to use this information to evaluate the relative likelihood between
the models, in our case between interior profiles with different locations of
discontinuous density. If we know the relative likelihoods we can combine subsets
from the individual models in proportion to their likelihood to obtain our
posterior sample.

Although this is a common and well studied task it is nevertheless a difficult
one, and there is no known best method or even useful error bounds.
\citet{Nelson2018} report on a thorough comparison of many different approaches to
this problem (often referred to as calculating the \emph{posterior odds} or the
\emph{Bayes factor} or the \emph{evidence integral} or simple the evidence),
including the method we chose which is based on calculating the Bayes Information
Criterion, or BIC:
\begin{equation}
\mathrm{BIC}(\V{z}) = -2\log(max_{\V{x}}(p_\V{z}(\V{x'}))) + \log{N},
\end{equation}
where $k$ is the number of model parameters and $N$ is the number of data points.
The relative likelihood is given by
\begin{equation}
\frac{p_\V{z_a}}{p_\V{z_b}} = \exp(-(\mathrm{BIC}_\V{z_b} -
\mathrm{BIC}_\V{z_a})/2).
\end{equation}

\begin{figure}[tb!]
\gridline{\fig{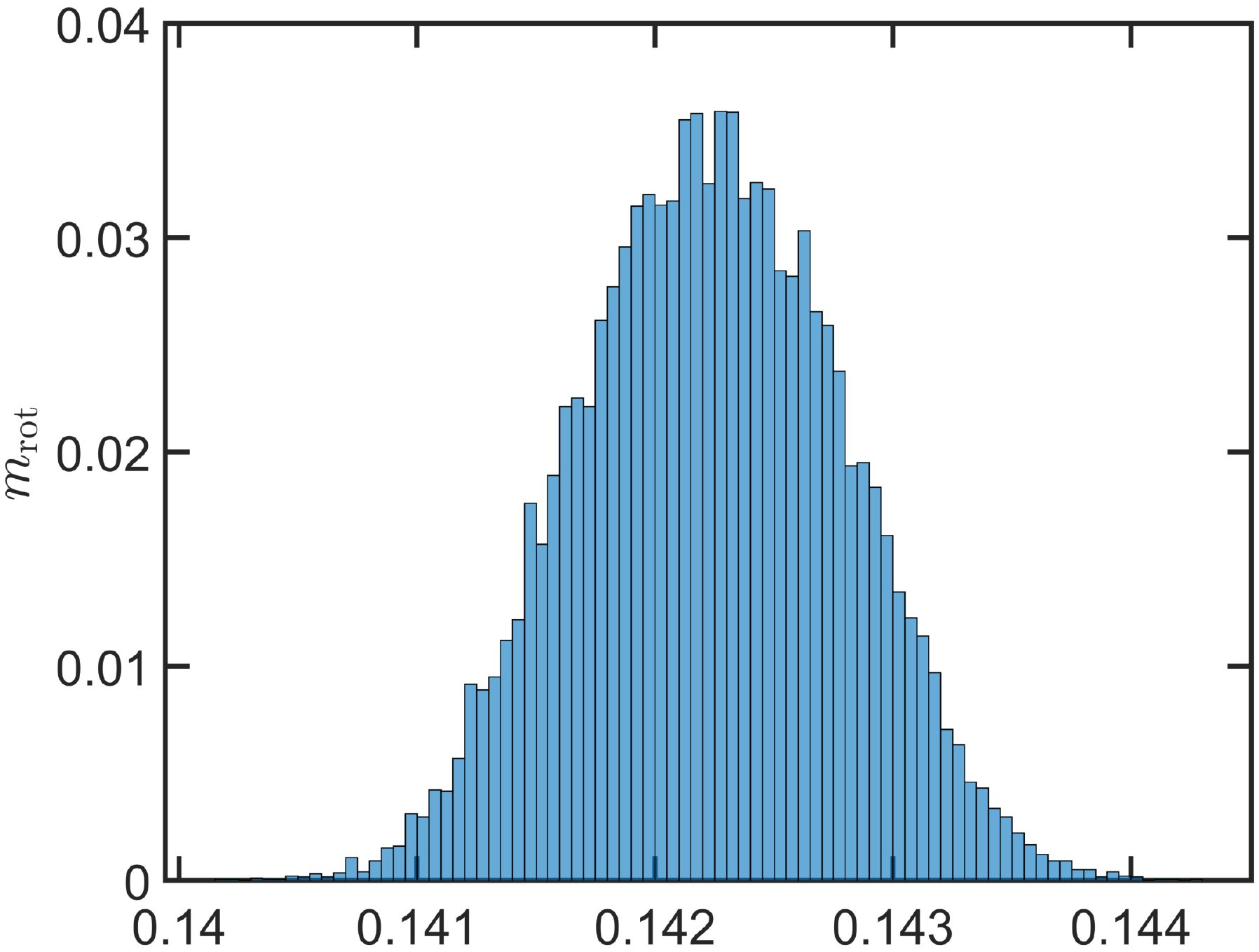}{0.3\textwidth}{}
          \fig{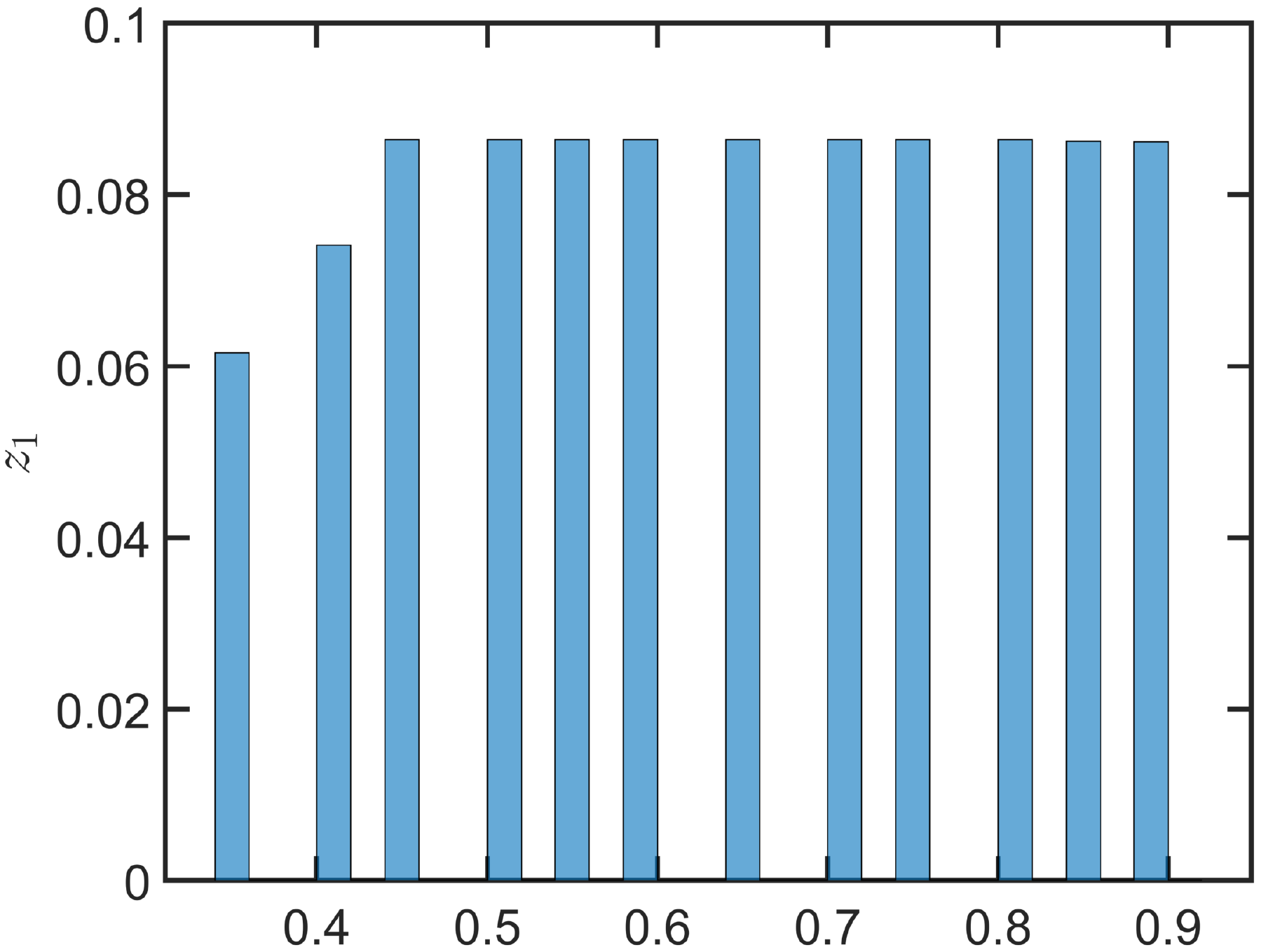}{0.3\textwidth}{}
          \fig{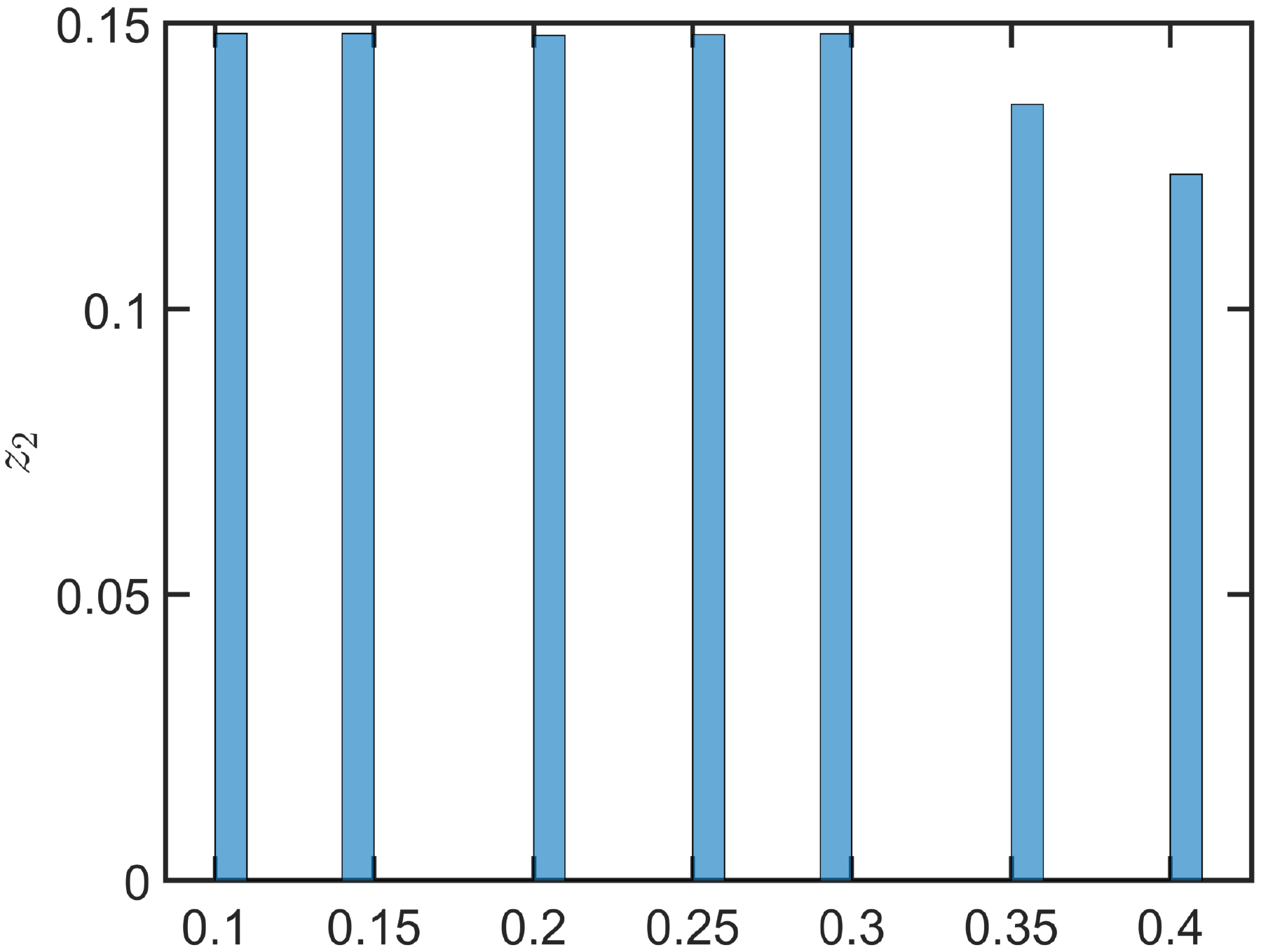}{0.3\textwidth}{}}
\gridline{\fig{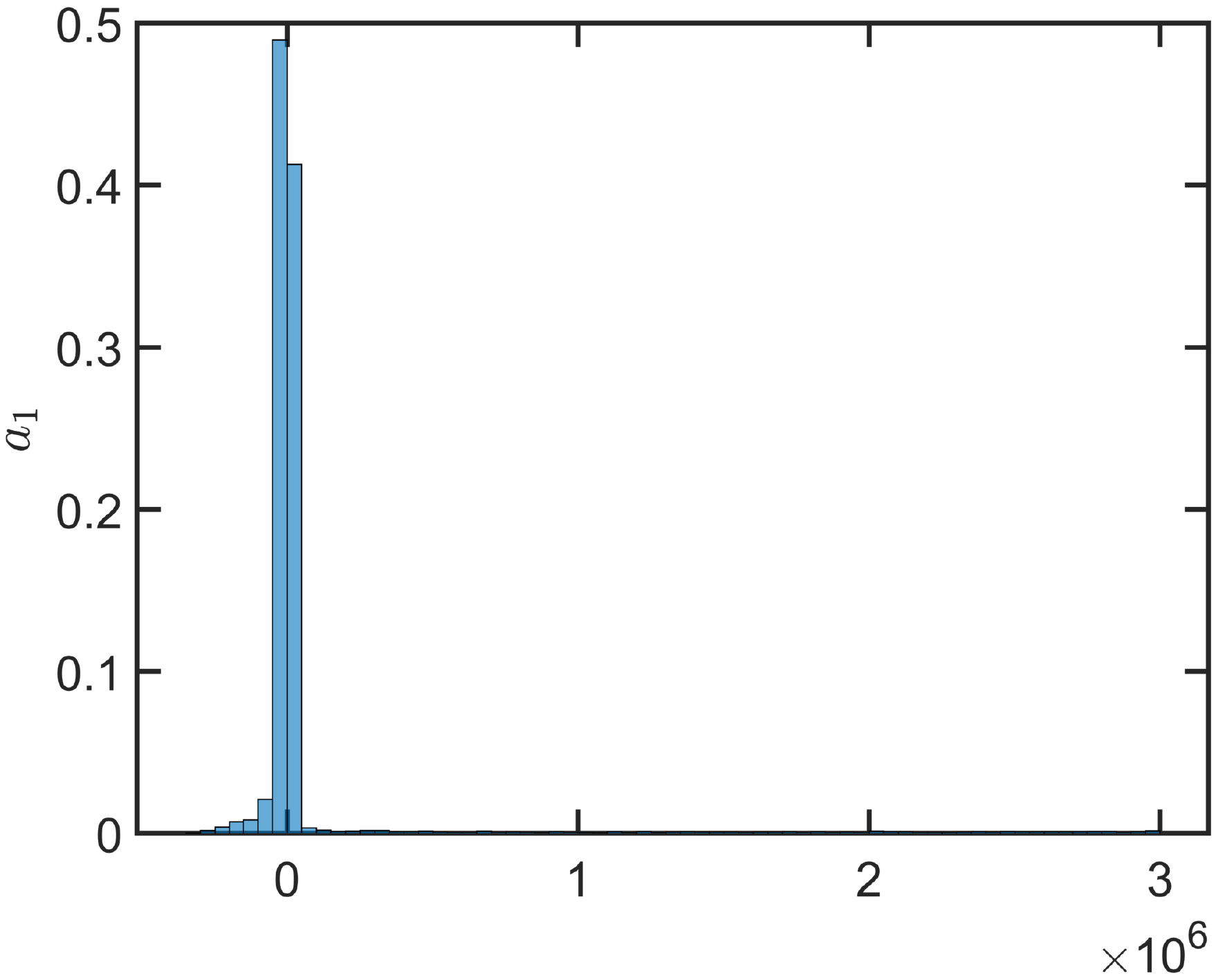}{0.3\textwidth}{}
          \fig{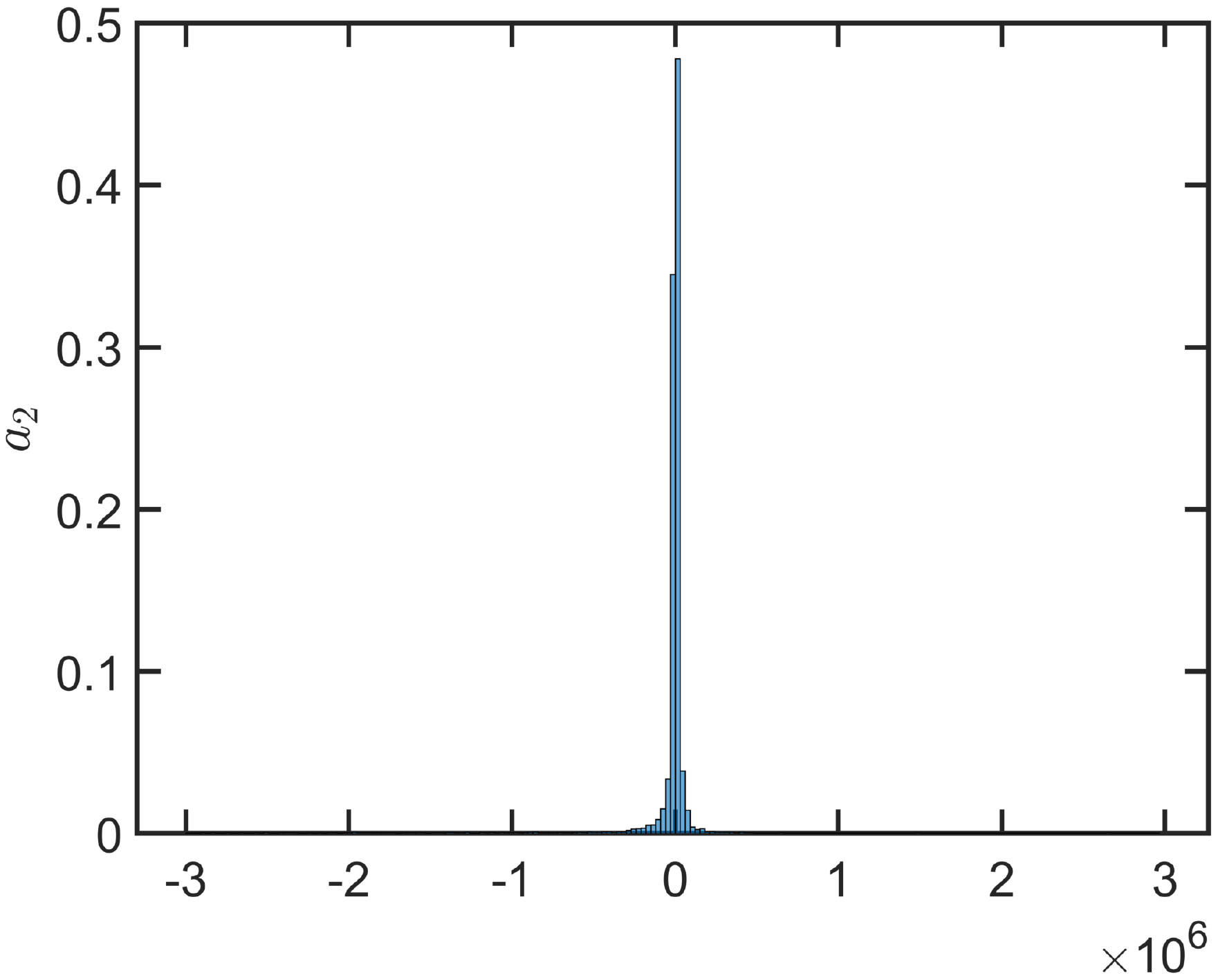}{0.3\textwidth}{}
          \fig{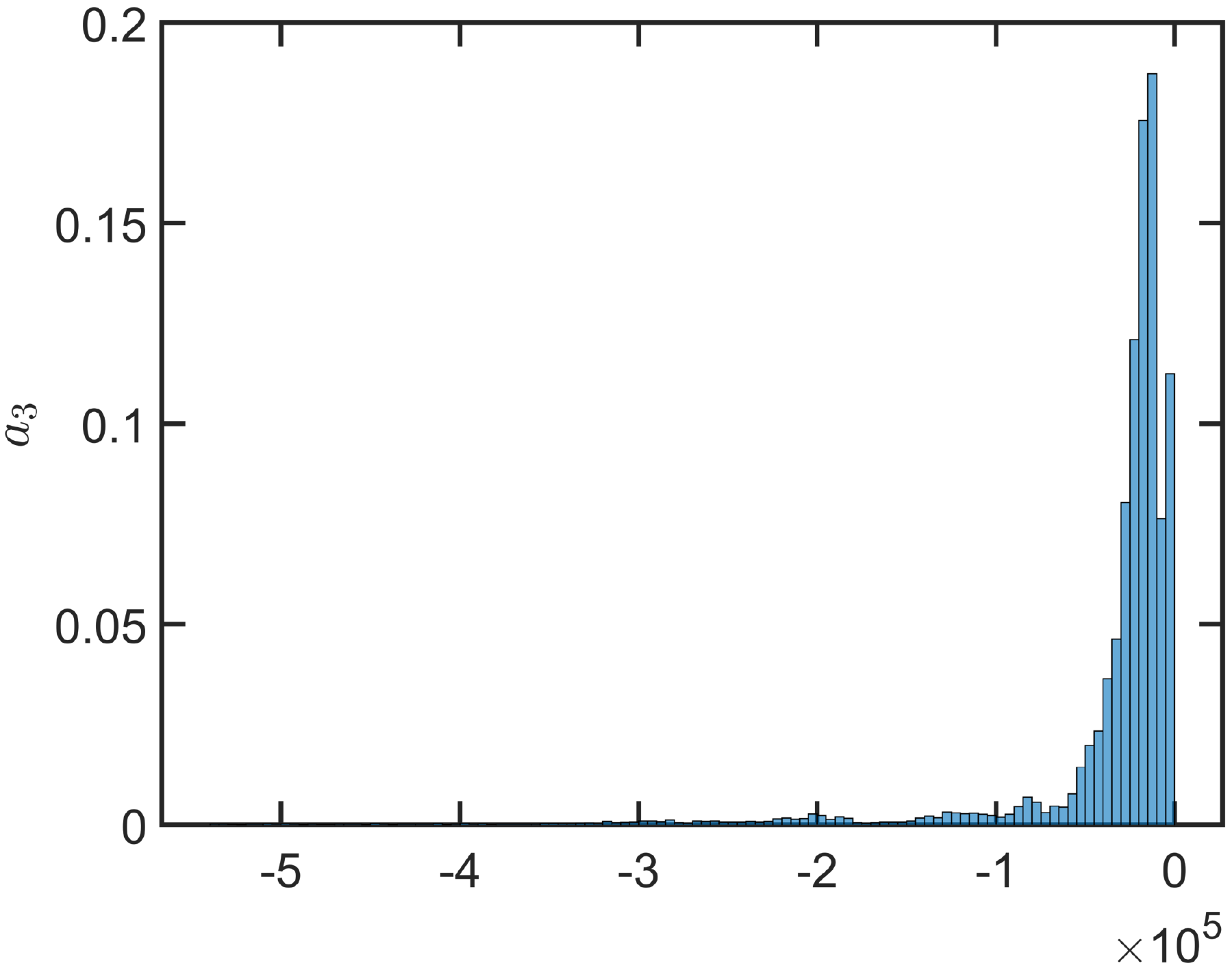}{0.3\textwidth}{}}
\gridline{\fig{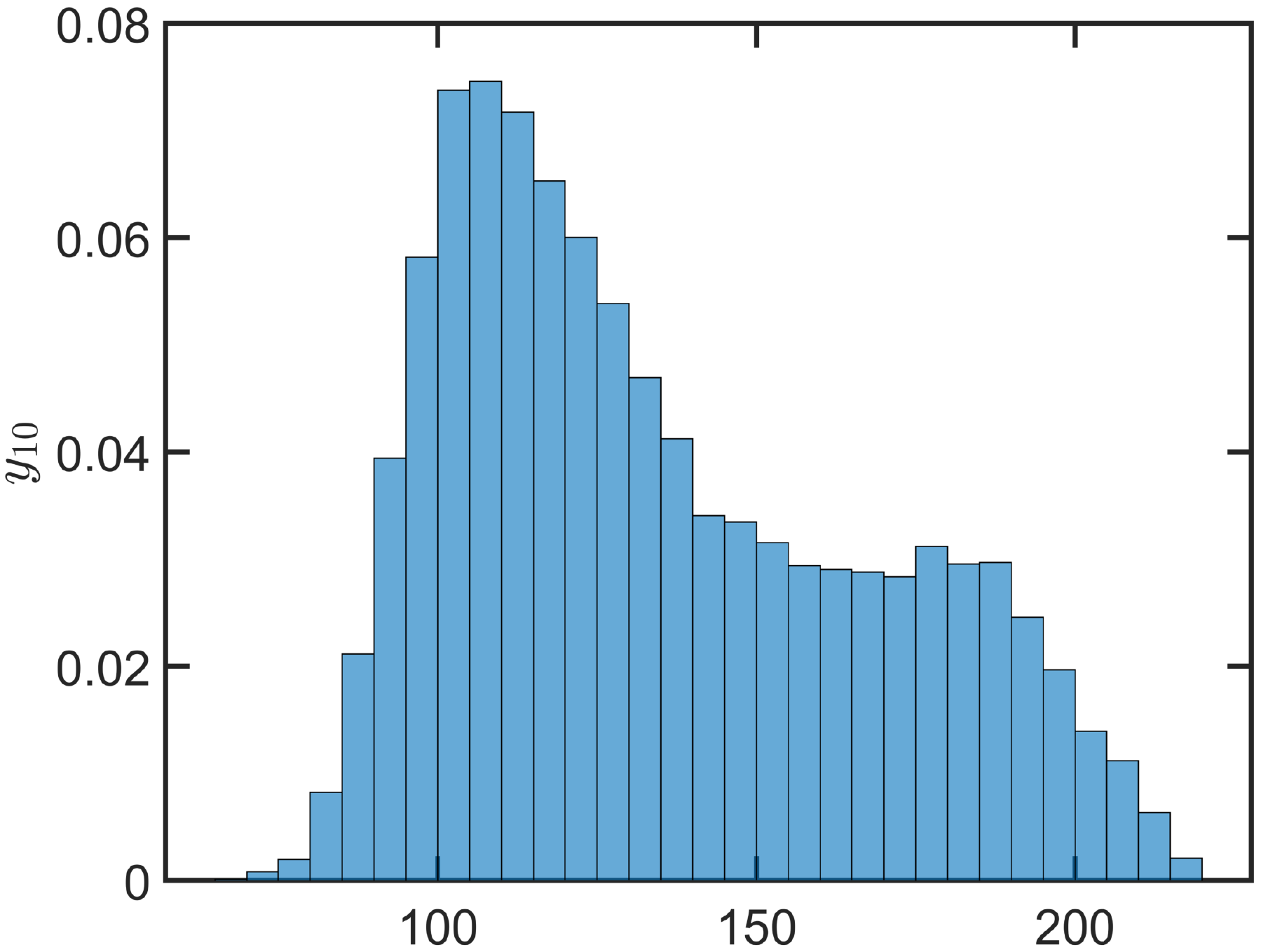}{0.3\textwidth}{}
          \fig{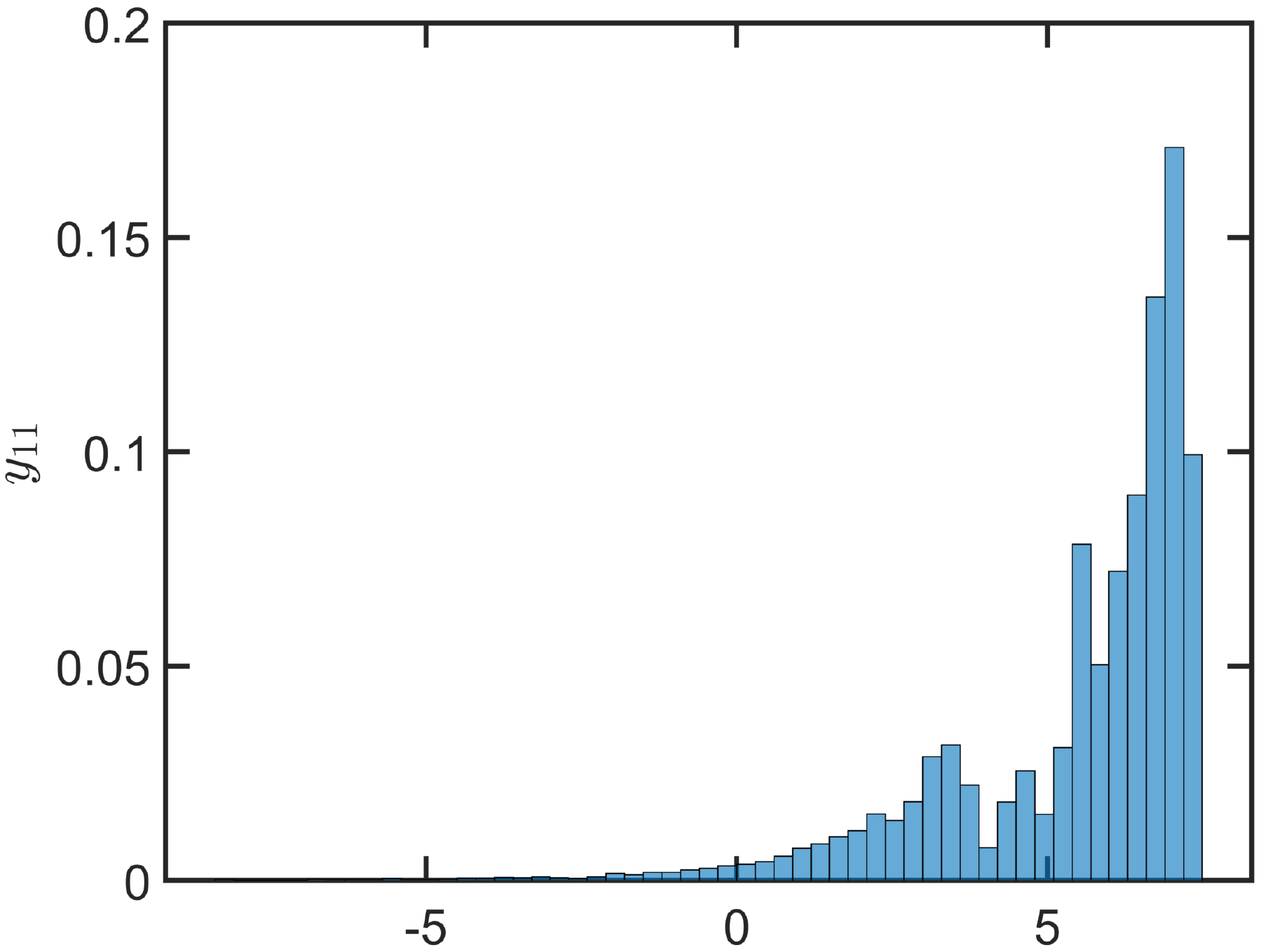}{0.3\textwidth}{}
          \fig{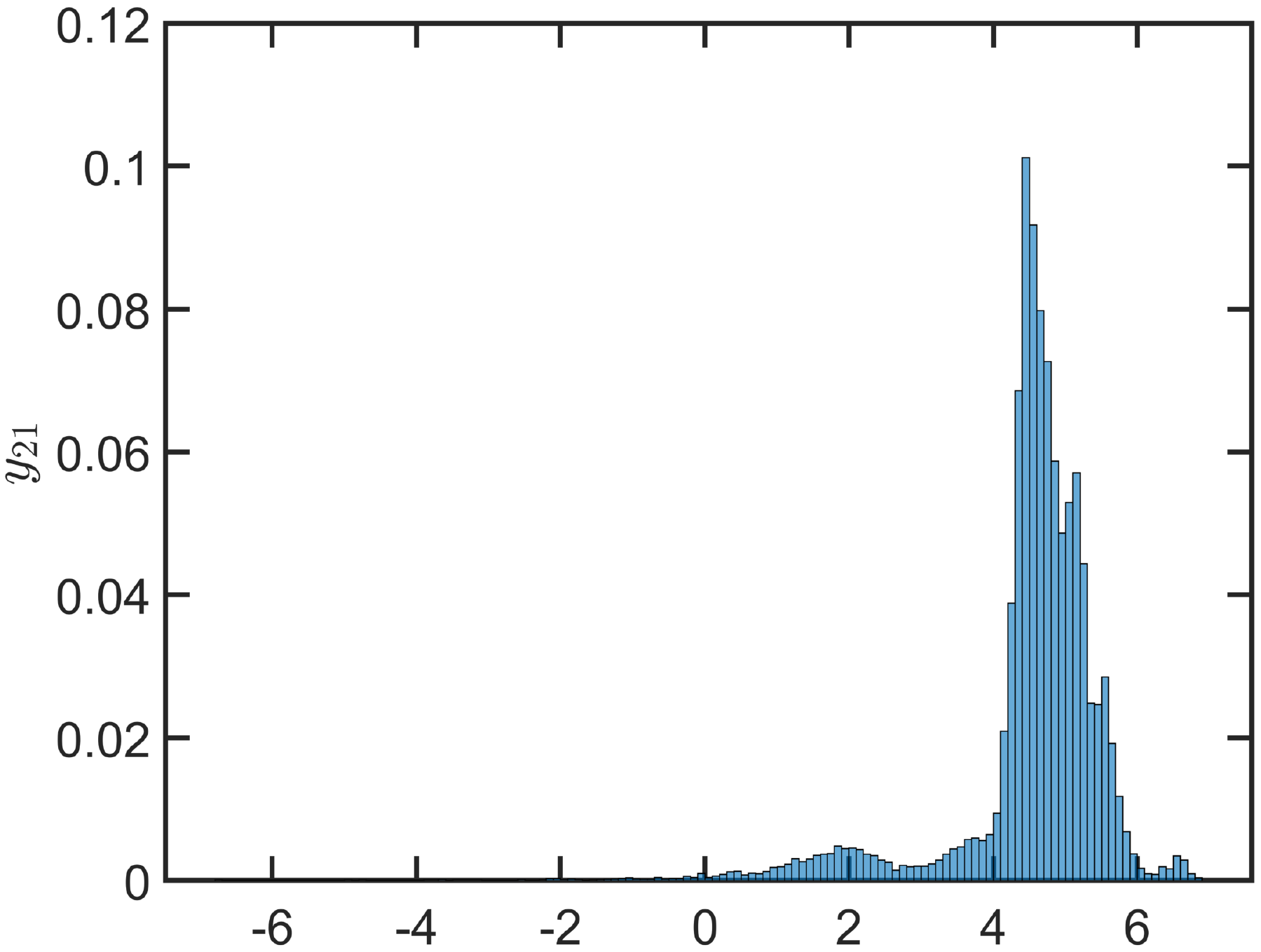}{0.3\textwidth}{}}
\gridline{\fig{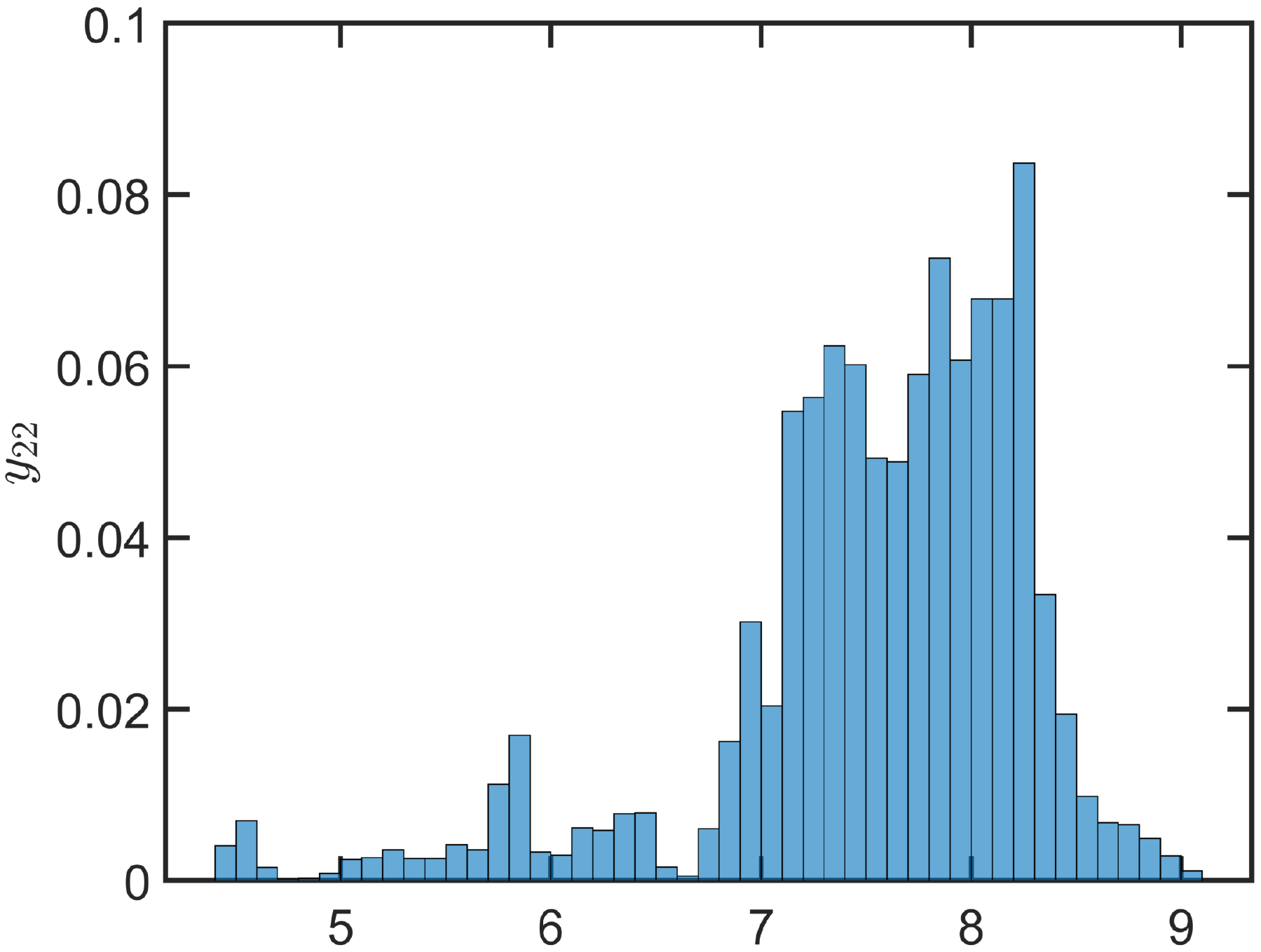}{0.3\textwidth}{}
          \fig{dim8_hist}{0.3\textwidth}{}
          \fig{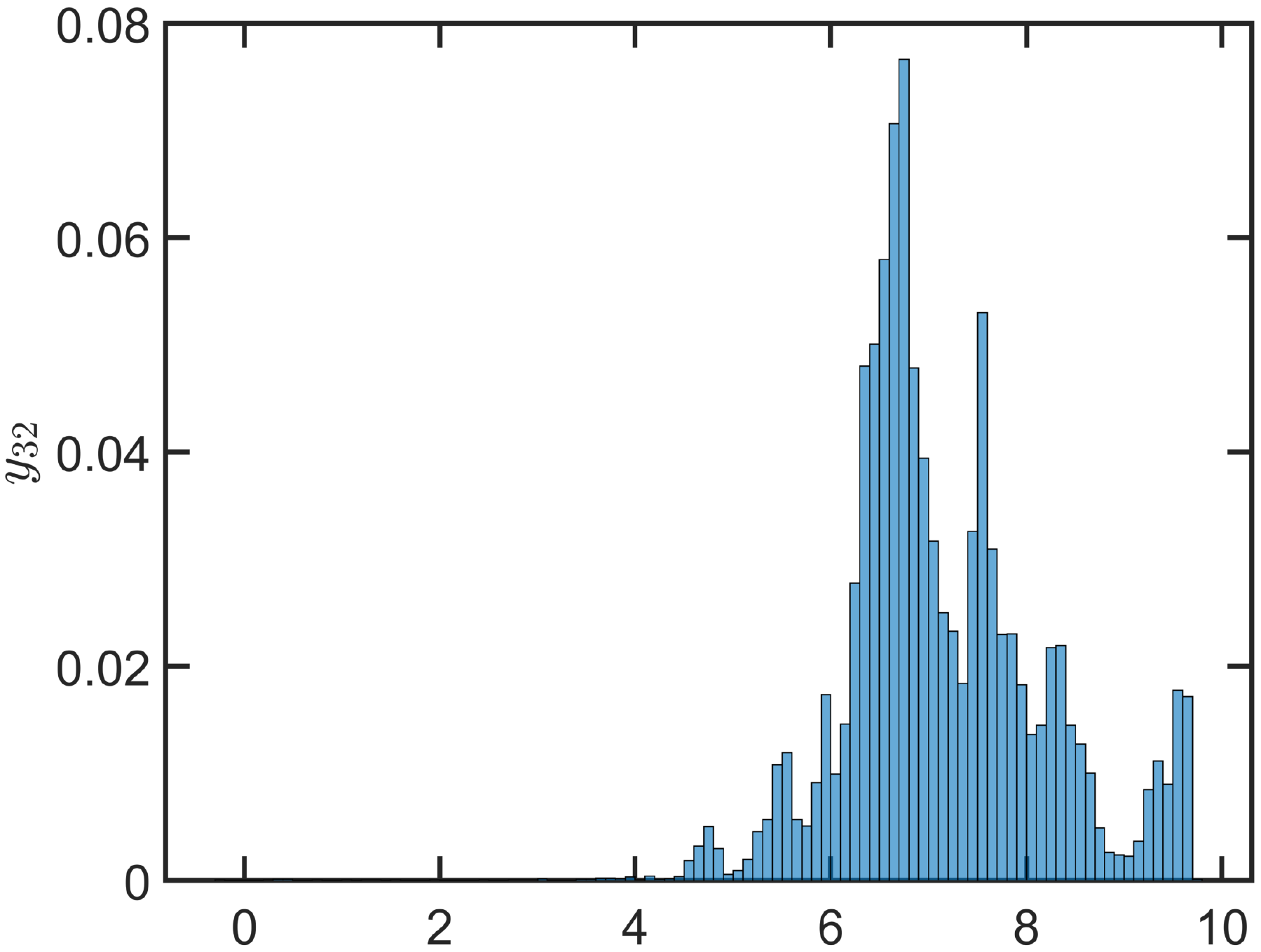}{0.3\textwidth}{}}
\caption{Histograms of parameter values used to construct the density profiles
used in this work.}
\label{fig:hists}
\end{figure}

In our case, $k=9$ always and $N\approx{10^4}$, and the maximum likelihood is
likewise very similar between all 81 conditional samples. So it happens that the
pairwise relative likelihood among all the conditional distributions is close to
one. We take random draws from the 81 conditional samples, in almost equal
proportions, to obtain a single set of 20,000 hopefully independent draws from the
unknown posterior, $p(\V{x}|\mathbf{OBS})$. Histograms of the 12 parameters
(including $a_3$ which is not sampled but uniquely determined by eq~\eqref{eq:a3})
are shown in Figure~\ref{fig:hists}. These are the parameters used to reconstruct
the density profiles shown in Figure~\ref{fig:rhoposterior} and to perform the
analysis in the rest of the paper.

\begin{figure*}[tb!]
\plottwo{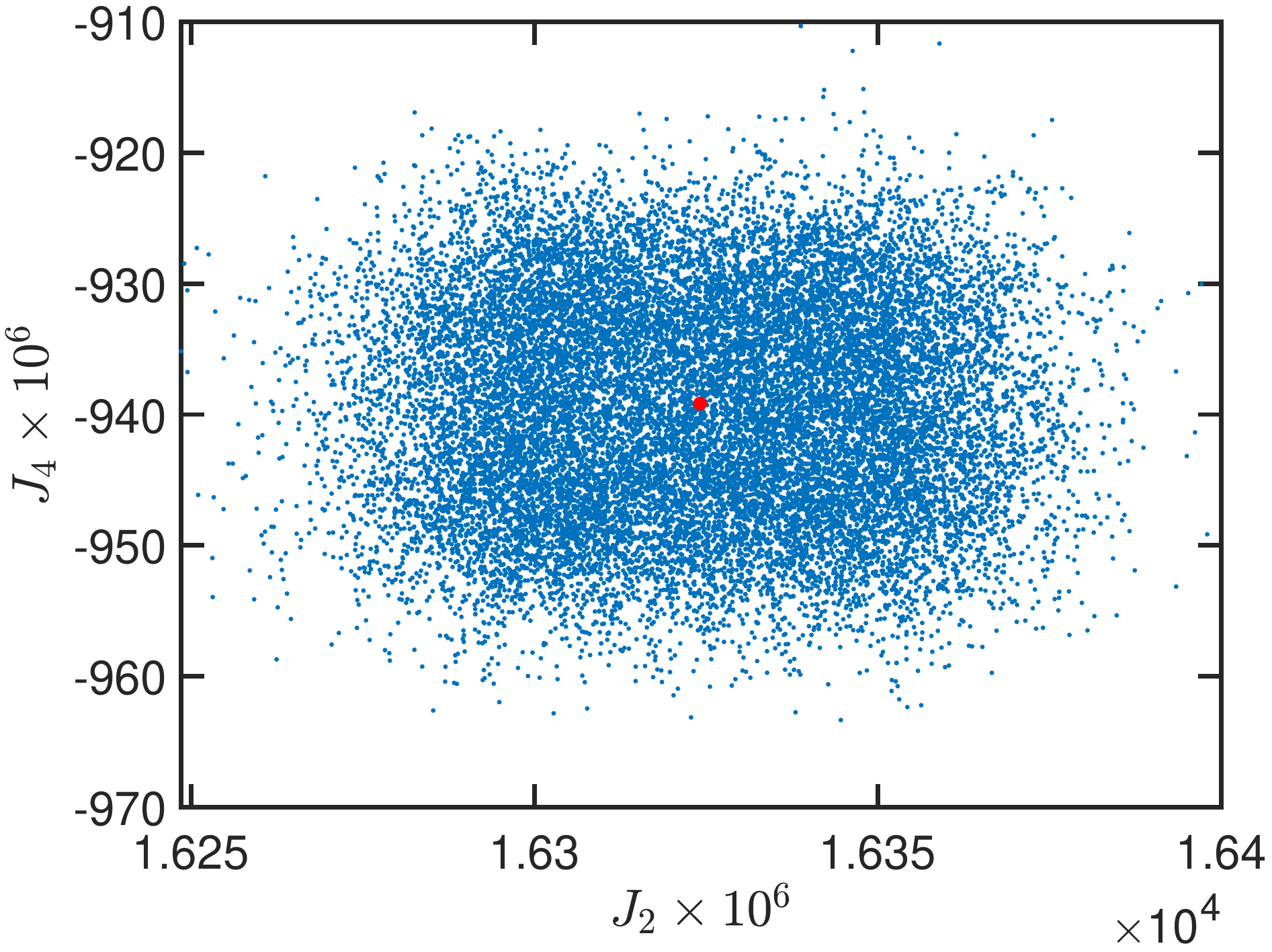}{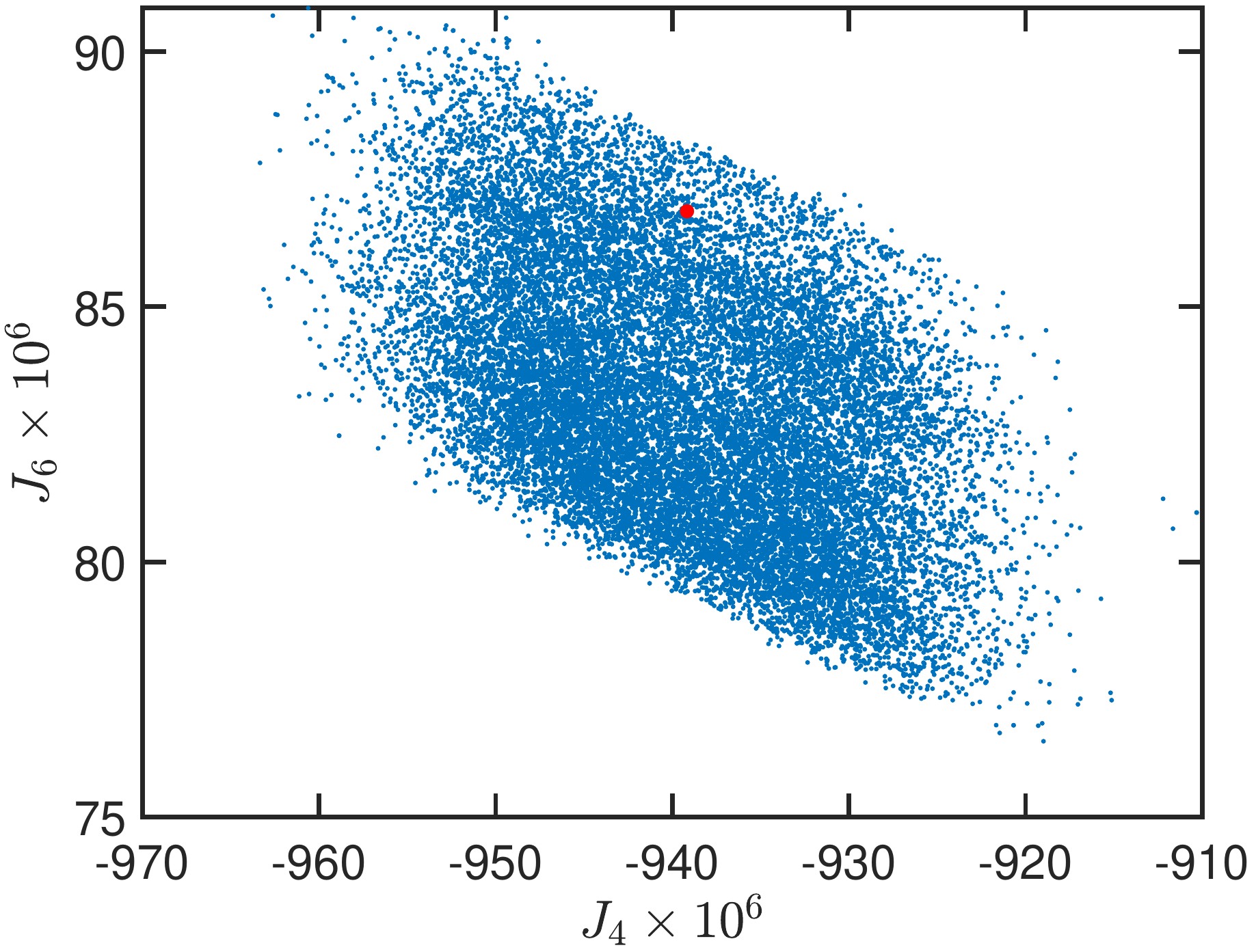}
\caption{Gravitational harmonics of the empirical models of
Fig.~\ref{fig:rhoposterior} of the main text. The coefficients for Saturn's
observed gravity \citep{Iess2019} are indicated by a red circle.}
\label{fig:JJscat}
\end{figure*}

Finally, the distribution of empirical models from our sample in the
$J_2\text{-}J_4$ and $J_4\text{-}J_6$ planes is shown in Figure~\ref{fig:JJscat}.
In many previous works that use the gravity field to study the planetary interior
this is a central result and a similar plot would be a prominent figure in the
main text. In the traditional modeling approach this is a useful indication of how
variation of model parameters (which in traditional models have important physical
meaning) translates to variation in the model's gravity. In our empirical,
MCMC-driven study however this distribution is much less informative. Recall that
the sampling algorithm is driven by a likelihood function that compares model
values of $J_i$ with observed values. Unless there is a bug in the implementation,
the $J_i$ distribution in the final sample is determined entirely by the choice of
likelihood function and tells us nothing about the underlying model. Nonetheless,
we include this figure to potentially help a direct comparison with past or future
investigations.

\bibliographystyle{aasjournal} 
\bibliography{library}
\end{document}